\documentclass[12pt,english]{article}
\usepackage{times}
\usepackage[T1]{fontenc}
\usepackage{geometry}
\usepackage{color}
\usepackage{graphicx}
\usepackage{setspace}
\usepackage{amssymb}

\makeatletter

\providecommand{\tabularnewline}{\\}

\usepackage{bm}

\usepackage{babel}
\makeatother
\begin{document}

\section*{\noindent \textcolor{black}{Pareto-Optimal Domino-Tiling of Orthogonal
Polygon Phased Arrays}}

\noindent \textcolor{black}{\vfill}

\noindent \textcolor{black}{P. Rocca,$^{(1)(2)}$} \textcolor{black}{\emph{Senior
Member, IEEE}}\textcolor{black}{, N. Anselmi,$^{(1)}$} \textcolor{black}{\emph{Member,
IEEE}}\textcolor{black}{, A. Polo,$^{(1)}$} \textcolor{black}{\emph{Member,
IEEE}}\textcolor{black}{, and A. Massa,$^{(1)(3)(4)}$} \textcolor{black}{\emph{Fellow,
IEEE}}

\noindent \textcolor{black}{\vfill}

\noindent \textcolor{black}{\footnotesize $^{(1)}$} \textcolor{black}{\emph{\footnotesize CNIT}}
\textcolor{black}{\footnotesize - \char`\"{}University of Trento\char`\"{}
Research Unit}{\footnotesize \par}

\noindent \textcolor{black}{\footnotesize Via Sommarive 9, 38123 Trento
- Italy}{\footnotesize \par}

\noindent \textit{\textcolor{black}{\emph{\footnotesize E-mail:}}}
\textcolor{black}{\footnotesize \{}\textcolor{black}{\emph{\footnotesize nicola.anselmi.1,
paolo.rocca}}\textcolor{black}{\footnotesize ,} \textcolor{black}{\emph{\footnotesize andrea.massa}}\textcolor{black}{\footnotesize \}@}\textcolor{black}{\emph{\footnotesize unitn.it}}{\footnotesize \par}

\noindent \textcolor{black}{\footnotesize Website:} \textcolor{black}{\emph{\footnotesize www.eledia.org/eledia-unitn}}{\footnotesize \par}

\noindent \textcolor{black}{\footnotesize ~}{\footnotesize \par}

\noindent \textcolor{black}{\footnotesize $^{(2)}$} \textcolor{black}{\emph{\footnotesize ELEDIA
Research Center}} \textcolor{black}{\footnotesize (}\textcolor{black}{\emph{\footnotesize ELEDIA@XIDIAN}}
\textcolor{black}{\footnotesize - Xidian University)}{\footnotesize \par}

\noindent \textcolor{black}{\footnotesize P.O. Box 191, No.2 South
Tabai Road, 710071 Xi'an, Shaanxi Province - China}{\footnotesize \par}

\noindent \textcolor{black}{\footnotesize E-mail:} \textcolor{black}{\emph{\footnotesize paolo.rocca@xidian.edu.cn}}{\footnotesize \par}

\noindent \textcolor{black}{\footnotesize Website:} \textcolor{black}{\emph{\footnotesize www.eledia.org/eledia-xidian}}{\footnotesize \par}

\noindent \textcolor{black}{\footnotesize ~}{\footnotesize \par}

\noindent \textcolor{black}{\footnotesize $^{(3)}$} \textcolor{black}{\emph{\footnotesize ELEDIA
Research Center}} \textcolor{black}{\footnotesize (}\textcolor{black}{\emph{\footnotesize ELEDIA}}\textcolor{black}{\footnotesize @}\textcolor{black}{\emph{\footnotesize UESTC}}
\textcolor{black}{\footnotesize - UESTC)}{\footnotesize \par}

\noindent \textcolor{black}{\footnotesize School of Electronic Engineering,
Chengdu 611731 - China}{\footnotesize \par}

\noindent \textit{\textcolor{black}{\emph{\footnotesize E-mail:}}}
\textcolor{black}{\emph{\footnotesize andrea.massa@uestc.edu.cn}}{\footnotesize \par}

\noindent \textcolor{black}{\footnotesize Website:} \textcolor{black}{\emph{\footnotesize www.eledia.org/eledia}}\textcolor{black}{\footnotesize -}\textcolor{black}{\emph{\footnotesize uestc}}{\footnotesize \par}

\noindent \textcolor{black}{\footnotesize ~}{\footnotesize \par}

\noindent \textcolor{black}{\footnotesize $^{(4)}$} \textcolor{black}{\emph{\footnotesize ELEDIA
Research Center}} \textcolor{black}{\footnotesize (}\textcolor{black}{\emph{\footnotesize ELEDIA@TSINGHUA}}
\textcolor{black}{\footnotesize - Tsinghua University)}{\footnotesize \par}

\noindent \textcolor{black}{\footnotesize 30 Shuangqing Rd, 100084
Haidian, Beijing - China}{\footnotesize \par}

\noindent \textcolor{black}{\footnotesize E-mail:} \textcolor{black}{\emph{\footnotesize andrea.massa@tsinghua.edu.cn}}{\footnotesize \par}

\noindent \textcolor{black}{\footnotesize Website:} \textcolor{black}{\emph{\footnotesize www.eledia.org/eledia-tsinghua}}{\footnotesize \par}

\noindent \textcolor{black}{~}

\noindent \textcolor{black}{\vfill}

This work has been submitted to the IEEE for possible publication.
Copyright may be transferred without notice, after which this version
may no longer be accessible.

\newpage
\section*{\textcolor{black}{Pareto-Optimal Domino-Tiling of Orthogonal Polygon
Phased Arrays}}

\textcolor{black}{~}

\textcolor{black}{~}

\textcolor{black}{~}

\begin{flushleft}\textcolor{black}{P. Rocca, N. Anselmi, A. Polo,
and A. Massa}\end{flushleft}

\textcolor{black}{\vfill}

\begin{abstract}
\noindent \textcolor{black}{The modular design of planar phased arrays
arranged on orthogonal polygon-shaped apertures is addressed and a
new method is proposed to synthesize domino-tiled arrays fitting multiple,
generally conflicting, requirements. Starting from an analytic procedure
to check the domino-tileability of the aperture, two multi-objective
optimization techniques are derived to efficiently and effectively
deal with small and medium/large arrays depending on the values of
the bounds for the cardinality of the solution space of the admissible
clustered solutions. A set of representative numerical examples is
reported to assess the effectiveness of the proposed synthesis approach
also through full-wave simulations when considering non-ideal models
for the radiating elements of the array.}

\textcolor{black}{\vfill}
\end{abstract}
\noindent \textbf{\textcolor{black}{Key words}}\textcolor{black}{:
Phased Array Antenna, Planar Array, Orthogonal Polygon-Shaped Apertures,
Irregular Tiling, Domino Tiles, Multi-Objective Optimization.}

\newpage
\section{\textcolor{black}{Introduction}}

\noindent \textcolor{black}{Modern wireless applications need antenna
systems with fast scanning capabilities, accurate beam pointing, interference
rejection, and multiple beams. Phased arrays (}\textcolor{black}{\emph{PA}}\textcolor{black}{s)
fit these requirements \cite{Elliott.2003}-\cite{Mailloux.2018}
being able to simultaneously accomplish different and complex tasks
in a reliable manner as well as to quickly adapt to the surrounding
environment. Therefore, they represent a key-enabling technology for
developing a variety of civil applications like 5G communications}
\textbf{\textcolor{black}{}}\textcolor{black}{\cite{Oliveri.2017}-\cite{Zhang.2020},}
\textbf{\textcolor{black}{}}\textcolor{black}{autonomous driving}
\textbf{\textcolor{black}{}}\textcolor{black}{\cite{Ku.2014}-\cite{Hamberger.2019},
next generation weather and air traffic control radar systems \cite{Herd.2010}-\cite{Diaz.2018}
just to mention a few. Accordingly,} \textcolor{black}{\emph{PA}}\textcolor{black}{s
are currently deployed in cars, drones, and 5G handsets with severe
constraints on the antenna aperture, which is usually irregular, due
to the limited space and the integration with other co-located electronic
devices (e.g., microchips and micro-controllers that are usually used
to implement high levels of digitization). On the other hand, classical}
\textcolor{black}{\emph{PA}} \textcolor{black}{design techniques are
mainly concerned with regular apertures with rectangular, square,
or circular shapes. Consequently, the development of efficient and
reliable} \textcolor{black}{\emph{ad-hoc}} \textcolor{black}{methods
for synthesizing} \textcolor{black}{\emph{PA}}\textcolor{black}{s
arranged on irregular regions is strongly suggested to fully exploit
the available area and to guarantee the synthesis of optimal trade-offs
between radiation performance, complexity, and costs. In the last
years, several unconventional architectures have been proposed and
suitable} \textcolor{black}{\emph{PA}} \textcolor{black}{design methodologies
have been implemented (see \cite{Rocca.2016} and the reference therein)
to yield cost-effective and high-performance antenna array solutions
\cite{Herd.2016}. In such a framework, sub-arrayed arrangements of
the array elements have attracted a non-negligible attention thanks
to their capabilities to fit challenging radiation requirements by
implementing high-performance radiating systems with affordable costs.
Indeed, sub-arrayed} \textcolor{black}{\emph{PA}}\textcolor{black}{s
implement the antenna functionalities at the sub-array level by reducing
the number of radio-frequency chains and transmit-received modules
(}\textcolor{black}{\emph{TRM}}\textcolor{black}{s) with respect to
classical/fully-populated arrays having complete beam-forming networks.
Irregular clustering methods, which use either sub-arrays of arbitrary
shapes and/or sizes \cite{Lopez.2001}-\cite{Anselmi.2018.b} or tiles
with pre-defined shapes and sizes \cite{Mailloux.2009}-\cite{Ma.2019},
have been widely studied. Such techniques define an irregular organization
of the array clusters in the antenna aperture, which results in an
aperiodic distribution of the sub-array phase centers, to reduce the
presence and the level of undesired quantization lobes that severely
affect the performance of} \textcolor{black}{\emph{PA}}\textcolor{black}{s
when the beam scanning and the operation bandwidth increase \cite{Mailloux.2018}.
Recently, the synthesis of irregular tiled arrays that cover the antenna
aperture to maximize the antenna efficiency and its directivity, subject
to the antenna size, has been addressed with enumerative \cite{Xiong.2013}\cite{Dong.2019}
and analytic-driven optimization-based \cite{Anselmi.2017}-\cite{Rocca.2020.a}
strategies. Theorems on the tileability of the array aperture with
simple tile shapes, such as dominoes \cite{Anselmi.2017}, diamonds
\cite{Rocca.2020.a}, or two-sizes squares \cite{Rocca.2020.b}, have
been stated and succesfully exploited to derive effective uniform
clustering methods. However, they are limited to regular apertures
(e.g., rectangular \cite{Anselmi.2017}\cite{Rocca.2020.b} and hexagonal
\cite{Rocca.2020.a}) and tile shapes (e.g., dominoes \cite{Anselmi.2017},
diamonds \cite{Rocca.2020.a}, and squares \cite{Rocca.2020.b}). }

\noindent \textcolor{black}{This work faces with the first limitation
of current tiling methods by proposing a novel method for designing
domino-tiled planar} \textcolor{black}{\emph{PA}}\textcolor{black}{s
arranged on arbitrary orthogonal polygon apertures. First, the} \textcolor{black}{\emph{height
function}}\textcolor{black}{, which has been introduced in \cite{Thurston.1990}
for efficiently coding a domino-based array arrangement, is here exploited
to derive the condition on the tileability of the antenna aperture
by recurring to a proper application of the theorems in \cite{Fournier.1996}\cite{Fournier.1997}.
It is worth pointing out the relevance of this result for modern wireless
systems where there are hard limitations on the antenna dimensions
and thus the need of guaranteeing the maximum aperture efficiency.
Furthermore, starting from the approach in \cite{Anselmi.2017}, a
new design method based on two multi-objective optimization techniques
is here proposed to synthesize domino-tiled configurations of planar}
\textcolor{black}{\emph{PA}}\textcolor{black}{s with arbitrary orthogonal
polygon apertures fitting multiple conflicting antenna requirements
\cite{Deb.2001}-\cite{Nagar.2018}.}

\noindent \textcolor{black}{The paper is organized as follows. Section
2 describes the formulation of the tiling synthesis problem at hand
and the corresponding tileability theorems. Two versions of a multi-objective
design method are then presented in Sect. 3, while Section 4 is devoted
to a numerical validation of the proposed synthesis approach along
with a performance assessment. Eventually, some conclusions are drawn
(Sect. 5).}

\section{\noindent \textcolor{black}{Mathematical Formulation\label{sec:Mathematical-Formulation}}}

\noindent \textcolor{black}{Let us consider a planar (}\textcolor{black}{\emph{2D}}\textcolor{black}{)}
\textcolor{black}{\emph{PA}} \textcolor{black}{with $P$ elements,
each one belonging to a square unit cell (i.e., a} \textcolor{black}{\emph{pixel}}\textcolor{black}{)}%
\footnote{\noindent \textcolor{black}{For the sake of notation simplicity, each
pixel is hereinafter supposed to include only one radiating element.
However, it is worth noticing that this is not an hypothesis for the
proposed tiling method since each {}``cell'' is a logical unit and,
physically, it can include one or more radiating elements without
any loss of generality of the underlying theoretical formulation.}%
}\textcolor{black}{. The ensemble of the $P$ unit cells defines a
simply-connected region}%
\footnote{\noindent \textcolor{black}{A simply-connected region is a domain
in which any simple closed curve can continuously shrink into a point
while remaining in the same domain. As for the} \textcolor{black}{\emph{2D}}
\textcolor{black}{case, a simply-connected domain is an area without
holes.}%
} \textcolor{black}{in the $x-y$ plane so that the array aperture
$\mathcal{O}$ turns out to be an orthogonal polygon defined as a
polygon where every edge of the contour is either horizontal (i.e.,
along the $x$ direction) or vertical (i.e., along the $y$ direction)
\cite{Su.2005}. Moreover, the inter-element spacing between the centers
of two adjacent pixel cells along the $x$-axis and the $y$-axis
is equal to $d_{x}$ ($x$-axis) and $d_{y}$ ($y$-axis) {[}Fig.
1(}\textcolor{black}{\emph{a}}\textcolor{black}{){]}, respectively,
while the maximum number of unit cells along the same orthogonal directions
is $N$ and $M$ {[}Fig. 1(}\textcolor{black}{\emph{b}}\textcolor{black}{){]}.
The array elements are clustered in $Q$ ($Q\triangleq\frac{P}{2}$)
elementary {}``bricks'', $\bm{\sigma}=\left\{ \sigma_{q};\, q=1,...,Q\right\} $,
which are denoted as} \textcolor{black}{\emph{tiles}}\textcolor{black}{,
having domino shapes and grouping two neighbouring pixels sharing
one side (Fig. 1). Since the manufacturing of a single tile is doable
only if the} \textcolor{black}{\emph{EM}} \textcolor{black}{field
generated by the $p$-th ($p=1,...,P$) elementary radiator, $E_{p}\left(\theta,\phi\right)$
, is invariant with respect to a $90$ {[}deg{]} rotation with respect
to the $z$-axis, two types of dominoes are used. More specifically,
the tile shapes are a vertical domino, $\sigma_{q}=\sigma^{V}$, and
a horizontal one, $\sigma_{q}=\sigma^{H}$, that combine two adjacent
pixels sharing a horizontal or a vertical edge, respectively. Each
$q$-th ($q=1,...,Q$) tile is controlled by an amplifier and a phase
shifter to weight in amplitude, $\alpha_{q}$, and/or phase, $\beta_{q}$,
the received/transmitted signal at the sub-array level {[}Fig. 1(}\textcolor{black}{\emph{a}}\textcolor{black}{){]}.}

\noindent \textcolor{black}{The far-field pattern radiated by the
array is then given by\begin{equation}
P\left(\theta,\phi\right)=\left|\sum_{q=1}^{Q}\alpha_{q}\left\{ \sum_{p=1}^{P}\delta_{c_{p}q}E_{p}\left(\theta,\phi\right)e^{j\frac{2\pi}{\lambda}\left(x_{p}\sin\theta\cos\phi+y_{p}\sin\theta\sin\phi\right)}\right\} e^{j\beta_{q}}\right|^{2}\label{eq:_array.factor}\end{equation}
where $E_{p}\left(\theta,\phi\right)$ is the} \textcolor{black}{\emph{embedded/active}}\textcolor{black}{-}\textcolor{black}{\emph{element
pattern}} \textcolor{black}{\cite{Haupt.2010}\cite{Mailloux.2018}
of the $p$-th ($p=1,...,P$) array element, $\lambda$ is the wavelength,
$\left(\theta,\phi\right)$ are the angular coordinates, $\left(x_{p},\, y_{p}\right)$
are the Cartesian coordinates of the center of the $p$-th ($p=1,...,P$)
unit cell of the array, while $\delta_{c_{p}q}$ is the Kronecker
delta function ($\delta_{c_{p}q}=1$ if $c_{p}=q$ or $\delta_{c_{p}q}=0$
when $c_{p}\neq q$), $c_{p}$ being the integer index ($c_{p}\in\left[1,\, Q\right]$)
whose value indicates the membership of the $p$-th ($p=1,...,P$)
array element to the $q$-th ($q=1,...,Q$) tile {[}Fig. 1(}\textcolor{black}{\emph{b}}\textcolor{black}{){]}.
Moreover, let us define the vectors $\bm{\alpha}=\left\{ \alpha_{q};\, q=1,...,Q\right\} $,
$\bm{\beta}=\left\{ \beta_{q};\, q=1,...,Q\right\} $, and $\mathbf{c}=\left\{ c_{p};\, p=1,...,P\right\} $
for notation simplicity. Accordingly, the array tiling problem addressed
in this work can be formulated as follows:}

\begin{quote}
\textcolor{black}{\emph{Array Tiling Synthesis}} \textcolor{black}{-
Given an array of $M\times N$ elements displaced on an antenna aperture
$\mathcal{O}$ with orthogonal polygon shape, find the optimal clustering
of the array elements, $\mathbf{c}^{opt}$ (i.e., a complete tessellation
of $\mathcal{O}$), by using $Q$ domino tiles and the corresponding
sub-array amplitudes, $\bm{\alpha}^{opt}$, and phases, $\bm{\beta}^{opt}$,
such that the radiated power pattern $P^{opt}\left(\theta,\phi\right)$
fits a set of $K$ user-defined requirements.}
\end{quote}
\noindent \textcolor{black}{To address such a} \textcolor{black}{\emph{Synthesis
Problem}}\textcolor{black}{, an innovative design strategy is proposed
(Sect. 3) that makes use of an analytic procedure to check the domino-tileability
of the aperture $\mathcal{O}$ as well as to derive useful bounds
on the cardinality of the solution space of the admissible clustering
configurations. Those latter items are detailed in the following.}

\subsection{\textcolor{black}{Tileability Theorem and Cardinality Bounds\label{sub:Tileability-Theorem-and}}}

\noindent \textcolor{black}{Given an orthogonal polygon-shaped aperture
$\mathcal{O}$, the analytic procedure for} \textcolor{black}{\emph{a-priori}}
\textcolor{black}{determining whether the aperture is fully domino-tileable
is based on the} \textcolor{black}{\emph{height function}}\textcolor{black}{,
which has been firstly introduced in \cite{Thurston.1990} and then
used in \cite{Desreux.2006} to develop an efficient algorithm for
the exhaustive generation of all the domino-tiles configurations completely
covering a surface. Such an approach has been exploited in \cite{Anselmi.2017}
for the domino-tiling of rectangular arrays, but here it is non-trivially
extended and generalized to more complex and arbitrary orthogonal
polygon antenna apertures. }

\noindent \textcolor{black}{Towards this end, let us consider the
set of the vertices of the $P$ unit cells of the aperture $\mathcal{O}$,
$\bm{v}=\left\{ v_{g};\, g=1,...,G\right\} $, {[}Fig. 2(}\textcolor{black}{\emph{a}}\textcolor{black}{){]}
composed by boundary, $\bm{b}=\left\{ b_{s};\, s=1,...,S\right\} $
($b_{s}\in\partial\mathcal{O}$, $\partial\mathcal{O}$ being the
boundary of $\mathcal{O}$), and internal, $\bm{a}=\left\{ a_{l};\, l=1,...,L\right\} $
($a_{l}\notin\partial\mathcal{O}$), elements {[}Fig. 2(}\textcolor{black}{\emph{b}}\textcolor{black}{){]}
so that $\bm{v}=\bm{b}\cup\bm{a}$. Moreover, a chessboard pattern
for the $P$ pixels in $\mathcal{O}$ is considered {[}Fig. 2(}\textcolor{black}{\emph{c}}\textcolor{black}{){]}
and the notation $v_{g}\rightarrow v_{e}$ ($v_{g}\leftarrow v_{e}$)
is used to indicate an edge oriented from $v_{g}$ towards $v_{e}$
(vice-versa) and connecting the two adjacent vertices $v_{g}$ and
$v_{e}$ of the set $\bm{v}$ {[}Fig. 2(}\textcolor{black}{\emph{c}}\textcolor{black}{){]}. }

\noindent \textcolor{black}{The value of the height function, $\psi$,
for the boundary vertices ($\bm{\psi}\triangleq\left\{ \psi\left(b_{s}\right);\, s=1,...,S\right\} $)
is then computed by setting $\psi\left(b_{1}\right)=0$ {[}Fig. 2(}\textcolor{black}{\emph{c}}\textcolor{black}{){]}
and applying the following rule\begin{equation}
\psi\left(b_{s}\right)=\left\{ \begin{array}{ll}
\psi\left(b_{s-1}\right)+1 & \;\;\; if\;\left(b_{s-1}\rightarrow b_{s}\right)\\
\psi\left(b_{s-1}\right)-1 & \;\;\; if\;\left(b_{s-1}\leftarrow b_{s}\right)\end{array}\right.\label{eq:_heigh_function_on_boundary}\end{equation}
for the remaining external entries ($s>1$).}

\noindent \textcolor{black}{Furthermore, the boundary set $\bm{b}$
is also described by the corresponding $S$ values of the} \textcolor{black}{\emph{auxiliary
function}} \textcolor{black}{$\chi$ \cite{Fournier.1996}, $\left\{ \chi\left(b_{s}\right);\, s=1,...,S\right\} $,
being\begin{equation}
\chi\left(b_{s}\right)=\min_{b_{g}\in\partial\mathcal{O}}\left[\gamma\left(b_{g},b_{s}\right)\right]\label{eq:_height_function_auxiliary}\end{equation}
where $\gamma\left(b_{g},b_{s}\right)\triangleq\psi\left(b_{g}\right)+\Delta\left(b_{g},b_{s}\right)$
and $\Delta\left(b_{g},b_{s}\right)$ is the number of edges of the
shortest path from the boundary edge $b_{g}$ to the other one $b_{s}$
through the graph $\mathcal{G}$ {[}Fig. 2(}\textcolor{black}{\emph{c}}\textcolor{black}{){]},
$\mathcal{G}$ being a directed graph built by connecting the neighboring
elements of the vertices set $\bm{v}$ and with edges that are oriented
clockwise/counter-clockwise if they belong to a white/black pixel
{[}Fig. 2(}\textcolor{black}{\emph{c}}\textcolor{black}{){]}.}

\noindent \textcolor{black}{The array aperture $\mathcal{O}$ turns
out to be fully domino-tileable if the following condition (}\textcolor{black}{\emph{Orthogonal-Polygon
Tileability}}\textcolor{black}{) holds true \cite{Fournier.1996}:\begin{equation}
\psi\left(b_{s}\right)=\chi\left(b_{s}\right),\,\, s=1,...,S.\label{eq:_tilability_condition}\end{equation}
For instance, the array aperture $\mathcal{O}$ in Fig. 2(}\textcolor{black}{\emph{c}}\textcolor{black}{)
fulfils (\ref{eq:_tilability_condition}) as pictorially shown in
Fig. 1, while the one in Fig. 3 does not comply the tileability condition
even though it has an equal number of white and black pixels. As a
representative example, let us check (\ref{eq:_tilability_condition})
for the border vertex $b_{8}$. Since the element of $\bm{b}$ that
provides the minimum value of $\gamma$ is the boundary vertex $v_{5}=b_{17}$,
it turns out that $\chi\left(b_{8}\right)=1$ being $\psi\left(b_{17}\right)=-2$
and $\Delta\left(b_{17},b_{8}\right)=3$. Differently, $\psi\left(b_{8}\right)=5$
and therefore the region $\mathcal{O}$ of Fig. 3 is not fully-tileable
with dominoes.}

\noindent \textcolor{black}{Of course, the tileability condition is
a pre-requisite for starting the design of an irregular subarrayed}
\textcolor{black}{\emph{PA}} \textcolor{black}{with a user-defined
tile shape, but certainly an estimate of the dimension of the solution
space of the admissible set of complete tiled arrangements is very
important, as well. Indeed, the {}``knowledge'' of the number $T$
of existing domino tilings is a key-stone for defining/choosing the
most suitable synthesis strategy to determine the optimal solution
fitting the} \textcolor{black}{\emph{PA}} \textcolor{black}{requirements.
Unlike rectangular apertures \cite{Anselmi.2017}\cite{Kasteleyn.1961},
there is no closed-form relation for $T$ when dealing with arbitrary
apertures, but it is still possible to estimate the problem cardinality
by defining an upper, $T_{u}$, and a lower, $T_{l}$, bound of the
number of domino arrangements that fully cover the aperture $\mathcal{O}$.}

\noindent \textcolor{black}{The upper bound value, $T_{u}$, is equal
to the number of tiling configurations fully covering the smallest
regular $M\times N$ rectangle including the orthogonal polygon $\mathcal{O}$
\cite{Anselmi.2017}, $\tau\left(M,N\right)$, which depends on the
array dimensions, $M$ and $N$, as follows \begin{equation}
\tau\left(M,N\right)=2^{\frac{MN}{2}}\prod_{m=1}^{M}\prod_{n=1}^{N}\left[cos^{2}\left(\frac{\pi m}{M+1}\right)+cos^{2}\left(\frac{\pi n}{N+1}\right)\right]^{1/4},\label{eq:_number_tilings_upper_bound}\end{equation}
while the lower bound, $T_{l}$, is given by\begin{equation}
T_{l}=\sum_{j=1}^{J}\tau\left(M_{j},N_{j}\right)\label{eq:_number_tilings_lower_bound}\end{equation}
$M_{j}$ and $N_{j}$ ($j=1,...,J$) being the dimensions of $J$
disjoint rectangles covering $\mathcal{O}$ {[}Fig. 4(}\textcolor{black}{\emph{a}}\textcolor{black}{){]}.
To give a pictorial indication on the dimension of the solution space
of the tiling problem at hand, the behaviour of the $T$ bounds for
the orthogonal polygon shape in Fig. 4(}\textcolor{black}{\emph{a}}\textcolor{black}{)
when setting $N=12$ and varying $M$ within the range $6\le M\le30$
is shown in Fig. 4(}\textcolor{black}{\emph{b}}\textcolor{black}{).}

\section{\textcolor{black}{Orthogonal Polygon Array Design Methods\label{sec:Orthogonal-Polygon-Array}}}

\noindent \textcolor{black}{The array tiling problem formulated in
Sect. \ref{sec:Mathematical-Formulation} is intrinsically a multi-objective
synthesis problem (}\textcolor{black}{\emph{MOP}}\textcolor{black}{)
where $K$, generally conflicting, requirements have to be satisfied.
Mathematically, {}``the solution'' of such a} \textcolor{black}{\emph{MOP}}
\textcolor{black}{turns out to be a Pareto-front (}\textcolor{black}{\emph{PF}}\textcolor{black}{)
of optimal (i.e.,} \textcolor{black}{\emph{non-dominated}}\textcolor{black}{)
trade-off solutions \cite{Pareto.1986} fitting the design specifications
coded into $K$ cost function terms, $\Phi_{k}\left(\mathbf{c},\bm{\alpha},\bm{\beta}\right)$,
$k=1,...,K$. Towards this end, a synthesis strategy, which is based
on two multi-objective optimization techniques, for the tiling of
orthogonal polygon arrays is here presented.}

\subsection{\textcolor{black}{\emph{EPF}} \textcolor{black}{Method (}\textcolor{black}{\emph{EPFM}}\textcolor{black}{)\label{sub:EPF-Method-(EPFM)}}}

\textcolor{black}{Once checked the} \textcolor{black}{\emph{Orthogonal
Polygon Tileability condition}} \textcolor{black}{to guarantee the
tileability of the aperture $\mathcal{O}$, the first tiling method
allows one to fully determine the} \textcolor{black}{\emph{PF}} \textcolor{black}{of
the trade-off solutions when the synthesis problem at hand is computationally
affordable (i.e., the} \textcolor{black}{\emph{CPU}}\textcolor{black}{-time
for processing at most $T_{u}$ admissible domino arrangements is
reasonable/moderate) thanks to an exhaustive generation of all ($T$)
possible tiling configurations. More in detail, the} \textcolor{black}{\emph{Exact
PF Method}} \textcolor{black}{(}\textcolor{black}{\emph{EPFM}}\textcolor{black}{)
implements the following procedural steps:}

\begin{itemize}
\item \textbf{\textcolor{black}{Step 0}} \textcolor{black}{-} \textcolor{black}{\emph{Reference
Array Definition}} \textcolor{black}{- Determine the set of the amplitude,
$\bm{\alpha}^{ref}$ $=$ \{$\alpha_{p}^{ref}$; $p=1,...,P$\}, and
the phase, $\bm{\beta}^{ref}$ $=$ \{$\beta_{p}^{ref}$; $p=1,...,P$\},
coefficients of a fully-populated array affording a pattern compliant
with the design requirements;}
\item \textbf{\textcolor{black}{Step 1}} \textcolor{black}{-} \textcolor{black}{\emph{Tilings
Generation}} \textcolor{black}{- Starting from the knowledge of the
$S$-size height function vector, $\bm{\psi}$, determine the} \textcolor{black}{\emph{minimal}}
\textcolor{black}{tiling $\mathbf{c}^{\left(1\right)}$ \cite{Anselmi.2017}.
Generate the whole set of tilings, $\mathbf{C}$ ($\mathbf{C}\triangleq\left\{ \mathbf{c}^{\left(t\right)};\, t=1,...,T\right\} $),
by using the Enumerative Tiling Method (}\textcolor{black}{\emph{ETM}}\textcolor{black}{)
\cite{Anselmi.2017} to iteratively yield the $t$-th ($t=2,...,T$)
domino arrangement, $\mathbf{c}^{\left(t\right)}$, from the previous
one, $\mathbf{c}^{\left(t-1\right)}$;}
\item \textbf{\textcolor{black}{Step 2}} \textcolor{black}{-} \textcolor{black}{\emph{Excitations
Computation}} \textcolor{black}{- For each $t$-th ($t=1,...,T$)
tiling configuration compute the sub-array amplitudes, $\bm{\alpha}^{\left(t\right)}$,
and phase, $\bm{\beta}^{\left(t\right)}$, weights as follows\begin{equation}
\left(\begin{array}{c}
\alpha_{q}^{\left(t\right)}\\
\beta_{q}^{\left(t\right)}\end{array}\right)=\frac{1}{2}\sum_{p=1}^{P}\left(\begin{array}{c}
\alpha_{p}^{ref}\\
\beta_{p}^{ref}\end{array}\right)\delta_{c_{p}q}\label{eq:_sub-array.amplitudes-phases_EM}\end{equation}
($q=1,...,Q$);}
\item \textbf{\textcolor{black}{Step 3}} \textcolor{black}{-} \textcolor{black}{\emph{Tilings
Evaluation}} \textcolor{black}{- For each $t$-th ($t=1,...,T$) array
clustering, evaluate the $K$-size cost function vector $\bm{\Phi}^{\left(t\right)}$
($\bm{\Phi}^{\left(t\right)}$ $\triangleq$ \{ $\Phi_{k}\left(\mathbf{c}^{\left(t\right)},\,\bm{\alpha}^{\left(t\right)},\:\bm{\beta}^{\left(t\right)}\right)$;
$k=1,...,K$\};}
\item \textbf{\textcolor{black}{Step 4}} \textcolor{black}{-} \textcolor{black}{\emph{EPF
Definition}} \textcolor{black}{- For each ($t$,$z$)-th ($t,\, z=1,...,T$,
$z\neq t$) couple of tilings, $\left(\mathbf{c}^{\left(t\right)},\,\mathbf{c}^{\left(z\right)}\right)$,
select the} \textcolor{black}{\emph{non-dominated}} \textcolor{black}{\cite{Deb.2001}
solution $\mathbf{c}^{\left(f\right)}$, that is, $\mathbf{c}^{\left(f\right)}=\mathbf{c}^{\left(t\right)}$
if }%
\footnote{\noindent \textcolor{black}{Without loss of generality, the} \textcolor{black}{\emph{dominance
condition}} \textcolor{black}{is here referred to a design problem
where all the $K$ cost function terms, \{$\Phi_{k}\left(\mathbf{c}^{\left(t\right)}\right)$;
$k=1,...,K$\}, have to be minimized.}%
}

\textcolor{black}{\[
\Phi_{k}\left(\mathbf{c}^{\left(t\right)},\,\bm{\alpha}^{\left(t\right)},\:\bm{\beta}^{\left(t\right)}\right)\leq\Phi_{k}\left(\mathbf{c}^{\left(z\right)},\,\bm{\alpha}^{\left(z\right)},\:\bm{\beta}^{\left(z\right)}\right)\]
($k=1,...,K$) and it exists a $h$-th ($h\in[1,K]$) cost function
term such that\begin{equation}
\Phi_{h}\left(\mathbf{c}^{\left(t\right)},\,\bm{\alpha}^{\left(t\right)},\:\bm{\beta}^{\left(t\right)}\right)<\Phi_{h}\left(\mathbf{c}^{\left(z\right)},\,\bm{\alpha}^{\left(z\right)},\:\bm{\beta}^{\left(z\right)}\right)\label{eq:_Pareto_dominance}\end{equation}
or $\mathbf{c}^{\left(f\right)}=\mathbf{c}^{\left(z\right)}$, otherwise;}

\item \textbf{\textcolor{black}{Step 5}} \textcolor{black}{-} \textcolor{black}{\emph{Optimal
Tiling Computation}} \textcolor{black}{- Starting from the knowledge
of the set of $F$ optimal trade-off tilings belonging to the} \textcolor{black}{\emph{EPF,}}
\textcolor{black}{$\mathbf{C}^{PF}$ ($\mathbf{C}^{PF}\triangleq\left\{ \mathbf{c}^{\left(f\right)};\, f=1,...,F\right\} $),
choose as $\mathbf{c}^{opt}$ the domino arrangement $\mathbf{c}^{MMD}$
that minimizes the} \textcolor{black}{\emph{Minimum Manhattan Distance}}
\textcolor{black}{(}\textcolor{black}{\emph{MMD}}\textcolor{black}{)
\cite{Deb.2001} ($\mathbf{c}^{opt}=\mathbf{c}^{MMD}$)\begin{equation}
\left(\mathbf{c}^{MMD},\bm{\alpha}^{MMD},\bm{\beta}^{MMD}\right)=\arg\left[\min_{f=1,...,F}\left\{ \left\Vert \bm{\Phi}^{\left(f\right)}-\bm{\Phi}^{\left(ideal\right)}\right\Vert _{1}\right\} \right]\label{eq:_MMD.1}\end{equation}
where $\left\Vert \cdot\right\Vert _{1}$ stands for the L1 norm,
while $\bm{\Phi}^{\left(f\right)}$ ($\bm{\Phi}^{\left(f\right)}\triangleq\left\{ \hat{\Phi}_{k}^{\left(f\right)};\, k=1,...,K\right\} $)
and ${\Phi}^{\left(ideal\right)}$ (${\Phi}^{\left(ideal\right)}\triangleq\left\{ \Phi_{k}^{\left(ideal\right)};\, k=1,...,K\right\} $)
are two $K$-size vectors whose $k$-th ($k=1,...,K$) elements are\begin{equation}
\hat{\Phi}_{k}^{\left(f\right)}=\frac{\Phi_{k}\left(\mathbf{c}^{(f)},\bm{\alpha}^{(f)},\bm{\beta}^{(f)}\right)}{{\displaystyle \max_{f=1,...,F}}\left[\Phi_{k}\left(\mathbf{c}^{\left(f\right)},\bm{\alpha}^{\left(f\right)},\bm{\beta}^{\left(f\right)}\right)\right]-{\displaystyle \min_{f=1,...,F}}\left[\Phi_{k}\left(\mathbf{c}^{\left(f\right)},\bm{\alpha}^{\left(f\right)},\bm{\beta}^{\left(f\right)}\right)\right]}\,,\label{eq:_MMD.2}\end{equation}
and\begin{equation}
\Phi_{k}^{\left(ideal\right)}=\min_{f=1,...,F}\left[\hat{\Phi}_{k}\left(\mathbf{c}^{\left(f\right)},\bm{\alpha}^{\left(f\right)},\bm{\beta}^{\left(f\right)}\right)\right],\label{eq:}\end{equation}
respectively.}
\end{itemize}
\textcolor{black}{It is worth pointing out that the} \textcolor{black}{\emph{Step
5}} \textcolor{black}{is not mandatory, since {}``the solution''
of a} \textcolor{black}{\emph{MOP}} \textcolor{black}{cannot be, by
definition, a single one if the project requirements are conflicting.
Indeed, the outcome of a multi-objective optimization is a} \textcolor{black}{\emph{PF}}
\textcolor{black}{of trade-off solutions (}\textcolor{black}{\emph{Step
4}}\textcolor{black}{) where the designer could freely choose a} \textcolor{black}{\emph{PF}}
\textcolor{black}{element as the optimal one according to the design
requirements and the other constraints such as manufacturing issues
or costs, which are not involved here in the synthesis process. In
this work,} \textcolor{black}{\emph{Step 5}} \textcolor{black}{has
been added to allow an easier and general (i.e., not subjective and
not on a user-needs case-by-case basis) analysis of the outcomes of
the proposed tiling method (Sect. \ref{sec:Numerical-Validation-and}).}

\subsection{\textcolor{black}{\emph{APF}} \textcolor{black}{Method (}\textcolor{black}{\emph{APFM}}\textcolor{black}{)\label{sub:APF-Method-(APFM)}}}

\noindent \textcolor{black}{Whether the exhaustive generation of all
$T$ tiling configurations can be executed in an acceptable time (e.g.,
hours of computations) for small apertures, the evaluation time of
multiple cost function terms becomes quickly unfeasible (e.g., years
of computations) for larger arrays. Therefore, the second tiling technique,
which is denoted as} \textcolor{black}{\emph{Approximate PF Method}}
\textcolor{black}{(}\textcolor{black}{\emph{APFM}}\textcolor{black}{),
is aimed at approximating the} \textcolor{black}{\emph{PF}} \textcolor{black}{and
it is suitable for medium/large arrays when the evaluation of the
fitness of $T_{l}$ tiling solutions is unfeasible. In particular,
the} \textcolor{black}{\emph{APFM}} \textcolor{black}{is based on
the customization to the tiling problem at hand of a} \textcolor{black}{\emph{Multi
Objective Evolutionary Algorithm}} \textcolor{black}{(}\textcolor{black}{\emph{MOEA}}\textcolor{black}{)
\cite{Deb.2001}\cite{Deb.2002} that exploits the} \textcolor{black}{\emph{Non-dominated
Sorting Genetic Algorithm II}} \textcolor{black}{(}\textcolor{black}{\emph{NSGA-II}}\textcolor{black}{)
\cite{Deb.2002} for sampling the solution space.}

\noindent \textcolor{black}{Towards this end, the $t$-th ($t=1,...,T$)
trial tiling (i.e., an} \textcolor{black}{\emph{individual}} \textcolor{black}{when
dealing with global optimization) is univocally represented by a tiling
word, $\mathbf{w}^{\left(t\right)}$, of $L$ letters ($\mathbf{w}^{\left(t\right)}\triangleq\left\{ w_{l}^{\left(t\right)};\, l=1,...,L\right\} $),
the $l$-th one being a function of the values of the height function
at the internal vertices of $\mathcal{O}$, $\bm{a}$, given by \cite{Anselmi.2017}\begin{equation}
w_{l}=\frac{\psi\left(a_{l}^{\left(t\right)}\right)-\psi\left(a_{l}^{\left(1\right)}\right)}{4}\label{eq:_HF_to_word}\end{equation}
where $a_{l}^{\left(t\right)}$ is the $l$-th ($l=1,...,L$) internal
($a_{l}^{\left(t\right)}\notin\partial\mathcal{O}$) vertex of the
$t$-th tiled array, while $a_{l}^{\left(1\right)}$ refers to the
$l$-th ($l=1,...,L$) vertex of the} \textcolor{black}{\emph{minimal
tiling}} \textcolor{black}{$\mathbf{c}^{\left(1\right)}$. Consequently,
$\mathbf{w}^{\left(1\right)}=\mathbf{0}$ and $0\leq w_{l}\leq w_{l}^{\left(T\right)}$,
$w_{l}^{\left(T\right)}$ being the $l$-th ($l=1,...,L$) letter
of the $T$-th tiling word corresponding to the} \textcolor{black}{\emph{maximal
tiling}} \textcolor{black}{$\mathbf{c}^{\left(T\right)}$ available
in closed-form analogously to $\mathbf{c}^{\left(1\right)}$ \cite{Anselmi.2017}.
Subject to such a word-coding, the} \textcolor{black}{\emph{APF}}
\textcolor{black}{is generated according to the following iterative
($i$ being the iteration index)} \textcolor{black}{\emph{NSGA-II}}\textcolor{black}{-based
procedure:}

\begin{itemize}
\item \textbf{\textcolor{black}{Step 1}} \textcolor{black}{\emph{- Initialization}}
\textcolor{black}{($i=0$) - Generate the initial alphabet of $U$
words, $\left.\mathcal{P}_{i}\right\rfloor _{i=0}=\left\{ \left.\mathbf{w}_{i}^{\left(u\right)}\right\rfloor _{i=0};\, u=1,...,U\right\} $
using the schemata-driven initialization in \cite{Anselmi.2017} to
efficiently sample the solution-space. For each $u$-th ($u=1,...,U$)
word, compute the corresponding vector $\left.\bm{\Phi}_{i}^{\left(u\right)}\right\rfloor _{i=0}$
by considering the power pattern $\left.P_{i}^{\left(u\right)}\left(\theta,\phi\right)\right\rfloor _{i=0}$
radiated by the corresponding tiling $\left.\mathbf{c}_{i}^{\left(u\right)}\right\rfloor _{i=0}$
and the sub-array amplitudes, $\left.\bm{\alpha}_{i}^{\left(u\right)}\right\rfloor _{i=0}$,
and phases, $\left.\bm{\beta}_{i}^{\left(u\right)}\right\rfloor _{i=0}$,
computed with (\ref{eq:_sub-array.amplitudes-phases_EM});}
\item \textbf{\textcolor{black}{Step 2}} \textcolor{black}{-} \textcolor{black}{\emph{Pareto
Ranking}} \textcolor{black}{- Rank the words according to the Pareto
dominance criterion (\ref{eq:_Pareto_dominance}) defining $R$-levels}
\textcolor{black}{\emph{PF}}\textcolor{black}{s \cite{Deb.2002}.
More specifically, the} \textcolor{black}{\emph{PF}} \textcolor{black}{of
the first level ($r=1$) is yielded by applying (\ref{eq:_Pareto_dominance})
to the whole alphabet $\mathcal{P}_{i}$, while the successive $r$-th
($2\le r\le R$) level} \textcolor{black}{\emph{PF}} \textcolor{black}{is
derived still using (\ref{eq:_Pareto_dominance}), but on a reduced
population, $\mathcal{P}_{i}^{\left(r\right)}$, obtained by discarding
the words of the previous ($r-1$)-th} \textcolor{black}{\emph{PF}}
\textcolor{black}{levels ($\mathcal{P}_{i}^{\left(r\right)}=\mathcal{P}^{\left(r\right)}\bigcap\left\{ \bigcup_{j=1}^{r-1}\mathcal{P}_{i}^{\left(j\right)}\right\} $).
The crowding distance criterion is then used to rank the words within
the same $r$-th ($r=1,...,R$)} \textcolor{black}{\emph{PF}} \textcolor{black}{level
by increasing the relevance of the corresponding tilings, $\left\{ \mathbf{c}_{i}^{\left(u_{r}\right)};\, u_{r}=1,...,U_{r}\right\} $,
with higher distance, in the space of the cost functions, with respect
to the neighboring ones \cite{Deb.2002};}
\item \textbf{\textcolor{black}{Step 3}} \textcolor{black}{\emph{- Alphabet
Update}} \textcolor{black}{- Update the iteration index ($i\leftarrow i+1$)
and generate a new temporary alphabet $\mathcal{Q}_{i}$ of $U$ words/individuals
by applying the} \textcolor{black}{\emph{NSGA-II}} \textcolor{black}{generation
strategy. Compute the corresponding $U$-size cost function vector,
$\bm{\Phi}^{\left(u\right)}$ ($\bm{\Phi}^{\left(u\right)}$ $\triangleq$
\{$\bm{\Phi}_{i}^{\left(u\right)}$; $u=1,...,U$\}). Build a word
pool $\mathcal{R}_{i}$ of $2\times U$ elements by merging $\mathcal{P}_{i-1}$
and $\mathcal{Q}_{i}$ ($\mathcal{R}_{i}=\mathcal{Q}_{i}\bigcup\mathcal{P}_{i-1}$),
then rank the words of $\mathcal{R}_{i}$ according to} \textcolor{black}{\emph{Step
2}} \textcolor{black}{and compose the new alphabet, $\mathcal{P}_{i}$,
with the first $U$ ranked words;}
\item \textbf{\textcolor{black}{Step 4}} \textcolor{black}{\emph{- Stopping
Criterion}} \textcolor{black}{- If the index $i$ is smaller than
a user-defined maximum value, $I$, (i.e., $i<I$), then repeat} \textcolor{black}{\emph{Step
2}} \textcolor{black}{and} \textcolor{black}{\emph{Step 3}}\textcolor{black}{.
Otherwise, stop the iterative loop and return, as} \textcolor{black}{\emph{APF}}\textcolor{black}{,
the first ($r=1$) level} \textcolor{black}{\emph{PF}} \textcolor{black}{of
the last alphabet, $\mathcal{P}_{I}$; }
\item \textbf{\textcolor{black}{Step 5}} \textcolor{black}{-} \textcolor{black}{\emph{Optimal
Tiling Computation}} \textcolor{black}{- Eventually, set $\mathbf{c}^{opt}=\mathbf{c}^{MMD}$
(\ref{eq:_MMD.1}).}
\end{itemize}

\section{\noindent \textcolor{black}{Numerical Assessment and Method Validation\label{sec:Numerical-Validation-and}}}

\noindent \textcolor{black}{The first numerical example refers to
the orthogonal polygon array in Fig. 5(}\textcolor{black}{\emph{a}}\textcolor{black}{)
with $P=40$ elements distributed on a uniform ($d_{x}=d_{y}=\frac{\lambda}{2}$)
lattice of $M\times N$ ($M=N=8$) unit cells. The goal is to synthesize
a domino-tiled configuration providing the maximum directivity (}\textcolor{black}{\emph{D}}\textcolor{black}{)
and affording a pattern that fulfils the mask $\Pi\left(u,v\right)$
in Fig. 5(}\textcolor{black}{\emph{c}}\textcolor{black}{). Mathematically,
it has been coded by defining the following $K=2$ cost function terms:}

\noindent \textcolor{black}{\begin{equation}
\Phi_{1}\left(\mathbf{c},\bm{\alpha},\bm{\beta}\right)\triangleq\max\left(\frac{4\pi P\left(u_{0},v_{0}\right)}{\int_{\Omega}\left[\frac{P\left(u,v\right)}{\sqrt{1-u^{2}-v^{2}}}\right]du\, dv}\right)\label{eq:_cost.function1}\end{equation}
and\begin{equation}
\begin{array}{r}
\Phi_{2}\left(\mathbf{c},\bm{\alpha},\bm{\beta}\right)\triangleq\min\left(\int_{\Omega}\left[P\left(u-u_{0},v-v_{0}\right)-\Pi\left(u-u_{0},v-v_{0}\right)\right]\right.\\
\left.\times\mathcal{H}\left\{ P\left(u-u_{0},v-v_{0}\right)-\Pi\left(u-u_{0},v-v_{0}\right)\right\} dudv\right)\end{array}\label{eq:_cost.function2}\end{equation}
$\mathcal{H}\left\{ \cdot\right\} $ and $\Omega$ being the Heaviside
function and the visible range ($\Omega=\left\{ \left(u,v\right):\, u^{2}+v^{2}\leq1\right\} $),
respectively, while $\left(u_{0},v_{0}\right)$ is the beam-pointing
direction.}

\noindent \textcolor{black}{When applying the proposed tiling strategy,
the tileability has been firstly checked through (\ref{eq:_tilability_condition})
once computed the values of the height function at the external-vertices,
$\bm{\psi}$, {[}Fig. 5(}\textcolor{black}{\emph{b}}\textcolor{black}{){]}.
As for the cardinality of the solution space, it turns out that $T_{u}=12.989\times10^{6}$
and $T_{l}=54$. As for the upper bound $T_{u}$, it is related to
the light-blue square of $M\times N$ pixels in Fig. 5(}\textcolor{black}{\emph{b}}\textcolor{black}{)
enclosing $\mathcal{O}$, while the value of $T_{l}$ is given by
the sum of the number of domino tilings of the $J=4$ rectangles {[}i.e.,
the yellow, the magenta, the blue, and the green ones in Fig. 5(}\textcolor{black}{\emph{b}}\textcolor{black}{){]}
within the aperture $\mathcal{O}$. Since the time for evaluating
the two cost function terms (\ref{eq:_cost.function1})-(\ref{eq:_cost.function2})
of an array of ideal elements {[}i.e., $E_{p}\left(\theta,\phi\right)=1${]}
is equal to $\Delta t\simeq0.1$ {[}sec{]} on an Intel 2.10GHz Xeon
CPU with 64Gb of RAM, the retrieval of the} \textcolor{black}{\emph{PF}}
\textcolor{black}{would cost} \textcolor{black}{\emph{}}\textcolor{black}{at
most $15$ days wether considering the $T_{u}$-wide solution space.
However, one can notice that the orthogonal-polygon array in Fig.
5(}\textcolor{black}{\emph{a}}\textcolor{black}{) occupies only a
limited portion of the $M\times N$ square area. Therefore, the exhaustive}
\textcolor{black}{\emph{ETM}}\textcolor{black}{-based technique has
been applied and the actual number of $T=9.521\times10^{3}$ configurations
has been exhaustively generated and evaluated in $\Delta t_{EPFM}=16$
{[}min{]}. Moreover, the corresponding sub-array weights have been
computed through (\ref{eq:_sub-array.amplitudes-phases_EM}) starting
from the reference fully-populated excitations, $\bm{\alpha}^{ref}$,
in Fig. 5(}\textcolor{black}{\emph{a}}\textcolor{black}{), which have
been computed with a convex programming (}\textcolor{black}{\emph{CP}}\textcolor{black}{)
optimization strategy \cite{Bucci.2002} to fit the power mask constraint
displayed in Fig. 5(}\textcolor{black}{\emph{c}}\textcolor{black}{).
Figure 6 shows the radiated reference power pattern along the principal,
$\phi=0$ {[}deg{]} ($v=0$) and the $\phi=90$ {[}deg{]} ($u=0$)
plane {[}Fig. 6(}\textcolor{black}{\emph{b}}\textcolor{black}{){]},
while its features are reported in Tab. I. Moreover, Figure 7 plots
the} \textcolor{black}{\emph{PF}} \textcolor{black}{of the domino-tiled
arrays in the (}\textcolor{black}{\emph{D}}\textcolor{black}{,} \textcolor{black}{\emph{SLL}}\textcolor{black}{)
plane along with the representative points of the whole set of admissible
tilings. As a representative example, the excitations of the} \textcolor{black}{\emph{MMD}}
\textcolor{black}{tiling are reported in Fig. 6(}\textcolor{black}{\emph{a}}\textcolor{black}{),
while Figure 6(}\textcolor{black}{\emph{b}}\textcolor{black}{) shows
the radiated power pattern with $SLL^{MMD}=-19.62$ {[}dB{]} and $D^{MMD}=19.98$
{[}dBi{]}, which are close to the values of the same pattern features
of the reference solution (Tab. I). For the sake of analysis, the
half-power beamwidth in the azimuth ($HPBW_{az}$) and in the elevation
($HPBW_{el}$) planes are given (Tab. I), as well.}

\noindent \textcolor{black}{The second example deals with the orthogonal
polygonal aperture shown in Fig. 8(}\textcolor{black}{\emph{a}}\textcolor{black}{)
that approximates the circular support of radius $\rho=2.0\lambda$
with $P=52$ elements spaced by $d_{x}=d_{y}=\frac{\lambda}{2}$.
The problem objective is here that of designing a tiled array pointing
the beam towards two directions, namely the broadside $\left(\theta_{1},\phi_{1}\right)=\left(0,0\right)$
{[}deg{]} {[}i.e., $\left(u_{1},v_{1}\right)=\left(0.0,0.0\right)${]}
and the angle $\left(\theta_{2},\phi_{2}\right)=\left(30,0\right)$
{[}deg{]} {[}i.e., $\left(u_{2},v_{2}\right)=\left(0.5,0.0\right)${]},
while fitting the} \textcolor{black}{\emph{SLL}} \textcolor{black}{requirements
stated by the broadside pattern mask in Fig. 8(}\textcolor{black}{\emph{b}}\textcolor{black}{)
and {}``translated'' for each steering direction (i.e., $SLL^{\Pi}=-20$
{[}dB{]}). According to the mathematical formulation in Sect. \ref{sec:Orthogonal-Polygon-Array},
this means to define the following $K=2$ cost functions terms: $\Phi_{k}\left(\mathbf{c},\bm{\alpha},\bm{\beta}\right)$
$\triangleq$ $\min\left(\int_{\Omega}\left[P\left(u-u_{k},v-v_{k}\right)\right.\right.$
$-$ $\left.\Pi\left(u-u_{k},v-v_{k}\right)\right]$ $\times$ $\mathcal{H}\left\{ P\left(u-u_{k},v-v_{k}\right)\right.$
$-$ $\left.\Pi\left(u-u_{k},v-v_{k}\right)\right\} $ $\left.dudv\right)$,
$k=1,...,K$. In order to address the synthesis problem at hand, first
the reference fully-populated solution, which is characterized by
the set of amplitudes $\bm{\alpha}^{ref}$ in Fig. 8(}\textcolor{black}{\emph{a}}\textcolor{black}{)
and radiating the power pattern with $SLL^{ref}=-20.60$ {[}dB{]}
{[}Fig. 8(}\textcolor{black}{\emph{c}}\textcolor{black}{){]}, has
been obtained with the} \textcolor{black}{\emph{CP}}\textcolor{black}{.
Once the aperture tileability has been verified as well as the possibility
to exhaustively sample the solution space of $T_{u}=12.989\times10^{6}$
configurations, the} \textcolor{black}{\emph{ETM}} \textcolor{black}{has
been used to generate $T=2.88\times10^{4}$ different tiles arrangements
then evaluated in $\Delta t_{EPFM}=48$ {[}min{]} to retrieve the}
\textcolor{black}{\emph{PF}} \textcolor{black}{shown in Fig. 9. As
it can be observed, the} \textcolor{black}{\emph{MMD}} \textcolor{black}{solution
coincides here with the one of the} \textcolor{black}{\emph{PF}} \textcolor{black}{that
better optimizes the $k=2$ cost function term (}\textcolor{black}{\emph{MOP-2}}
\textcolor{black}{- Fig. 9). The corresponding excitations and patterns
are reported in Figs. 10(}\textcolor{black}{\emph{a}}\textcolor{black}{)-10(}\textcolor{black}{\emph{b}}\textcolor{black}{)
and Figs. 10(}\textcolor{black}{\emph{c}}\textcolor{black}{)-10(}\textcolor{black}{\emph{d}}\textcolor{black}{),
respectively. More specifically, it turns out that the} \textcolor{black}{\emph{MMD}}\textcolor{black}{-tiled
array properly steers the beam along the desired directions with a
maximum} \textcolor{black}{\emph{SLL}} \textcolor{black}{degradation,
with respect to the ideal reference, of $\Delta_{SLL}^{MMD-ref}\approx1.5$
{[}dB{]} (Tab. II) and a maximum deviation from the target mask of
$\Delta_{SLL}^{MMD-\Pi}\approx0.8$ {[}dB{]} {[}Tab. II - Fig. 11(}\textcolor{black}{\emph{b}}\textcolor{black}{){]},
being $\Delta_{SLL}^{MMD-ref}=\max_{k}\left|SLL_{\left(u_{k},v_{k}\right)}^{MMD}-SLL_{\left(u_{k},v_{k}\right)}^{ref}\right|$
and $\Delta_{SLL}^{MMD-\Pi}=\max_{k}\left|SLL_{\left(u_{k},v_{k}\right)}^{MMD}-SLL_{\left(u_{k},v_{k}\right)}^{\Pi}\right|$.
For completeness, Figure 11 compares the patterns radiated by the}
\textcolor{black}{\emph{MMD}} \textcolor{black}{array with the reference
one and with that generated by the} \textcolor{black}{\emph{MOP-1}}
\textcolor{black}{arrangement together with the target mask $\Pi$
when $\left(u_{1},v_{1}\right)=\left(0.0,0.0\right)$ {[}Fig. 11(}\textcolor{black}{\emph{a}}\textcolor{black}{)
- $v=0$ plane; Fig. 11(}\textcolor{black}{\emph{b}}\textcolor{black}{)
- $u=0$ plane{]} and $\left(u_{2},v_{2}\right)=\left(0.5,0.0\right)$
{[}Fig. 11(}\textcolor{black}{\emph{c}}\textcolor{black}{) - $v=0$
plane{]}.}

\noindent \textcolor{black}{Successively, the same test case has been
dealt with the} \textcolor{black}{\emph{APFM}} \textcolor{black}{to
assess the reliability and the effectiveness of this latter in approximating
the} \textcolor{black}{\emph{PF}} \textcolor{black}{here available
thanks to the} \textcolor{black}{\emph{EPFM}}\textcolor{black}{. Accordingly,
the} \textcolor{black}{\emph{MOEA}} \textcolor{black}{parameters have
been set as follows: $U=L=37$ (size of the alphabet $\mathcal{P}$),
$I=100$ (maximum number of iterations), $p_{c}=0.9$ (crossover probability),
and $p_{m}=\frac{1}{L}$ (mutation probability)\cite{Nagar.2018}.
After} \textcolor{black}{\emph{}}\textcolor{black}{$\Delta t_{APFM}=220$
{[}sec{]}, the approximate} \textcolor{black}{\emph{PF}} \textcolor{black}{shown
in Fig. 9 (black circles) is derived. As it can be seen, all the}
\textcolor{black}{\emph{APFM}} \textcolor{black}{solutions (black
rings - Fig. 9) belong to the} \textcolor{black}{\emph{PF}} \textcolor{black}{(red
dots - Fig. 9) and the success rate is $70\%$, since $9$ among $F=13$}
\textcolor{black}{\emph{PF}} \textcolor{black}{solutions have been
faithfully recovered, despite the sampling of only $5\%$ (at most)
of the whole set of $T$ admissible tilings.}

\noindent \textcolor{black}{The third example is concerned with a
larger elliptically-shaped array {[}Fig. 12(}\textcolor{black}{\emph{a}}\textcolor{black}{){]}
composed by $P=224$ elements half-wavelength spaced ($d_{x}=d_{y}=\frac{\lambda}{2}$).
The synthesis is aimed at fitting the} \textcolor{black}{\emph{SLL}}
\textcolor{black}{mask in Fig. 12(}\textcolor{black}{\emph{b}}\textcolor{black}{)
when steering the beam at $\left(\theta_{1},\phi_{1}\right)=\left(8,0\right)$
{[}deg{]} {[}i.e., $\left(u_{1},v_{1}\right)=\left(0.1392,0.0\right)${]}
and $\left(\theta_{2},\phi_{2}\right)=\left(8,90\right)$ {[}deg{]}
{[}i.e., $\left(u_{2},v_{2}\right)=\left(0.0,0.1392\right)${]}, thus
the same two-terms cost function of the previous example has been
considered. Since the aperture shape satisfies the domino tileability
condition and, this time, the lower bound is $T_{l}>10^{16}$, the}
\textcolor{black}{\emph{EPFM}} \textcolor{black}{is avoided and the}
\textcolor{black}{\emph{APFM}} \textcolor{black}{only is used to determine
the solutions} \textcolor{black}{\emph{PF}}\textcolor{black}{. Starting
from the} \textcolor{black}{\emph{CP}} \textcolor{black}{reference
solution in Fig. 12(}\textcolor{black}{\emph{a}}\textcolor{black}{),
which radiates in broadside the pattern of Fig. 12(}\textcolor{black}{\emph{c}}\textcolor{black}{)
with $SLL^{ref}=-32.70$ {[}dB{]} (Tab. III), the} \textcolor{black}{\emph{NSGA-II}}
\textcolor{black}{loop has been executed by setting the optimization
parameters as before ($U=L=197$, $p_{c}=0.9$, $p_{m}=\frac{1}{L}$)
except for the number of iterations, now chosen equal to $I=1000$,
to keep the same percentage of sampling of the solution space due
to the wider array size ($P=224$ vs. $P=52$). Figure 13 shows the
estimated} \textcolor{black}{\emph{PF}} \textcolor{black}{together
with the best} \textcolor{black}{\emph{SOP}} \textcolor{black}{solution
drawn among $10$ independent runs of the} \textcolor{black}{\emph{GA}}\textcolor{black}{-based
approach in \cite{Anselmi.2017} aimed at optimizing the single-objective
cost function given by the linear combination of the two cost function
terms, which are independently optimized by the} \textcolor{black}{\emph{APFM}}\textcolor{black}{.
As it can be noticed, several} \textcolor{black}{\emph{PF}} \textcolor{black}{tilings
performs better than the} \textcolor{black}{\emph{SOP}} \textcolor{black}{one.
For instance, the} \textcolor{black}{\emph{MMD}} \textcolor{black}{solution
(Fig. 13) is ''described'' in Fig. 14. More in detail, the arrangement
of the domino tiles is shown in Fig. 14(}\textcolor{black}{\emph{a}}\textcolor{black}{)
along with the color level representation of the sub-array amplitudes,
while the phase distribution for steering the beam at $\left(u_{1},v_{1}\right)$
{[}Fig. 14(}\textcolor{black}{\emph{d}}\textcolor{black}{){]} and
$\left(u_{2},v_{2}\right)$ {[}Fig. 14(}\textcolor{black}{\emph{e}}\textcolor{black}{){]}
is given in Fig. 14(}\textcolor{black}{\emph{b}}\textcolor{black}{)
and Fig. 14(}\textcolor{black}{\emph{c}}\textcolor{black}{), respectively.
As for the radiation performance, it turns out that the maximum degradation
of the} \textcolor{black}{\emph{SLL}} \textcolor{black}{with respect
to the reference amounts to $\Delta_{SLL}^{MMD-ref}\approx2.19$ {[}dB{]}
(Tab. III) ($\Delta_{SLL}^{MMD-ref}=\Delta_{SLL}^{MMD-\Pi}$), but
halving the} \textcolor{black}{\emph{TRM}}\textcolor{black}{s, while
the directivity values are very close ($\Delta_{D}^{MMD-ref}\approx0.15$
{[}dBi{]} - Tab. III) also thanks to the full coverage of the aperture
guaranteed by the proposed synthesis strategy.}

\noindent \textcolor{black}{The last test case is aimed at investigating
the effectiveness of the domino-tiling method in minimizing the beam-pointing
error (}\textcolor{black}{\emph{BPE}}\textcolor{black}{) \cite{Skolnik.2008}}%
\footnote{\noindent \textcolor{black}{The beam-pointing error (}\textcolor{black}{\emph{BPE}}\textcolor{black}{)
is the deviation of the actual beam pointing, $\left(\theta^{act},\phi^{act}\right)$,
from the desired one, $\left(\theta^{des},\phi^{des}\right)$.}%
}\textcolor{black}{, due to the quantization of the phase distribution.
With reference to the same elliptically-shaped array of Fig. 12(}\textcolor{black}{\emph{a}}\textcolor{black}{),
the problem at hand is that of fitting the} \textcolor{black}{\emph{SLL}}
\textcolor{black}{mask in Fig. 12(}\textcolor{black}{\emph{b}}\textcolor{black}{)
and minimizing the} \textcolor{black}{\emph{BPE}} \textcolor{black}{when
scanning the beam within a given cone, \{$\theta_{0}\le\theta\le\theta_{0}+\theta_{max}$;
$0\le\phi\le2\pi$\}, $\theta_{max}$ being the maximum scan angle.
Accordingly, the following cost-function terms have been defined.
The former codes the} \textcolor{black}{\emph{BPE}} \textcolor{black}{requirement,
$\Phi_{1}\left(\mathbf{c},\,\bm{\alpha},\:\bm{\beta}\right)\triangleq\min\left\{ \frac{1}{R}\sum_{r=1}^{R}\xi_{r}\right\} ,$
$\xi\triangleq\sqrt{\left(\theta^{des}-\theta^{act}\right)^{2}+\left(\phi^{des}-\phi^{act}\right)^{2}}$
and $R$ being the} \textcolor{black}{\emph{BPE}} \textcolor{black}{and
the number of angular directions $\left\{ \left(\theta_{r},\phi_{r}\right);\, r=1,...,R\right\} $
($\theta_{1}=\theta_{0}+\theta_{max}$) that sample the scan range
(i.e., $\theta_{0}\le\theta_{r}\le\theta_{0}+\theta_{max}$ and $0\le\phi_{r}\le2\pi$)
(Fig. 15), while the other term is related to the} \textcolor{black}{\emph{SLL}}
\textcolor{black}{mask matching\begin{equation}
\begin{array}{r}
\Phi_{2}\left(\mathbf{c},\bm{\alpha},\bm{\beta}\right)\triangleq\min\left\{ \max_{r}\left(\int_{\Omega}\left[P\left(u-\left[u_{0}+u_{r}^{des}\right],v-\left[v_{0}+v_{r}^{des}\right]\right)-\right.\right.\right.\\
\left.\Pi\left(u-\left[u_{0}+u_{r}^{des}\right],v-\left[v_{0}+v_{r}^{des}\right]\right)\right]\times\mathcal{H}\left\{ P\left(u-\left[u_{0}+u_{r}^{des}\right],v-\left[v_{0}+v_{r}^{des}\right]\right)\right.-\\
\left.\left.\left.\Pi\left(u-\left[u_{0}+u_{r}^{des}\right],v-\left[v_{0}+v_{r}^{des}\right]\right)\right\} dudv\right)\right\} \end{array}\label{eq:_objecive_min_max_SLL}\end{equation}
 In particular, three elliptical scan cones with $\theta_{max}=\left\{ 30,\,45,\,60\right\} $
{[}deg{]} from broadside {[}i.e., $\left(\theta_{0},\phi_{0}\right)=\left(0,0\right)$
{[}deg{]} $\to$ $\left(u_{0},v_{0}\right)=\left(0.0,0.0\right)${]}
have been considered and sampled at the $R=12$ scanning directions
reported in Tab. IV. By applying the} \textcolor{black}{\emph{APFM}}\textcolor{black}{,
the} \textcolor{black}{\emph{PF}}\textcolor{black}{s in Fig. 16 are
obtained. With reference to the} \textcolor{black}{\emph{MMD}} \textcolor{black}{tilings,
it turns out that the beam-pointing is kept very accurate (e.g., $\left.\xi_{avg}^{MDD}\right\rfloor _{\theta_{max}=30\,[deg]}=0.32$
{[}deg{]}) as confirmed by the plots of the} \textcolor{black}{\emph{BPE}}
\textcolor{black}{in Fig. 17 being $\left.\xi^{MDD}\right\rfloor _{\theta_{max}=30\,[deg]}\le1.0$
{[}deg{]} {[}Fig. 17(}\textcolor{black}{\emph{a}}\textcolor{black}{){]},
$\left.\xi^{MDD}\right\rfloor _{\theta_{max}=45\,[deg]}\le1.5$ {[}deg{]}
{[}Fig. 17(}\textcolor{black}{\emph{b}}\textcolor{black}{){]}, and
$\left.\xi^{MDD}\right\rfloor _{\theta_{max}=60\,[deg]}\le2.7$ {[}deg{]}
{[}Fig. 17(}\textcolor{black}{\emph{c}}\textcolor{black}{){]}. On
the contrary, due to the simplicity of the architectural solution
(i.e., a tiled array instead of a fully-populated one) and the complexity
of the synthesis problem at hand (i.e., fitting the project requirements
on a complete angular cone instead of few steering directions), there
is a non-negligible degradation of the} \textcolor{black}{\emph{SLL}}
\textcolor{black}{even though the highest sidelobes do not occur close
to the mainbeam as shown in Fig. 18 where the beams generated at four
representative scanning directions of Tab. IV (i.e., $\theta_{max}=30$
{[}deg{]} and $r\in\left\{ 1,\,2,\,3,\,4\right\} $) by the} \textcolor{black}{\emph{MMD}}
\textcolor{black}{tiling synthesized when $\theta_{max}=30$ {[}deg{]}
are shown.}

\noindent \textcolor{black}{Finally, the reliability of the proposed
tiling strategy and its robustness against the non-idealities of real
arrays have been checked. Towards this end, the isotropic elementary
radiator of the previous example has been substituted with a rectangular
pin-fed patch antenna resonating at $f_{0}=77$ {[}GHz{]} and the
corresponding element pattern, $E_{p}\left(\theta,\phi\right)\neq1$
($p=1,...,P$), has been set to the embedded element pattern of the
central element of a neighbour of $5\times5$ identical elements to
include, in the element model, the mutual coupling effects of the
real architecture. Despite the difficulty of the synthesis problem
at hand, the pattern shape and the beam pointing as well as the relative}
\textcolor{black}{\emph{SLL}} \textcolor{black}{are not significantly
altered when introducing a real radiator in the place of the ideal
one (Fig. 18), the main differences arising in the far side-lobe region.}

\section{\noindent \textcolor{black}{Conclusions}}

\noindent \textcolor{black}{In this work, the design of orthogonal
polygon sub-arrayed arrays with domino tiles has been addressed by
means of an innovative design strategy. Starting from the assessment
of the full tileability of the polygonal aperture with an analytic
procedure, two multi-objective optimization methods have been proposed
to synthesize the clustering configuration and the sub-array weights.
More specifically, the former method is for compact apertures and
it is aimed at determining the Pareto front of the optimal trade-off
solutions. The other is suitable for larger arrays and it faithfully
approximates the Pareto front.}

\noindent \textcolor{black}{The key-features of the proposed design
strategy are:}

\begin{itemize}
\item \noindent \textcolor{black}{the design of high-efficiency (i.e., full
coverage of the aperture without holes) and arbitrary-shaped tiled
apertures thanks to the exploitation of suitable mathematical theorems
to} \textcolor{black}{\emph{a-priori}} \textcolor{black}{assess the
complete domino-tileability;}
\item \noindent \textcolor{black}{the computationally-effective and reliable
retrieval of the Pareto front of optimal trade-off tilings by means
of customized multi-objective methods based on the concept of height
function.}
\end{itemize}
\textcolor{black}{Future research activities, beyond the scope of
the current work, will be aimed at extending the proposed strategy
to different ''alphabets'' of tiles (i.e., different cardinality
and/or tile shapes) of interest for various commercial applications
ranging from wireless communications to remote sensing.}

\section*{\noindent \textcolor{black}{Acknowledgements}}

\textcolor{black}{This work has been partially supported by the Italian
Ministry of Education, University, and Research within the Program
PRIN 2017 (CUP: E64I19002530001) for the Project CYBER-PHYSICAL ELECTROMAGNETIC
VISION: Context-Aware Electromagnetic Sensing and Smart Reaction (EMvisioning)
(Grant no. 2017HZJXSZ) and by the Ministry of Education of China within
the Chang-Jiang Visiting Professor chair. A. Massa wishes to thank
E. Vico for her never-ending inspiration, support, guidance, and help.}

\newpage
\section*{\textcolor{black}{FIGURE CAPTIONS}}

\begin{itemize}
\item \textbf{\textcolor{black}{Figure 1.}} \textcolor{black}{\emph{Illustrative
Example}} \textcolor{black}{($P=30$) - Sketch of (}\textcolor{black}{\emph{a}}\textcolor{black}{)
the feeding network of (}\textcolor{black}{\emph{b}}\textcolor{black}{)
a domino-tiled clustering of an orthogonal polygon shaped aperture
with $P=30$ unit cells.}
\item \textbf{\textcolor{black}{Figure 2.}} \textcolor{black}{\emph{Illustrative
Example}} \textcolor{black}{($P=30$) - Sketch of (}\textcolor{black}{\emph{a}}\textcolor{black}{)
the ordered set of vertices of the $P$ unit cells of $\mathcal{O}$,
$\bm{v}=\left\{ v_{g};\, g=1,...,G\right\} $, consisting of (}\textcolor{black}{\emph{b}}\textcolor{black}{)
boundary, $\bm{b}=\left\{ b_{s};\, s=1,...,S\right\} $ ($b_{s}\in\partial\mathcal{O}$),
and internal, $\bm{a}=\left\{ a_{l};\, l=1,...,L\right\} $ ($a_{l}\notin\partial\mathcal{O}$),
elements ($\bm{v}=\bm{b}\cup\bm{a}$), and (}\textcolor{black}{\emph{c}}\textcolor{black}{)
a chessboard pattern for the $P$ pixels in $\mathcal{O}$ along with
the corresponding directed graph $\mathcal{G}$.}
\item \textbf{\textcolor{black}{Figure 3.}} \textcolor{black}{\emph{Illustrative
Example}} \textcolor{black}{($P=10$, $S=18$, $L=2$) - Values of
the} \textcolor{black}{\emph{heigth function}} \textcolor{black}{at
the boundary vertices ($\bm{\psi}\triangleq\left\{ \psi\left(b_{s}\right);\, s=1,...,S\right\} $)
along with the shortest path (red arrows) connecting the boundary
vertex $v_{5}=b_{17}$ to the other one $v_{16}=b_{8}$ {[}$\Delta\left(b_{17},b_{8}\right)=3${]}
through the graph $\mathcal{G}$.}
\item \textbf{\textcolor{black}{Figure 4.}} \textcolor{black}{\emph{Illustrative
Example}} \textcolor{black}{($N=12$) - Picture of (}\textcolor{black}{\emph{a}}\textcolor{black}{)
an orthogonal polygon partitioned into $J=3$ rectangles and (}\textcolor{black}{\emph{b}}\textcolor{black}{)
behavior of cardinality bounds when varying $M$ in the range $6\le M\le30$.}
\item \textbf{\textcolor{black}{Figure 5.}} \textcolor{black}{\emph{Numerical
Assessment}} \textcolor{black}{($M=8$, $N=8$, $P=40$, $SLL^{\Pi}=-20$
{[}dB{]}) - Plot of (}\textcolor{black}{\emph{a}}\textcolor{black}{)
the reference fully-populated excitations, $\bm{\alpha}^{ref}$, (}\textcolor{black}{\emph{b}}\textcolor{black}{)
the values of the height function at the external vertices, and (}\textcolor{black}{\emph{c}}\textcolor{black}{)
the power pattern mask, $\Pi\left(u,v\right)$.}
\item \textbf{\textcolor{black}{Figure 6.}} \textcolor{black}{\emph{Numerical
Assessment}} \textcolor{black}{($M=8$, $N=8$, $P=40$, $SLL^{\Pi}=-20$
{[}dB{]}) - Plot of (}\textcolor{black}{\emph{a}}\textcolor{black}{)
the} \textcolor{black}{\emph{MMD}} \textcolor{black}{tiling and clusters
amplitudes. Plot of (}\textcolor{black}{\emph{b}}\textcolor{black}{)
the reference and the} \textcolor{black}{\emph{MMD}} \textcolor{black}{power
patterns along the principal, $\phi=0$ {[}deg{]} ($v=0$) and the
$\phi=90$ {[}deg{]} ($u=0$) planes.}
\item \textbf{\textcolor{black}{Figure 7.}} \textcolor{black}{\emph{Numerical
Assessment}} \textcolor{black}{($M=8$, $N=8$, $P=40$, $SLL^{\Pi}=-20$
{[}dB{]}) - Plot of the Pareto front of the domino-tiled arrays in
the $\left(\Phi_{1},\Phi_{2}\right)$-plane along with the representative
points of the whole set of $T$ admissible tilings.}
\item \textbf{\textcolor{black}{Figure 8.}} \textcolor{black}{\emph{Numerical
Assessment}} \textcolor{black}{($M=8$, $N=8$, $P=52$, $SLL^{\Pi}=-20$
{[}dB{]}) - Plot of (}\textcolor{black}{\emph{a}}\textcolor{black}{)
the reference fully-populated excitations, $\bm{\alpha}^{ref}$, (}\textcolor{black}{\emph{b}}\textcolor{black}{)
the power pattern mask, $\Pi\left(u,v\right)$, and (}\textcolor{black}{\emph{c}}\textcolor{black}{)
the reference power pattern.}
\item \textbf{\textcolor{black}{Figure 9.}} \textcolor{black}{\emph{Numerical
Assessment}} \textcolor{black}{($M=8$, $N=8$, $P=52$, $SLL^{\Pi}=-20$
{[}dB{]}) - Plot of the Pareto front of the domino-tiled arrays in
the $\left(\Phi_{1},\Phi_{2}\right)$-plane along with the representative
points of the whole set of $T$ admissible tilings.}
\item \textbf{\textcolor{black}{Figure 10.}} \textcolor{black}{\emph{Numerical
Assessment}} \textcolor{black}{($M=8$, $N=8$, $P=52$, $SLL^{\Pi}=-20$
{[}dB{]}) - Plot of (}\textcolor{black}{\emph{a}}\textcolor{black}{)
the amplitude and (}\textcolor{black}{\emph{b}}\textcolor{black}{)
the phase of the excitations of the} \textcolor{black}{\emph{MMD}}
\textcolor{black}{tiling together with the radiated power patterns
at (}\textcolor{black}{\emph{c}}\textcolor{black}{) $\left(u_{1},v_{1}\right)=\left(0.0,0.0\right)$
and (}\textcolor{black}{\emph{d}}\textcolor{black}{) $\left(u_{2},v_{2}\right)=\left(0.5,0.0\right)$.}
\item \textbf{\textcolor{black}{Figure 11.}} \textcolor{black}{\emph{Numerical
Assessment}} \textcolor{black}{($M=8$, $N=8$, $P=52$, $SLL^{\Pi}=-20$
{[}dB{]}) - Plot of the power patterns radiated along (}\textcolor{black}{\emph{a}}\textcolor{black}{)(}\textcolor{black}{\emph{c}}\textcolor{black}{)
the $v=0$ and (}\textcolor{black}{\emph{b}}\textcolor{black}{) the
$u=0$ cuts when pointing the beam at (}\textcolor{black}{\emph{a}}\textcolor{black}{)(}\textcolor{black}{\emph{b}}\textcolor{black}{)
$\left(u_{1},\, v_{1}\right)=\left(0,\,0\right)$ and (}\textcolor{black}{\emph{c}}\textcolor{black}{)
$\left(u_{2},\, v_{2}\right)=\left(0.5,\,0\right)$.}
\item \textbf{\textcolor{black}{Figure 12.}} \textcolor{black}{\emph{Numerical
Assessment}} \textcolor{black}{($M=12$, $N=26$, $P=224$, $SLL^{\Pi}=-32.7$
{[}dB{]}) - Plot of (}\textcolor{black}{\emph{a}}\textcolor{black}{)
the reference fully-populated excitations, $\bm{\alpha}^{ref}$, (}\textcolor{black}{\emph{b}}\textcolor{black}{)
the power pattern mask, $\Pi\left(u,v\right)$, and (}\textcolor{black}{\emph{c}}\textcolor{black}{)
the reference power pattern.}
\item \textbf{\textcolor{black}{Figure 13.}} \textcolor{black}{\emph{Numerical
Assessment}} \textcolor{black}{($M=12$, $N=26$, $P=224$, $SLL^{\Pi}=-32.7$
{[}dB{]}) - Plot of the} \textcolor{black}{\emph{APFM}} \textcolor{black}{Pareto
front of the domino-tiled arrays in the $\left(\Phi_{1},\Phi_{2}\right)$-plane
along with the representative point of the} \textcolor{black}{\emph{SOP}}
\textcolor{black}{optimal arrangement.}
\item \textbf{\textcolor{black}{Figure 14.}} \textcolor{black}{\emph{Numerical
Assessment}} \textcolor{black}{($M=12$, $N=26$, $P=224$, $SLL^{\Pi}=-32.7$
{[}dB{]}) - Plot of (}\textcolor{black}{\emph{a}}\textcolor{black}{)
the amplitude and (}\textcolor{black}{\emph{b}}\textcolor{black}{)(}\textcolor{black}{\emph{c}}\textcolor{black}{)
the phase of the excitations of the} \textcolor{black}{\emph{MMD}}
\textcolor{black}{tiling together with the radiated power patterns
at (}\textcolor{black}{\emph{d}}\textcolor{black}{) $\left(u_{1},v_{1}\right)=\left(0.1392,0.0\right)$
and (}\textcolor{black}{\emph{e}}\textcolor{black}{) $\left(u_{2},v_{2}\right)=\left(0.0,0.1392\right)$.}
\item \textbf{\textcolor{black}{Figure 15.}} \textcolor{black}{Sketch of
the scan cone , \{$\theta_{0}\le\theta\le\theta_{0}+\theta_{max}$;
$0\le\phi\le2\pi$\}, $\theta_{max}$ being the maximum scan angle,
and of the set of $R$ angular samples,$\left\{ \left(\theta_{r},\phi_{r}\right);\, r=1,...,R\right\} $
($\theta_{1}=\theta_{0}+\theta_{max}$) being $\theta_{0}\le\theta_{r}\le\theta_{0}+\theta_{max}$
and $0\le\phi_{r}\le2\pi$.}
\item \textbf{\textcolor{black}{Figure 16.}} \textcolor{black}{\emph{Numerical
Assessment}} \textcolor{black}{($M=12$, $N=26$, $P=224$, $SLL^{\Pi}=-32.7$
{[}dB{]}) - Plot of the} \textcolor{black}{\emph{APFM}} \textcolor{black}{Pareto
front of the domino-tiled arrays in the $\left(\Phi_{1},\Phi_{2}\right)$-plane
for different values of the maximum scan angle, $\theta_{max}=\left\{ 30,\,45,\,60\right\} $
{[}deg{]}.}
\item \textbf{\textcolor{black}{Figure 17.}} \textcolor{black}{\emph{Numerical
Assessment}} \textcolor{black}{($M=12$, $N=26$, $P=224$, $SLL^{\Pi}=-32.7$
{[}dB{]}) - Plot of the} \textcolor{black}{\emph{BPE}} \textcolor{black}{of
the} \textcolor{black}{\emph{MMD}} \textcolor{black}{domino-tiled
array in the scan cone \{$\theta_{0}\le\theta\le\theta_{0}+\theta_{max}$;
$0\le\theta\le2\pi$\} when (}\textcolor{black}{\emph{a}}\textcolor{black}{)
$\theta_{max}=30$ {[}deg{]}, (}\textcolor{black}{\emph{b}}\textcolor{black}{)
$\theta_{max}=45$ {[}deg{]}, and (}\textcolor{black}{\emph{c}}\textcolor{black}{)
$\theta_{max}=60$ {[}deg{]}.}
\item \textbf{\textcolor{black}{Figure 18.}} \textcolor{black}{\emph{Numerical
Assessment}} \textcolor{black}{($M=12$, $N=26$, $P=224$, $SLL^{\Pi}=-32.7$
{[}dB{]}, $\theta_{max}=30$ {[}deg{]}) - Plot of the power patterns
radiated by the} \textcolor{black}{\emph{MMD}} \textcolor{black}{domino-tiled
array along the cuts at (}\textcolor{black}{\emph{a}}\textcolor{black}{)
$\phi=0$ {[}deg{]}, (}\textcolor{black}{\emph{b}}\textcolor{black}{)
$\phi=16.13$ {[}deg{]}, (}\textcolor{black}{\emph{c}}\textcolor{black}{)
$\phi=40.94$ {[}deg{]}, and (}\textcolor{black}{\emph{d}}\textcolor{black}{)
$\phi=90$ {[}deg{]}.}
\end{itemize}

\section*{\textcolor{black}{TABLE CAPTIONS}}

\begin{itemize}
\item \textbf{\textcolor{black}{Table I.}} \textcolor{black}{\emph{Numerical
Assessment}} \textcolor{black}{($M=8$, $N=8$, $P=40$, $SLL^{\Pi}=-20$
{[}dB{]}) - Pattern features.}
\item \textbf{\textcolor{black}{Table II.}} \textcolor{black}{\emph{Numerical
Assessment}} \textcolor{black}{($M=8$, $N=8$, $P=52$, $SLL^{\Pi}=-20$
{[}dB{]}) - Pattern features when steering the beam at $\left(\theta_{1},\,\phi_{1}\right)=\left(0,\,0\right)$
{[}deg{]} and $\left(\theta_{2},\,\phi_{2}\right)=\left(30,\,0\right)$
{[}deg{]}.}
\item \textbf{\textcolor{black}{Table III.}} \textcolor{black}{\emph{Numerical
Assessment}} \textcolor{black}{($M=12$, $N=26$, $P=224$, $SLL^{\Pi}=-32.7$
{[}dB{]}) - Pattern features when steering the beam at $\left(\theta_{1},\,\phi_{1}\right)=\left(8,\,0\right)$
{[}deg{]} and $\left(\theta_{2},\,\phi_{2}\right)=\left(8,\,90\right)$
{[}deg{]}.}
\item \textbf{\textcolor{black}{Table IV.}} \textcolor{black}{List of the
angular coordinates of the samples of the scan cone when $\theta_{max}=\left\{ 30,\,45,\,60\right\} $
{[}deg{]}.}\newpage

\end{itemize}
\begin{center}\textcolor{black}{~\vfill}\end{center}

\begin{center}\textcolor{black}{}\begin{tabular}{c}
\textcolor{black}{\includegraphics[%
  width=0.70\columnwidth]{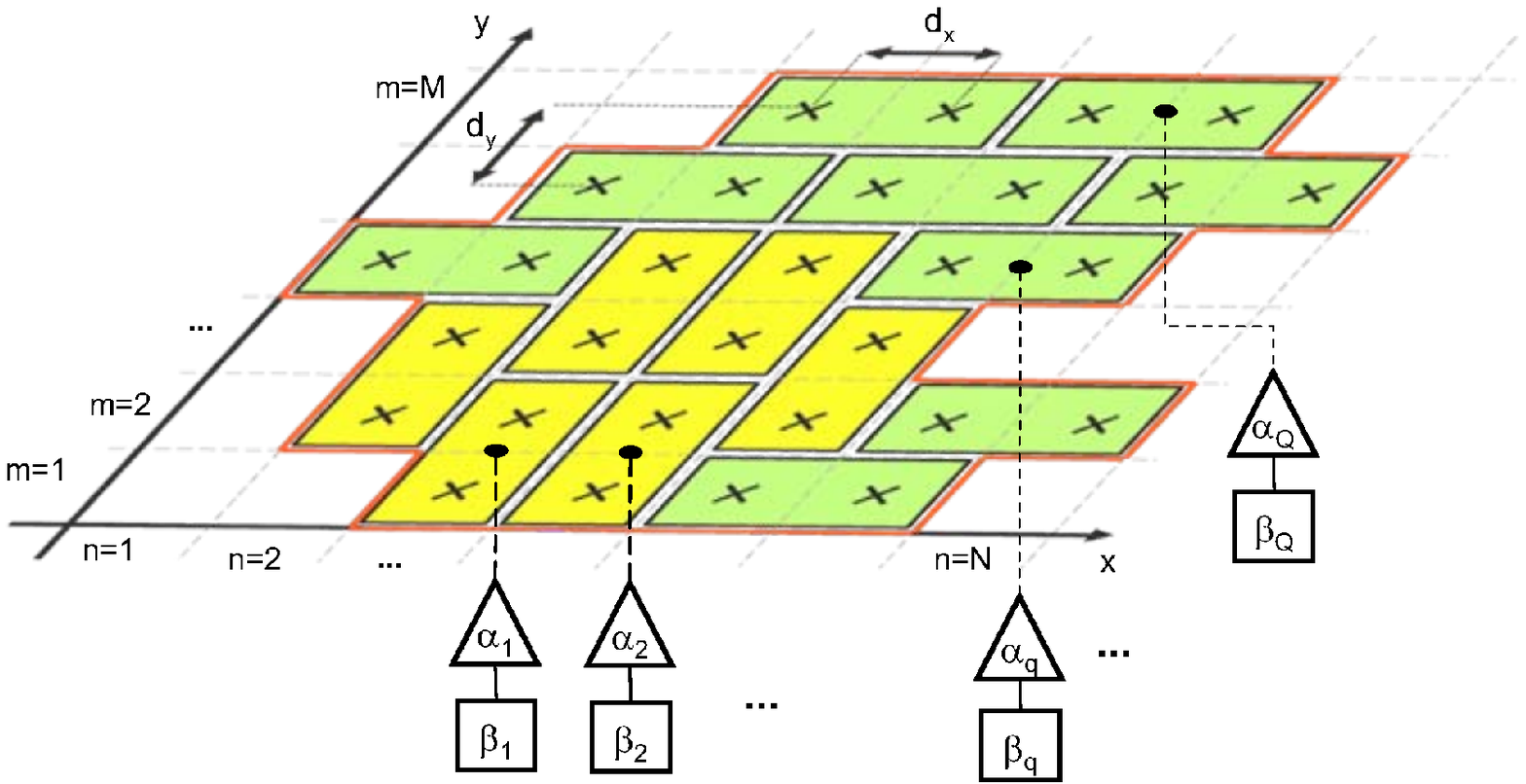}}\tabularnewline
\textcolor{black}{(}\textcolor{black}{\emph{a}}\textcolor{black}{)}\tabularnewline
\tabularnewline
\textcolor{black}{\includegraphics[%
  width=0.60\columnwidth]{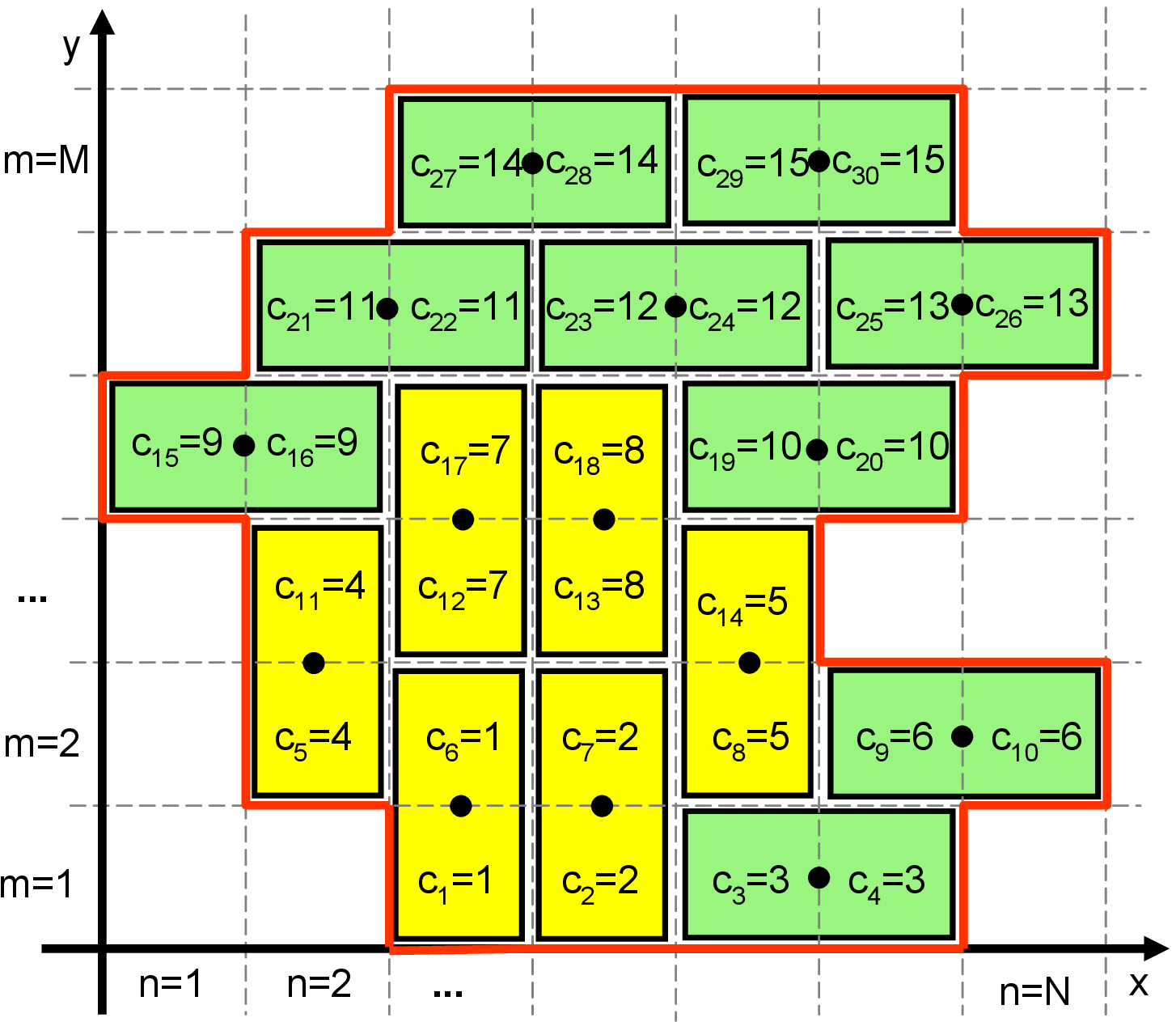}}\tabularnewline
\textcolor{black}{(}\textcolor{black}{\emph{b}}\textcolor{black}{)}\tabularnewline
\end{tabular}\end{center}

\begin{center}\textcolor{black}{~\vfill}\end{center}

\begin{center}\textbf{\textcolor{black}{Fig. 1 - P. Rocca et}} \textbf{\textcolor{black}{\emph{al.}}}\textbf{\textcolor{black}{,}}
\textbf{\textcolor{black}{\emph{{}``}}}\textcolor{black}{Pareto-Optimal
Domino-Tiling of ...''}\end{center}
\newpage

\begin{center}\textcolor{black}{}\begin{tabular}{c}
\textcolor{black}{\includegraphics[%
  width=0.38\columnwidth]{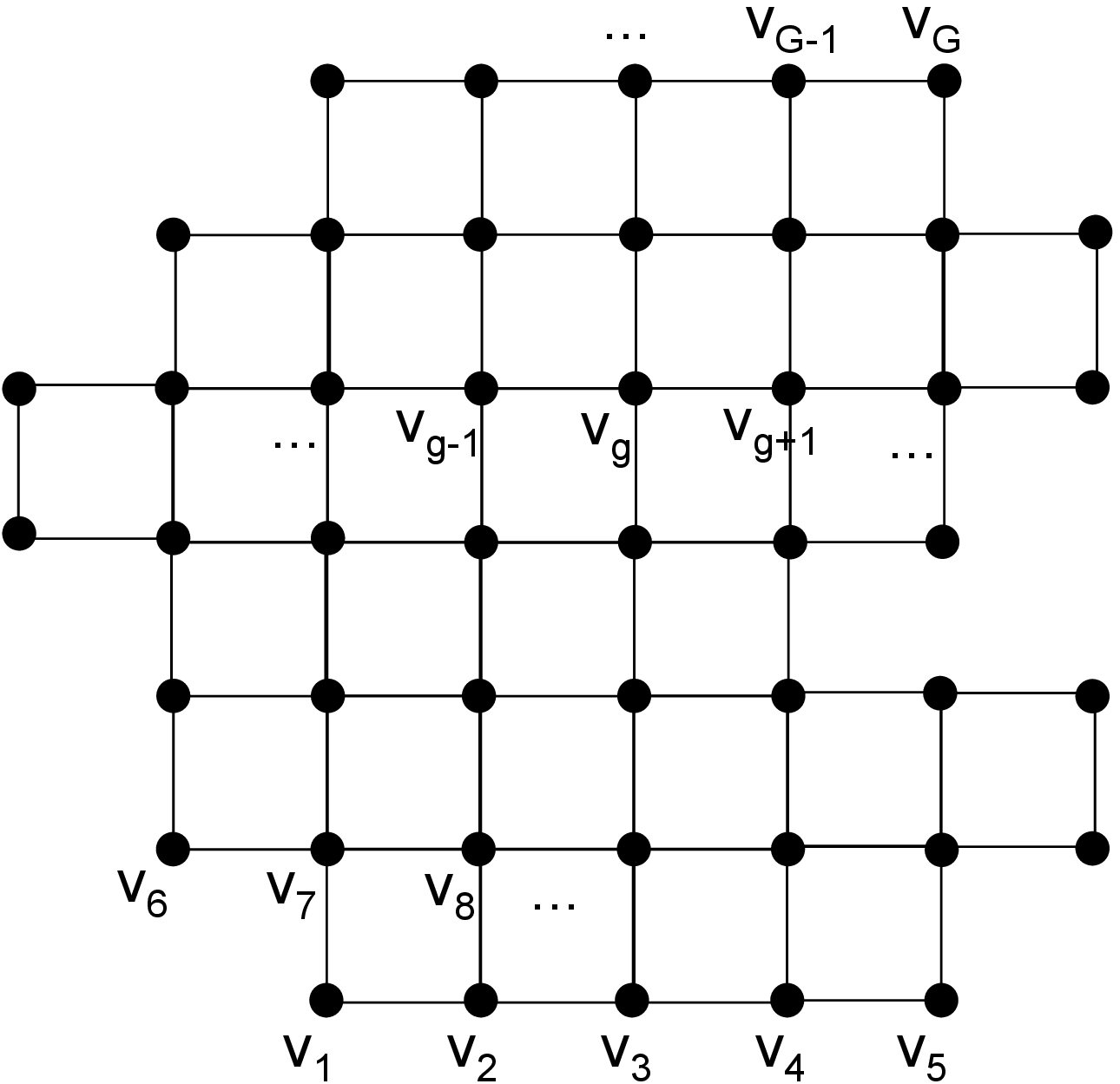}}\tabularnewline
\textcolor{black}{(}\textcolor{black}{\emph{a}}\textcolor{black}{)}\tabularnewline
\tabularnewline
\textcolor{black}{\includegraphics[%
  width=0.38\columnwidth]{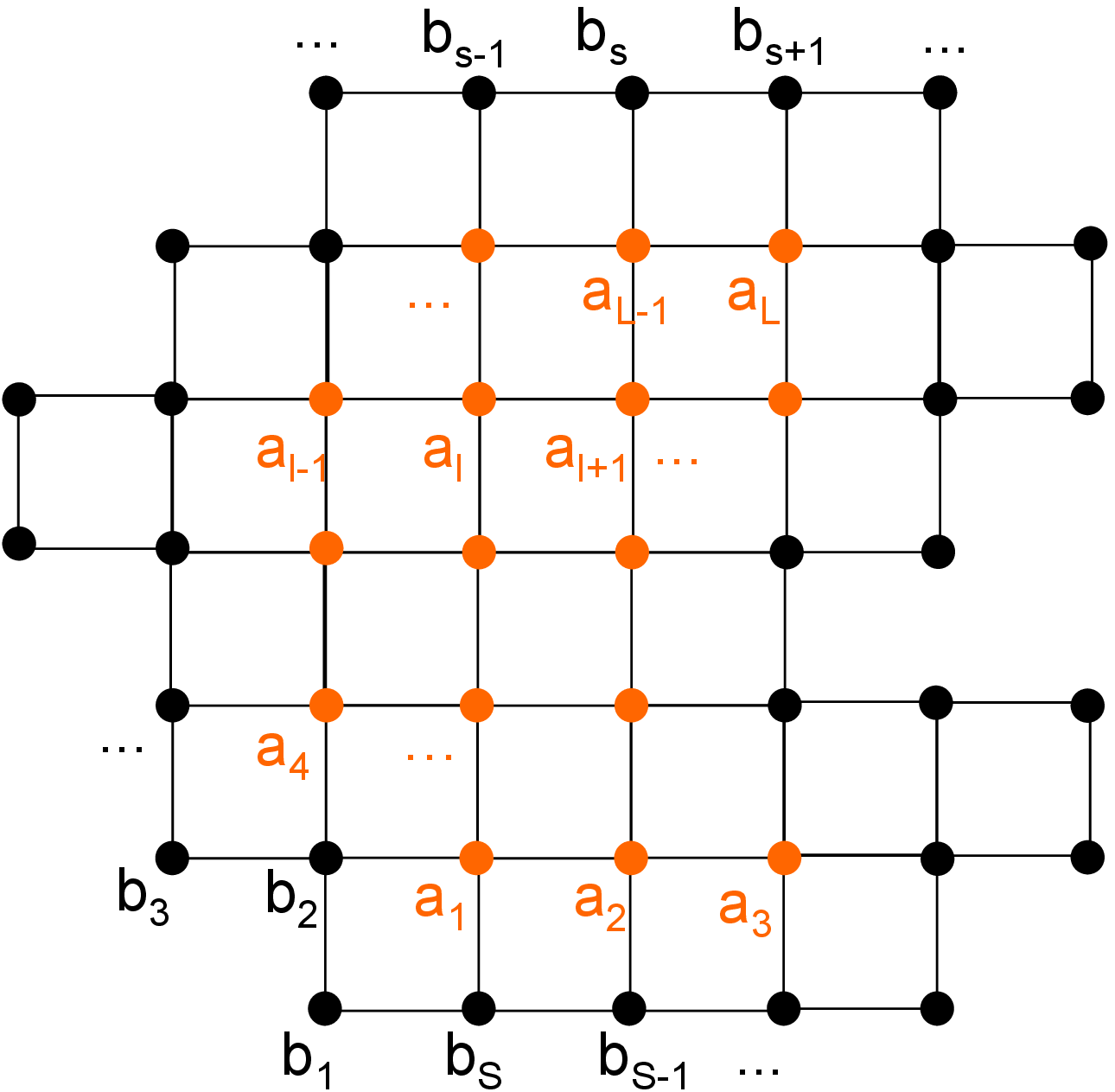}}\tabularnewline
\textcolor{black}{(}\textcolor{black}{\emph{b}}\textcolor{black}{)}\tabularnewline
\tabularnewline
\textcolor{black}{\includegraphics[%
  width=0.46\columnwidth]{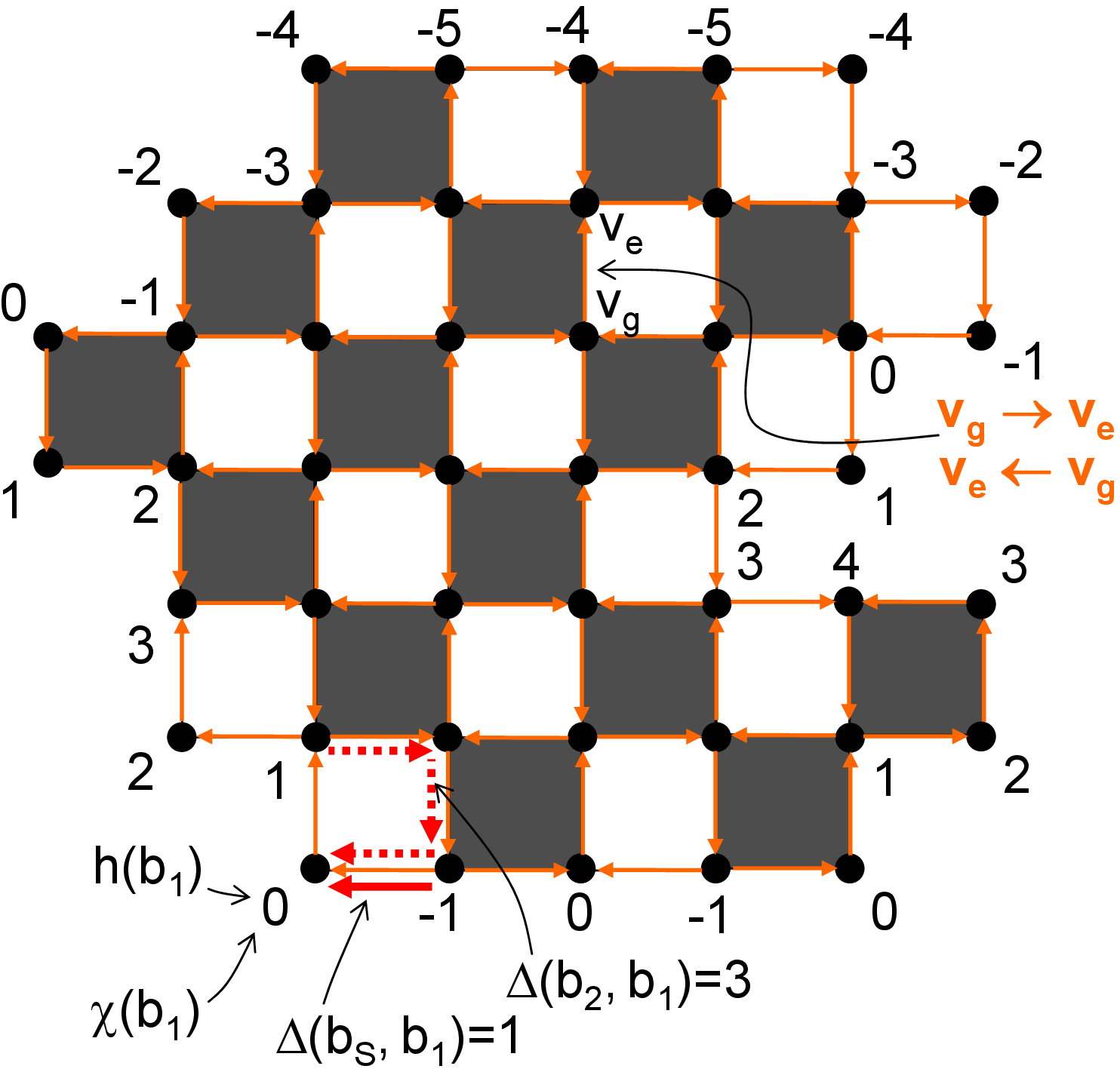}}\tabularnewline
\textcolor{black}{(}\textcolor{black}{\emph{c}}\textcolor{black}{)}\tabularnewline
\end{tabular}\end{center}

\begin{center}\textbf{\textcolor{black}{Fig. 2 - P. Rocca et}} \textbf{\textcolor{black}{\emph{al.}}}\textbf{\textcolor{black}{,}}
\textbf{\textcolor{black}{\emph{{}``}}}\textcolor{black}{Pareto-Optimal
Domino-Tiling of ...''}\end{center}
\newpage

\begin{center}\textcolor{black}{~\vfill}\end{center}

\begin{center}\textcolor{black}{}\begin{tabular}{c}
\textcolor{black}{\includegraphics[%
  width=0.60\columnwidth]{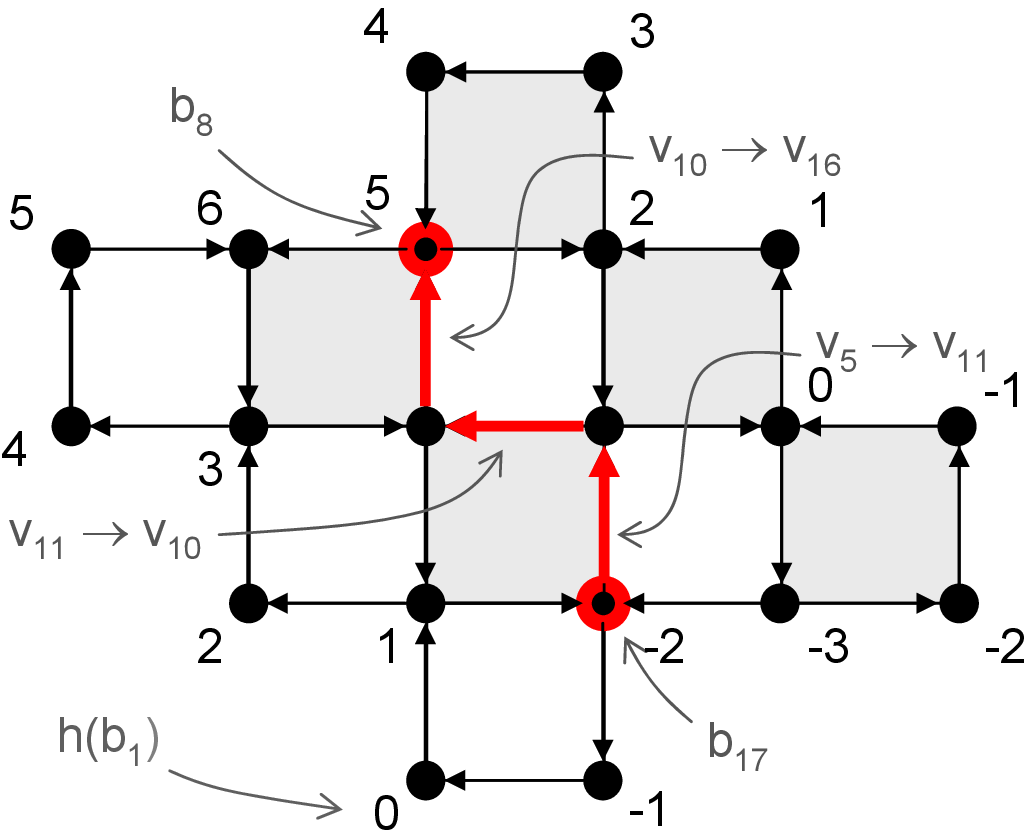}}\tabularnewline
\end{tabular}\end{center}

\begin{center}\textcolor{black}{~\vfill}\end{center}

\begin{center}\textbf{\textcolor{black}{Fig. 3 - P. Rocca et}} \textbf{\textcolor{black}{\emph{al.}}}\textbf{\textcolor{black}{,}}
\textbf{\textcolor{black}{\emph{{}``}}}\textcolor{black}{Pareto-Optimal
Domino-Tiling of ...''}\end{center}

\newpage
\textcolor{black}{~\vfill}

\begin{center}\textcolor{black}{}\begin{tabular}{c}
\textcolor{black}{\includegraphics[%
  scale=0.7]{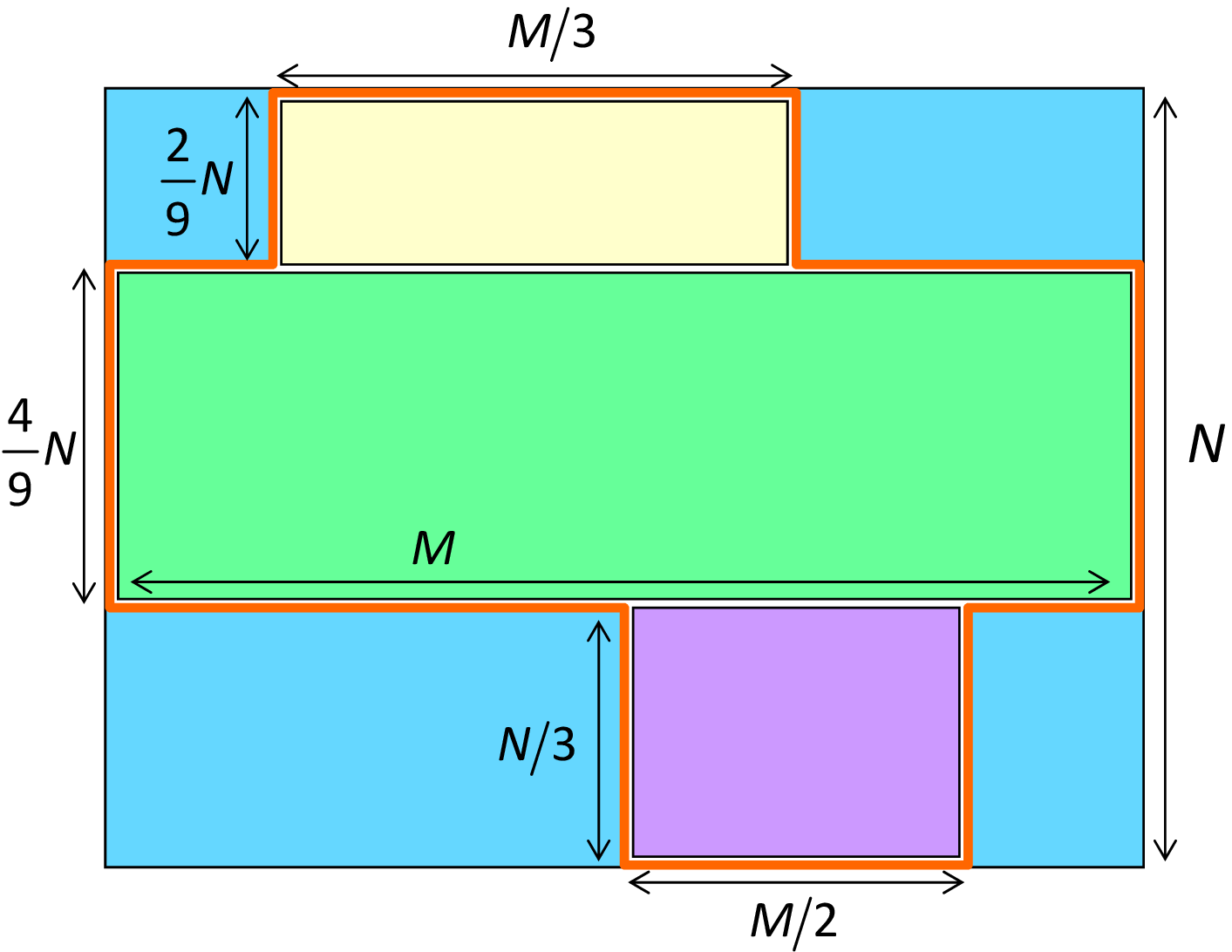}}\tabularnewline
\textcolor{black}{(}\textcolor{black}{\emph{a}}\textcolor{black}{)}\tabularnewline
\tabularnewline
\textcolor{black}{\includegraphics[%
  scale=0.8]{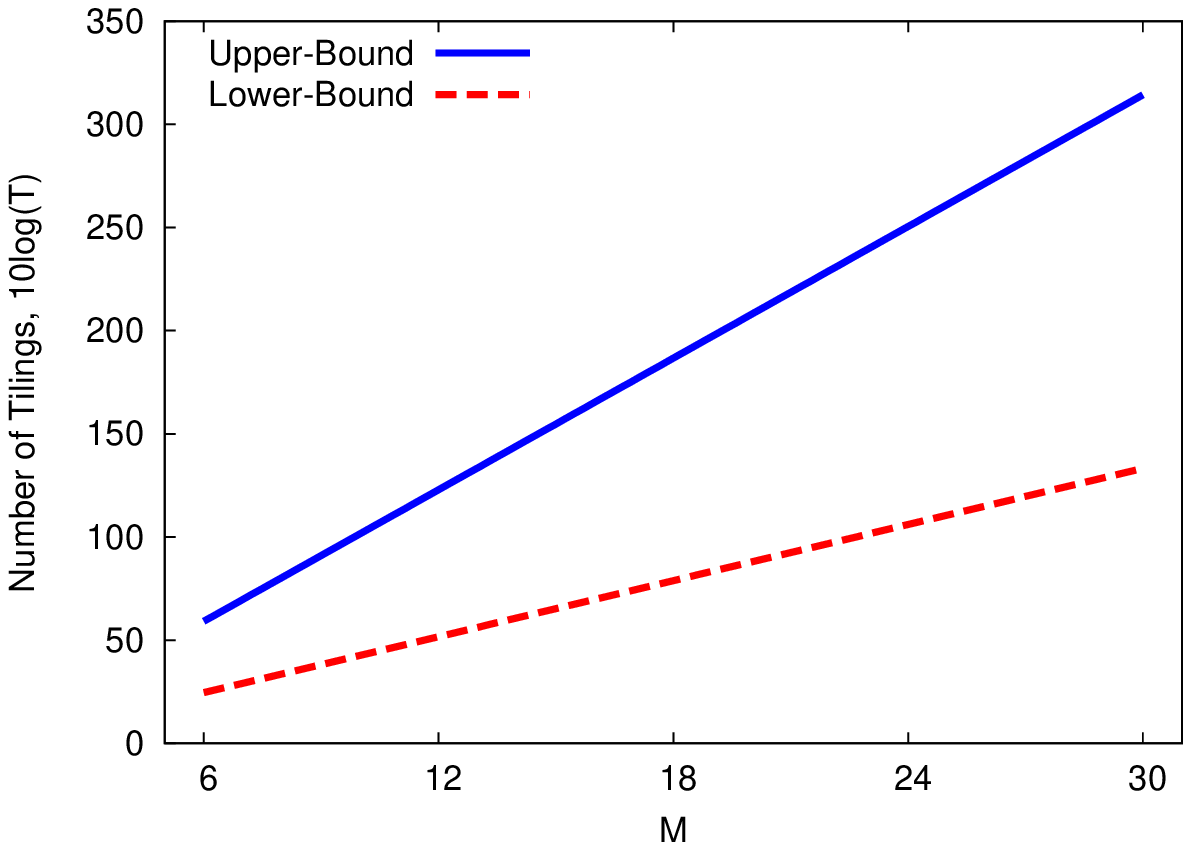}}\tabularnewline
\textcolor{black}{(}\textcolor{black}{\emph{b}}\textcolor{black}{)}\tabularnewline
\end{tabular}\end{center}

\begin{center}\textcolor{black}{~\vfill}\end{center}

\begin{center}\textbf{\textcolor{black}{Fig. 4 - P. Rocca et}} \textbf{\textcolor{black}{\emph{al.}}}\textbf{\textcolor{black}{,}}
\textbf{\textcolor{black}{\emph{{}``}}}\textcolor{black}{Pareto-Optimal
Domino-Tiling of ...''}\end{center}
\newpage

\begin{center}\textcolor{black}{}\begin{tabular}{c}
\textcolor{black}{\includegraphics[%
  width=0.48\columnwidth,
  keepaspectratio]{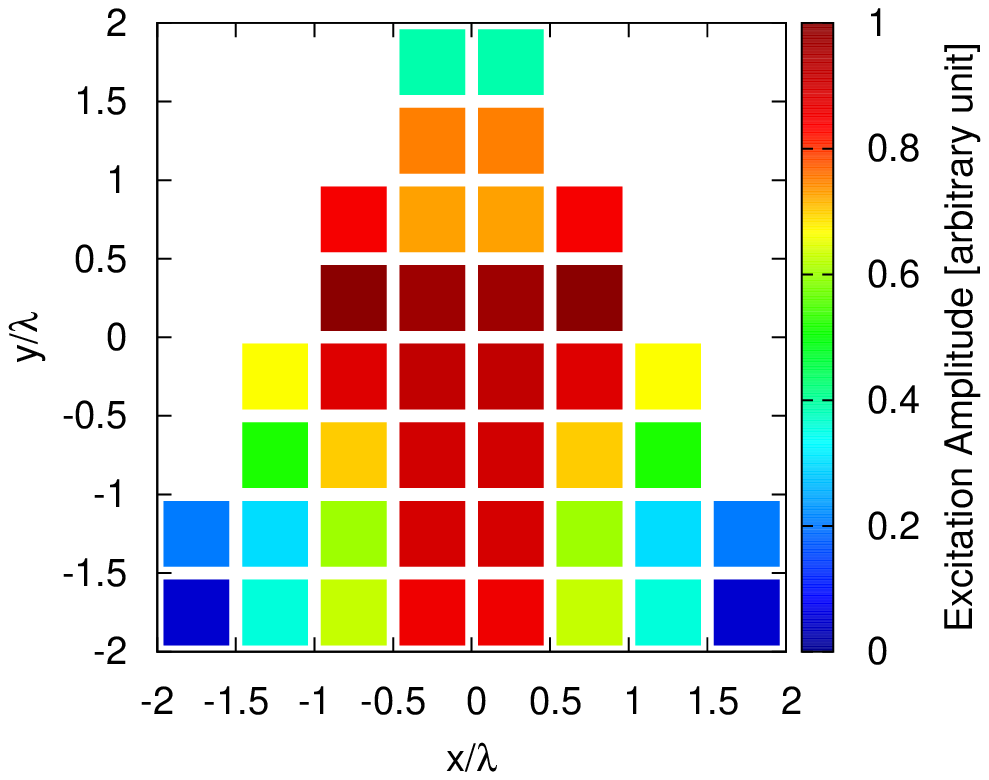}}\tabularnewline
\textcolor{black}{(}\textcolor{black}{\emph{a}}\textcolor{black}{)}\tabularnewline
\tabularnewline
\textcolor{black}{\includegraphics[%
  width=0.35\columnwidth,
  keepaspectratio]{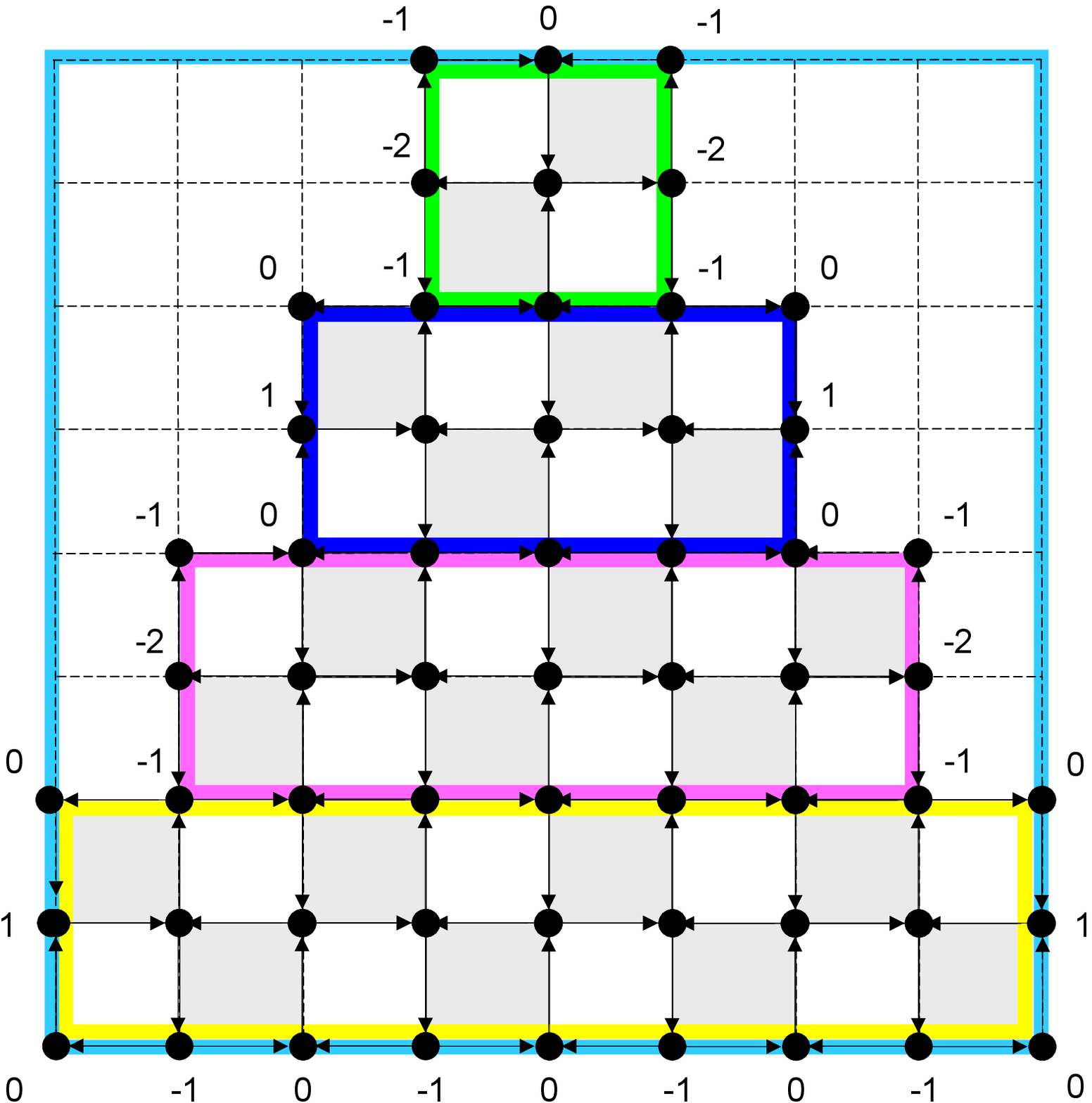}}\tabularnewline
\textcolor{black}{(}\textcolor{black}{\emph{b}}\textcolor{black}{)}\tabularnewline
\tabularnewline
\textcolor{black}{\includegraphics[%
  width=0.48\columnwidth,
  keepaspectratio]{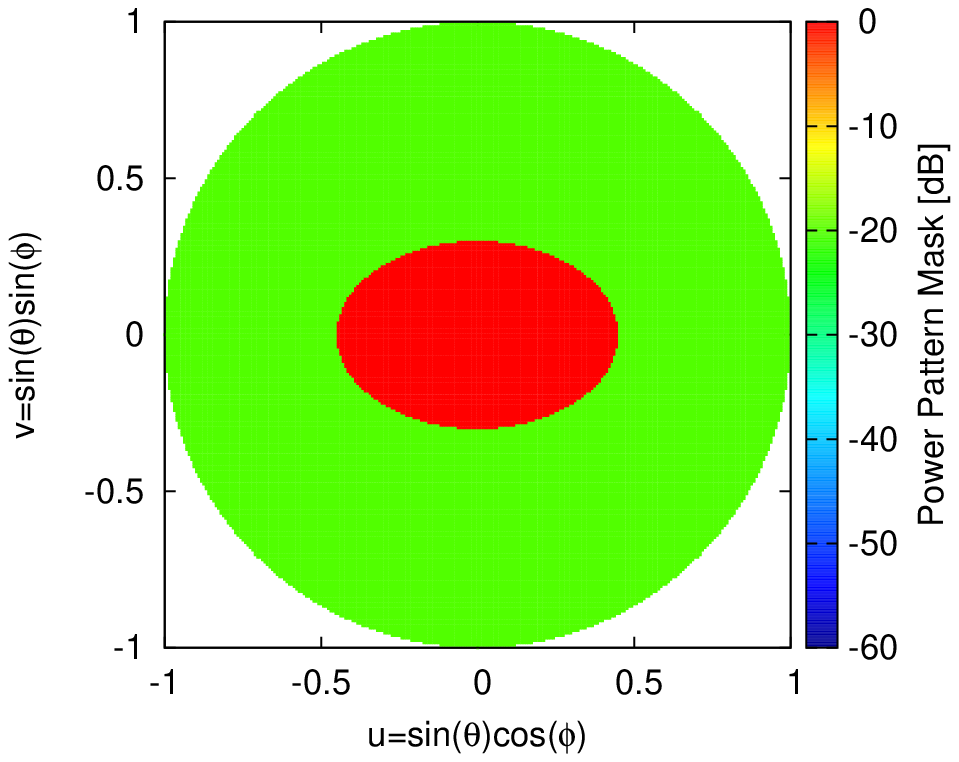}}\tabularnewline
\textcolor{black}{(}\textcolor{black}{\emph{c}}\textcolor{black}{)}\tabularnewline
\end{tabular}\end{center}

\begin{center}\textbf{\textcolor{black}{Fig. 5 - P. Rocca et}} \textbf{\textcolor{black}{\emph{al.}}}\textbf{\textcolor{black}{,}}
\textbf{\textcolor{black}{\emph{{}``}}}\textcolor{black}{Pareto-Optimal
Domino-Tiling of ...''}\end{center}
\newpage

\begin{center}\textcolor{black}{~\vfill}\end{center}

\begin{center}\textcolor{black}{}\begin{tabular}{c}
\textcolor{black}{\includegraphics[%
  width=0.50\columnwidth,
  keepaspectratio]{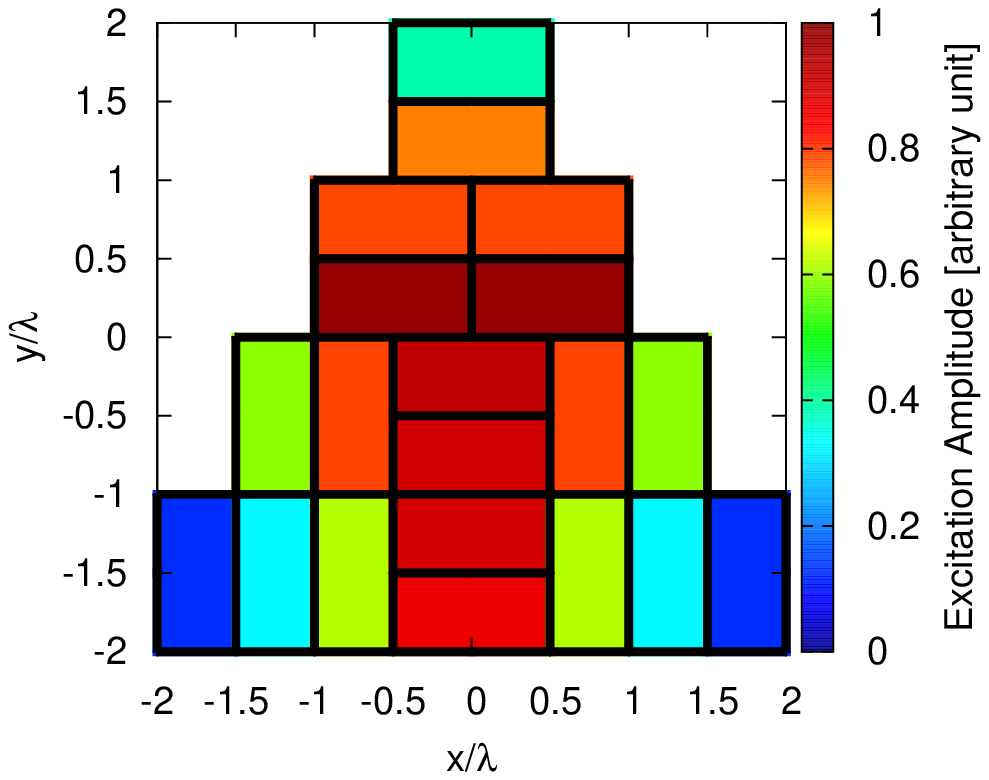}}\tabularnewline
\textcolor{black}{(}\textcolor{black}{\emph{a}}\textcolor{black}{)}\tabularnewline
\tabularnewline
\textcolor{black}{\includegraphics[%
  width=0.50\columnwidth,
  keepaspectratio]{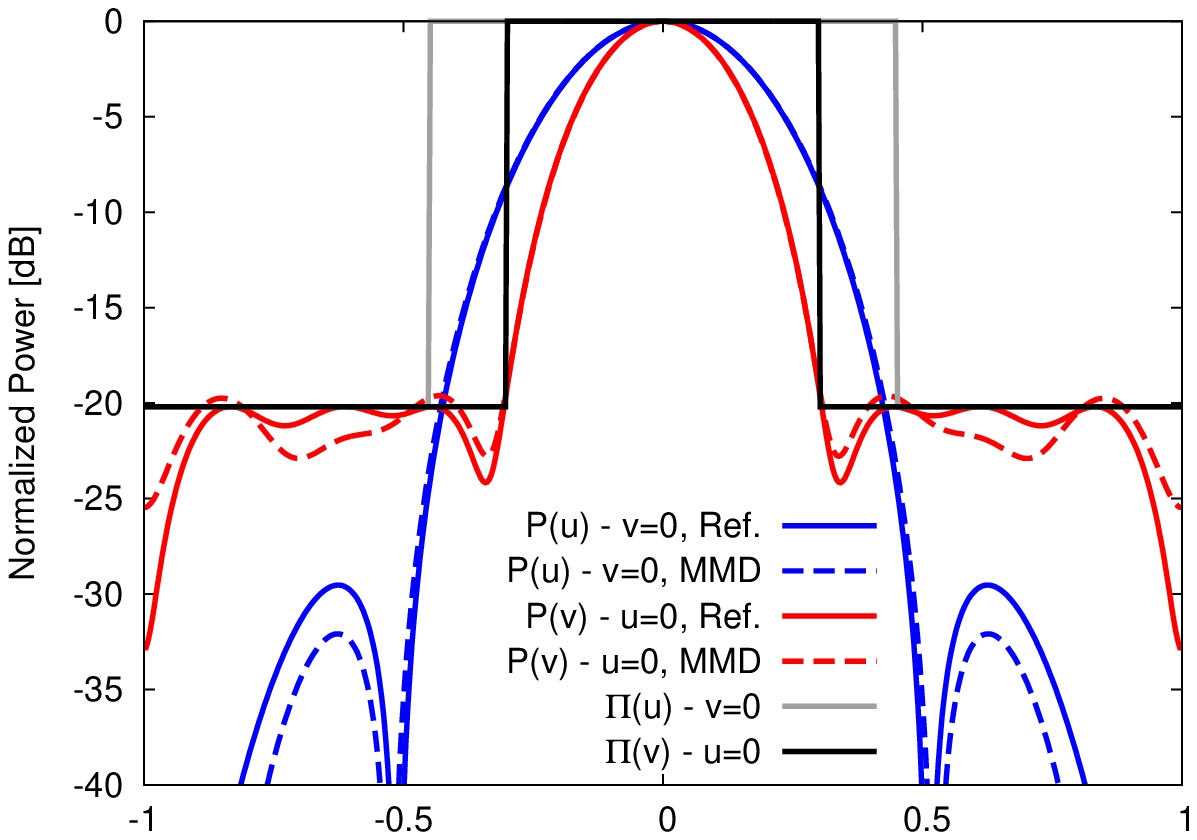}}\tabularnewline
\textcolor{black}{(}\textcolor{black}{\emph{b}}\textcolor{black}{)}\tabularnewline
\end{tabular}\end{center}

\begin{center}\textcolor{black}{~\vfill}\end{center}

\begin{center}\textbf{\textcolor{black}{Fig. 6 - P. Rocca et}} \textbf{\textcolor{black}{\emph{al.}}}\textbf{\textcolor{black}{,}}
\textbf{\textcolor{black}{\emph{{}``}}}\textcolor{black}{Pareto-Optimal
Domino-Tiling of ...''}\end{center}
\newpage

\begin{center}\textcolor{black}{~\vfill}\end{center}

\begin{center}\textcolor{black}{}\begin{tabular}{c}
\textcolor{black}{\includegraphics[%
  width=0.80\columnwidth,
  keepaspectratio]{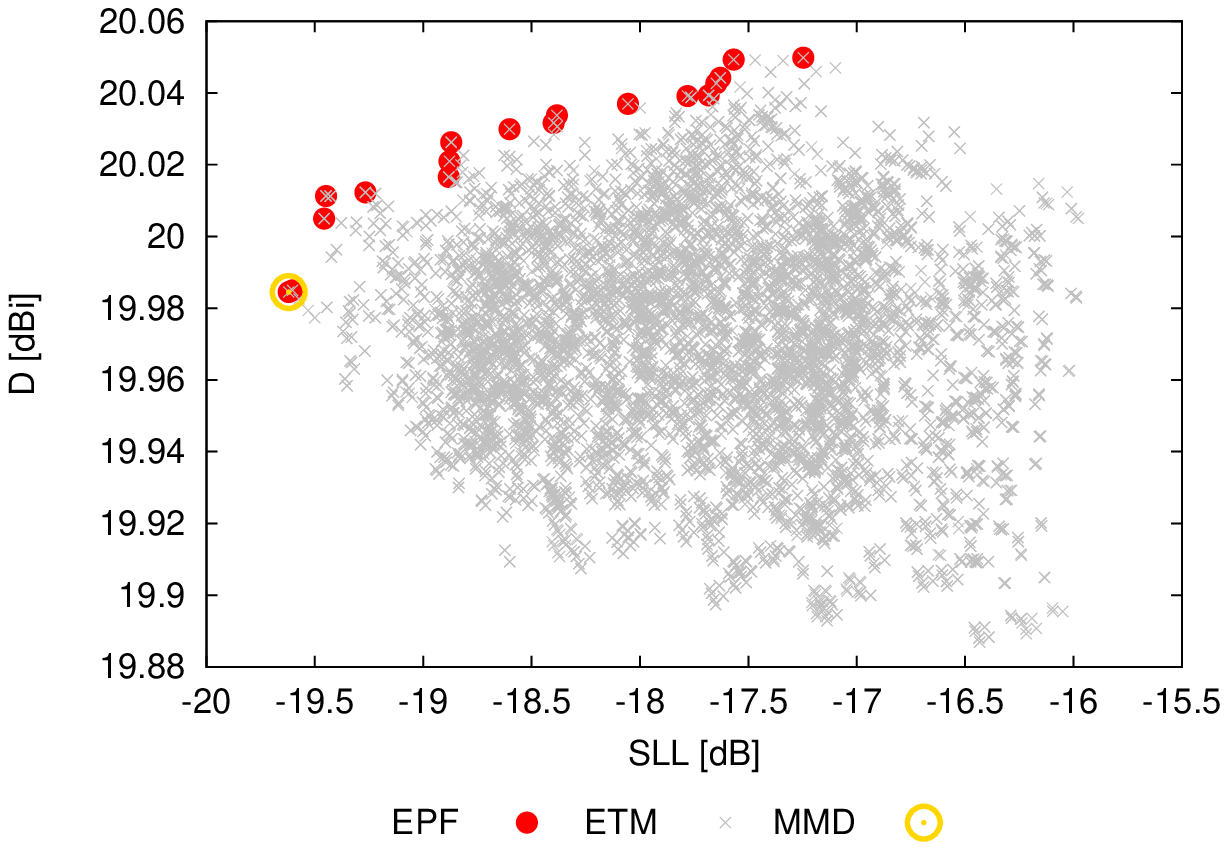}}\tabularnewline
\end{tabular}\end{center}

\begin{center}\textcolor{black}{~\vfill}\end{center}

\begin{center}\textbf{\textcolor{black}{Fig. 7 - P. Rocca et}} \textbf{\textcolor{black}{\emph{al.}}}\textbf{\textcolor{black}{,}}
\textbf{\textcolor{black}{\emph{{}``}}}\textcolor{black}{Pareto-Optimal
Domino-Tiling of ...''}\end{center}
\newpage

\begin{center}\textcolor{black}{}\begin{tabular}{c}
\textcolor{black}{\includegraphics[%
  width=0.48\columnwidth,
  keepaspectratio]{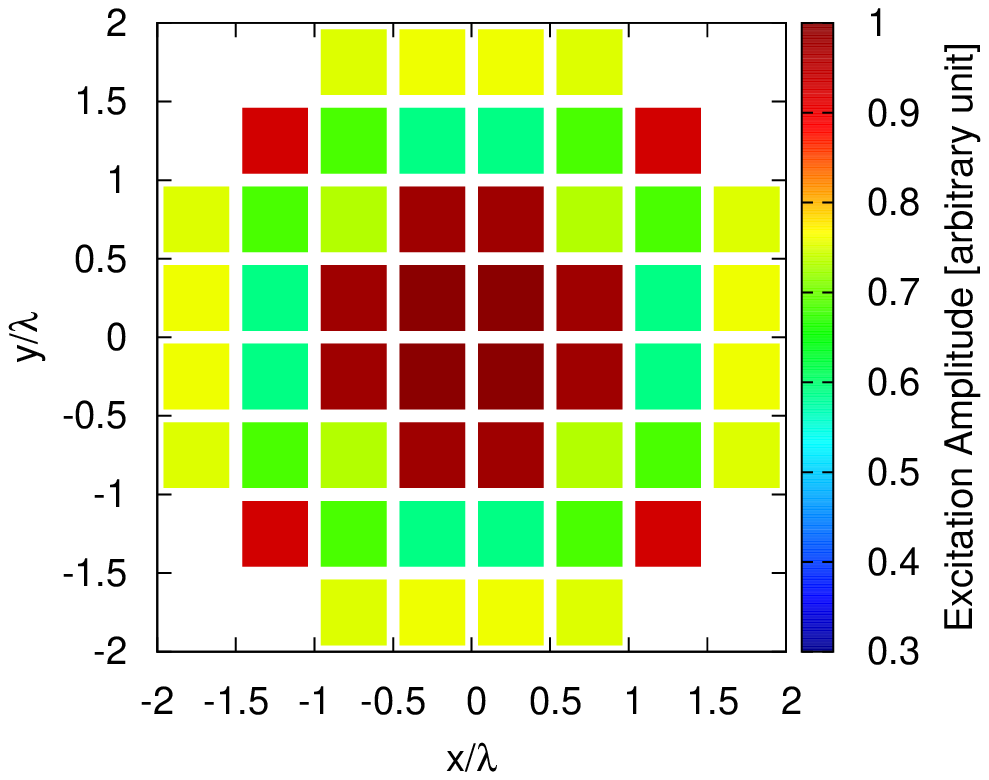}}\tabularnewline
\textcolor{black}{(}\textcolor{black}{\emph{a}}\textcolor{black}{)}\tabularnewline
\tabularnewline
\textcolor{black}{\includegraphics[%
  width=0.48\columnwidth,
  keepaspectratio]{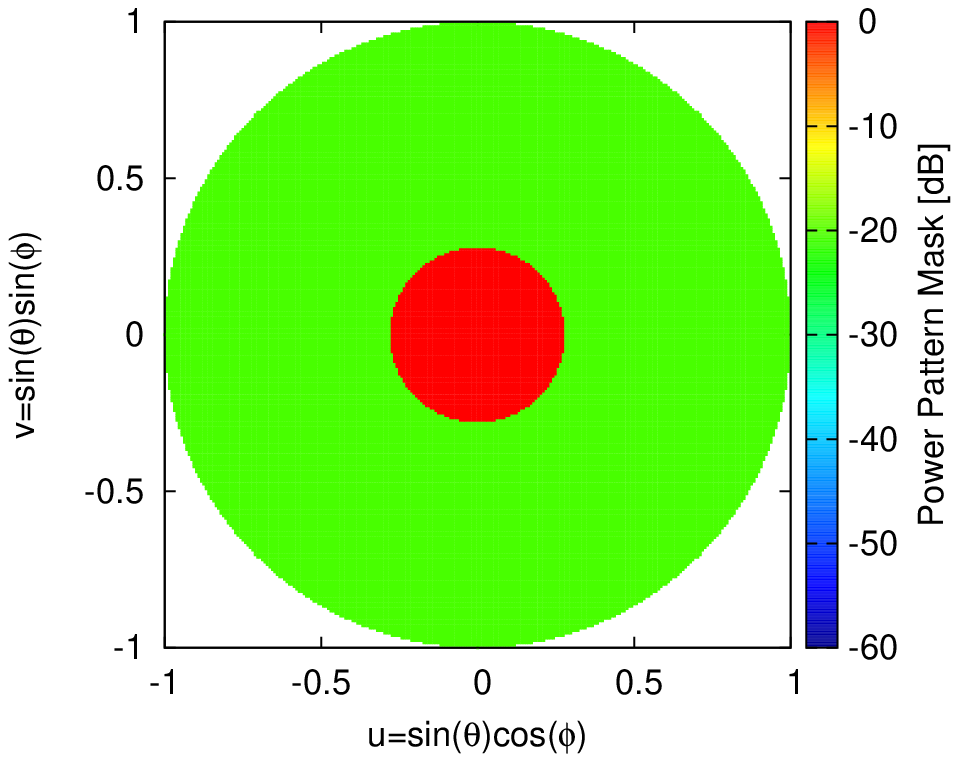}}\tabularnewline
\textcolor{black}{(}\textcolor{black}{\emph{b}}\textcolor{black}{)}\tabularnewline
\tabularnewline
\textcolor{black}{\includegraphics[%
  width=0.48\columnwidth,
  keepaspectratio]{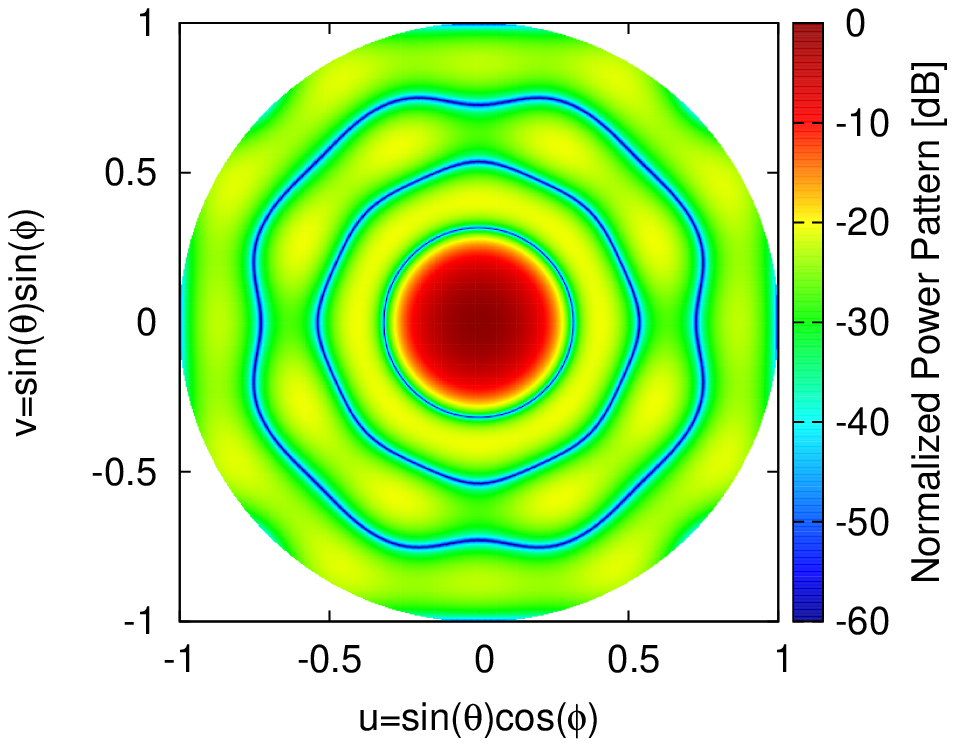}}\tabularnewline
\textcolor{black}{(}\textcolor{black}{\emph{c}}\textcolor{black}{)}\tabularnewline
\end{tabular}\end{center}

\begin{center}\textbf{\textcolor{black}{Fig. 8 - P. Rocca et}} \textbf{\textcolor{black}{\emph{al.}}}\textbf{\textcolor{black}{,}}
\textbf{\textcolor{black}{\emph{{}``}}}\textcolor{black}{Pareto-Optimal
Domino-Tiling of ...''}\end{center}
\newpage

\begin{center}\textcolor{black}{~\vfill}\end{center}

\begin{center}\textcolor{black}{}\begin{tabular}{c}
\textcolor{black}{\includegraphics[%
  width=0.80\columnwidth,
  keepaspectratio]{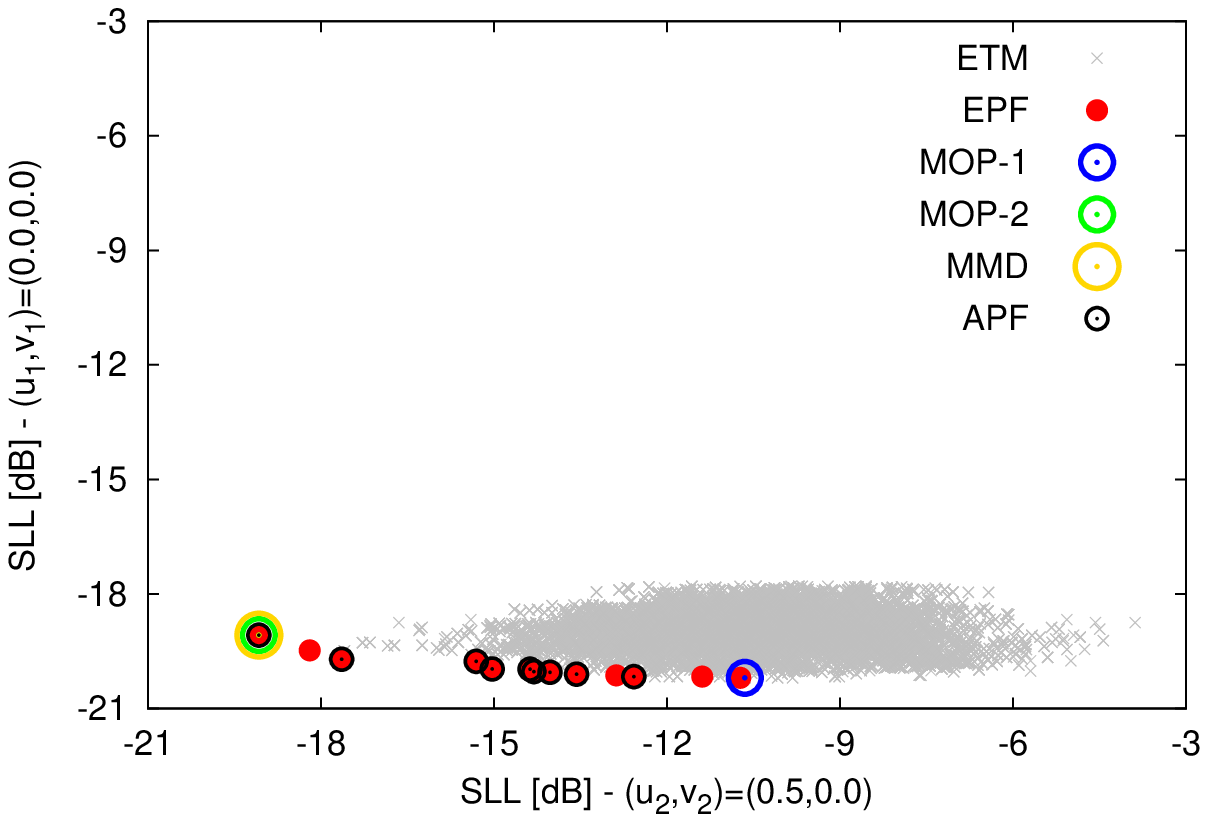}}\tabularnewline
\end{tabular}\end{center}

\begin{center}\textcolor{black}{~\vfill}\end{center}

\begin{center}\textbf{\textcolor{black}{Fig. 9 - P. Rocca et}} \textbf{\textcolor{black}{\emph{al.}}}\textbf{\textcolor{black}{,}}
\textbf{\textcolor{black}{\emph{{}``}}}\textcolor{black}{Pareto-Optimal
Domino-Tiling of ...''}\end{center}
\newpage

\begin{center}\textcolor{black}{~\vfill}\end{center}

\begin{center}\textcolor{black}{}\begin{tabular}{cc}
\textcolor{black}{\includegraphics[%
  width=0.48\columnwidth,
  keepaspectratio]{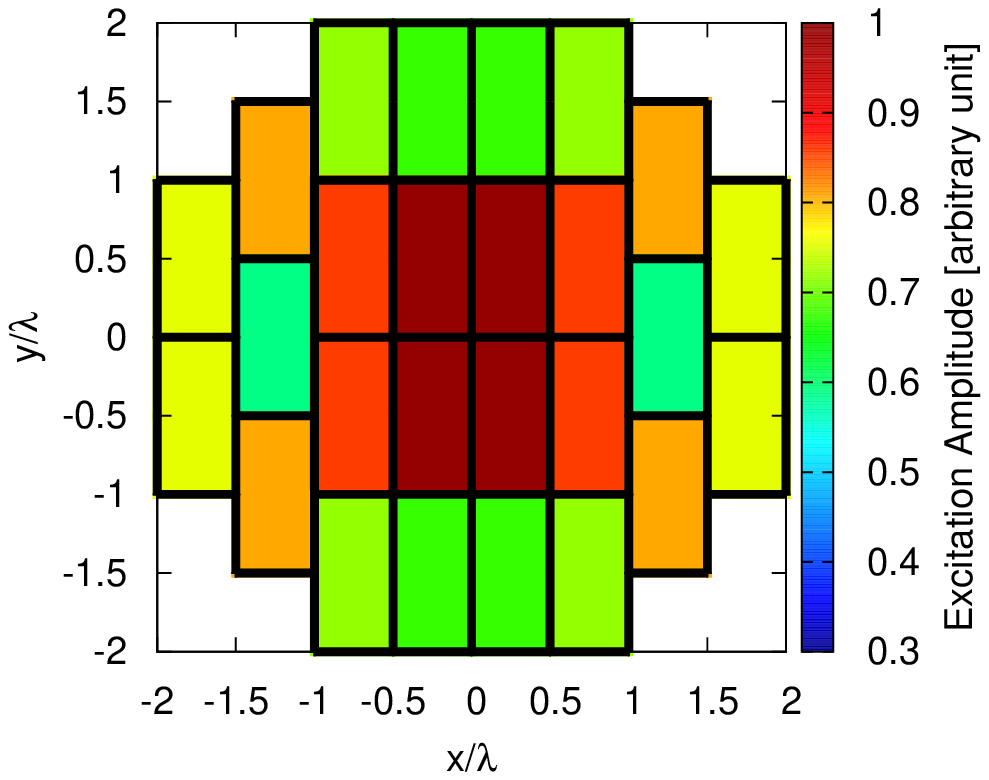}}&
\textcolor{black}{\includegraphics[%
  width=0.48\columnwidth,
  keepaspectratio]{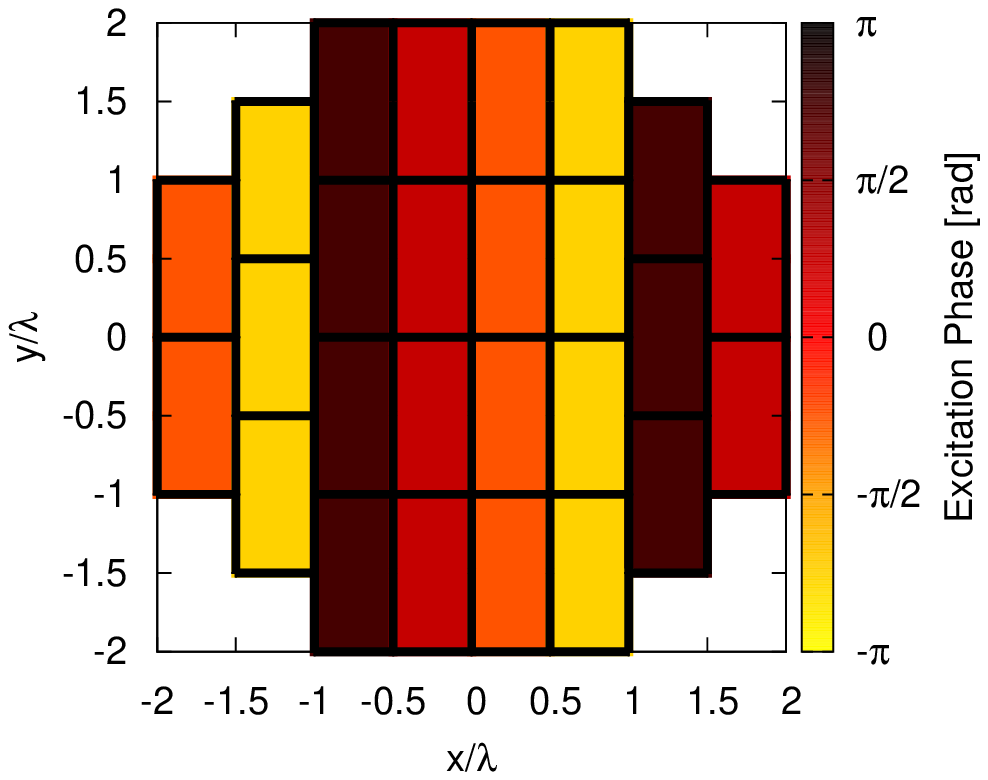}}\tabularnewline
\textcolor{black}{(}\textcolor{black}{\emph{a}}\textcolor{black}{)}&
\textcolor{black}{(}\textcolor{black}{\emph{b}}\textcolor{black}{)}\tabularnewline
&
\tabularnewline
\textcolor{black}{\includegraphics[%
  width=0.48\columnwidth,
  keepaspectratio]{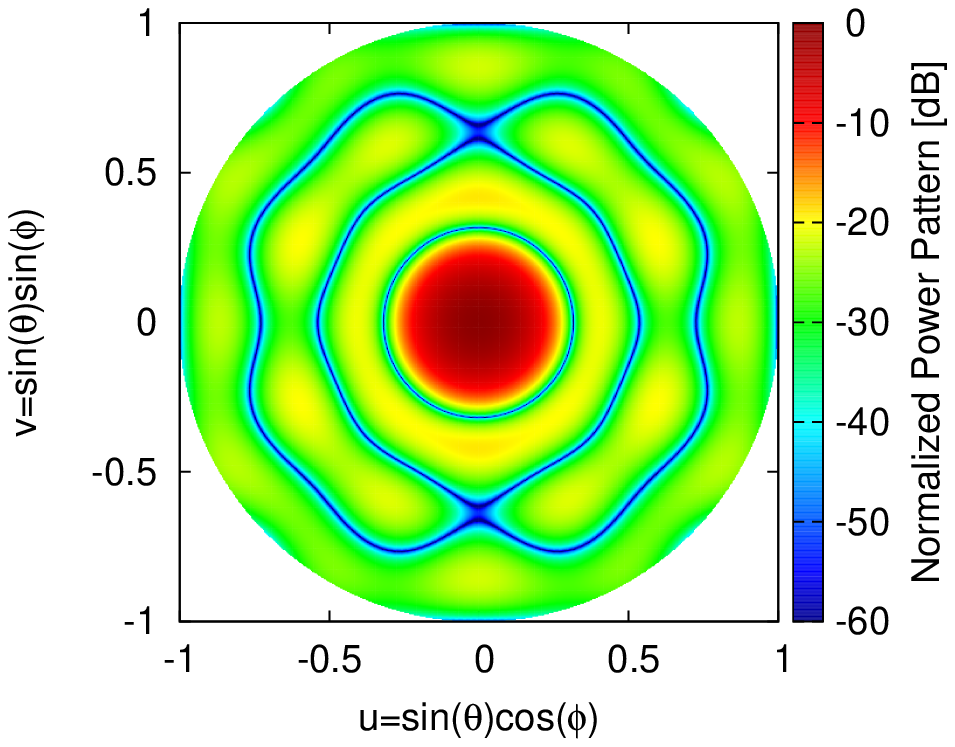}}&
\textcolor{black}{\includegraphics[%
  width=0.48\columnwidth,
  keepaspectratio]{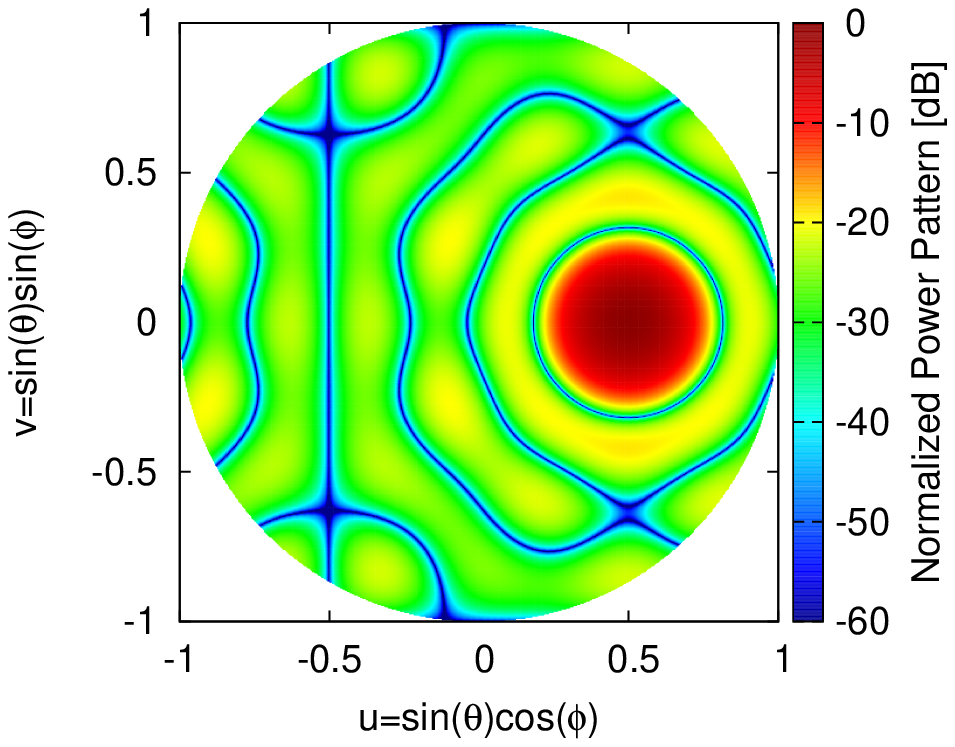}}\tabularnewline
\textcolor{black}{(}\textcolor{black}{\emph{c}}\textcolor{black}{)}&
\textcolor{black}{(}\textcolor{black}{\emph{d}}\textcolor{black}{)}\tabularnewline
\end{tabular}\end{center}

\begin{center}\textcolor{black}{~\vfill}\end{center}

\begin{center}\textbf{\textcolor{black}{Fig. 10 - P. Rocca et}} \textbf{\textcolor{black}{\emph{al.}}}\textbf{\textcolor{black}{,}}
\textbf{\textcolor{black}{\emph{{}``}}}\textcolor{black}{Pareto-Optimal
Domino-Tiling of ...''}\end{center}
\newpage

\begin{center}\textcolor{black}{~\vfill}\end{center}

\begin{center}\textcolor{black}{}\begin{tabular}{cc}
\textcolor{black}{\includegraphics[%
  width=0.48\columnwidth,
  keepaspectratio]{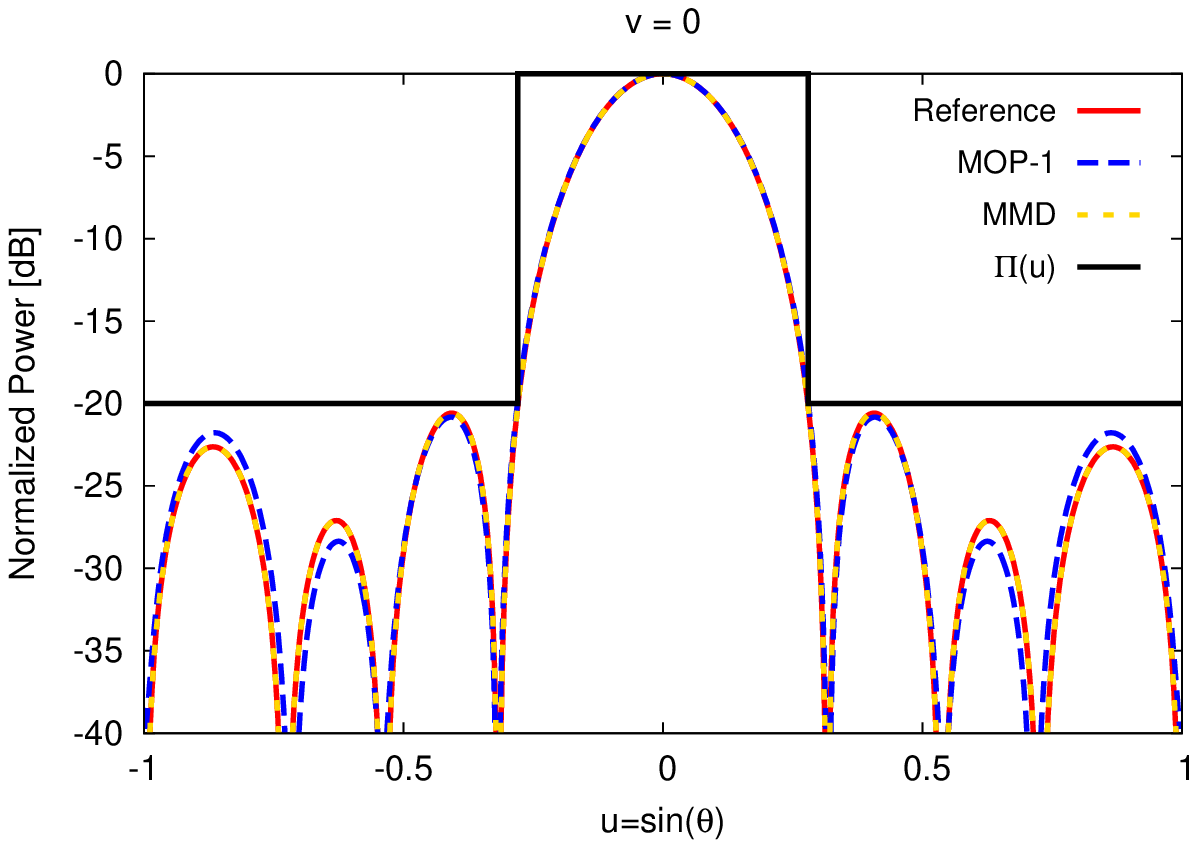}}&
\textcolor{black}{\includegraphics[%
  width=0.48\columnwidth,
  keepaspectratio]{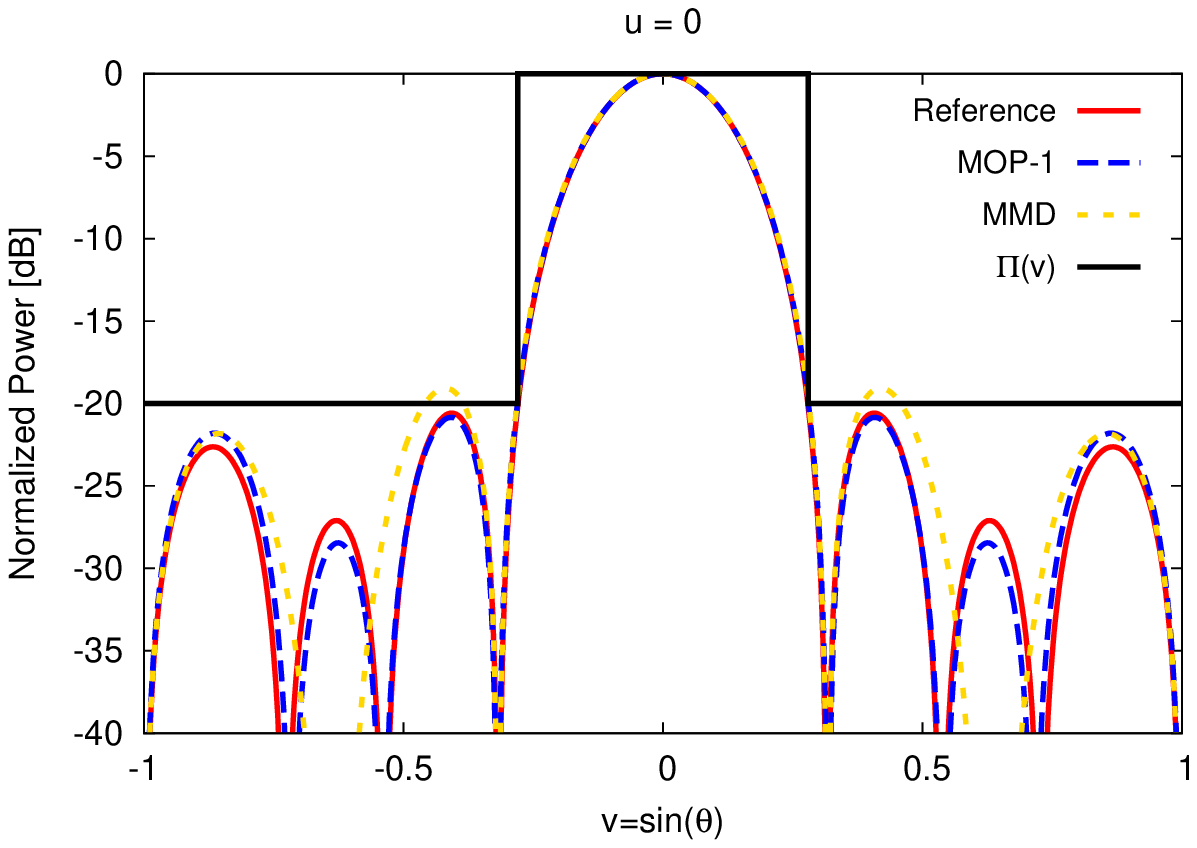}}\tabularnewline
\textcolor{black}{(}\textcolor{black}{\emph{a}}\textcolor{black}{)}&
\textcolor{black}{(}\textcolor{black}{\emph{b}}\textcolor{black}{)}\tabularnewline
&
\tabularnewline
\multicolumn{2}{c}{\textcolor{black}{\includegraphics[%
  width=0.48\columnwidth,
  keepaspectratio]{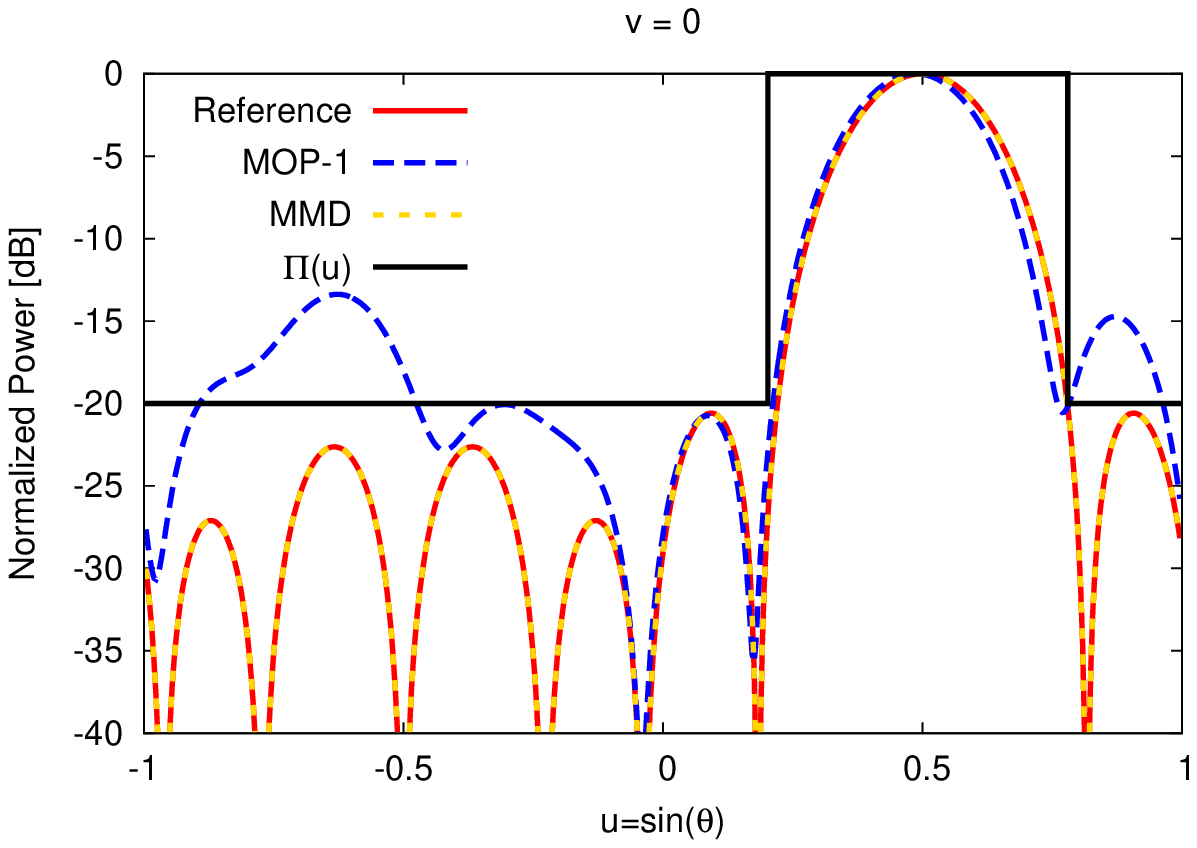}}}\tabularnewline
\multicolumn{2}{c}{\textcolor{black}{(}\textcolor{black}{\emph{c}}\textcolor{black}{)}}\tabularnewline
\end{tabular}\end{center}

\begin{center}\textcolor{black}{~\vfill}\end{center}

\begin{center}\textbf{\textcolor{black}{Fig. 11 - P. Rocca et}} \textbf{\textcolor{black}{\emph{al.}}}\textbf{\textcolor{black}{,}}
\textbf{\textcolor{black}{\emph{{}``}}}\textcolor{black}{Pareto-Optimal
Domino-Tiling of ...''}\end{center}
\newpage

\begin{center}\textcolor{black}{~\vfill}\end{center}

\begin{center}\textcolor{black}{}\begin{tabular}{c}
\textcolor{black}{\includegraphics[%
  width=0.48\columnwidth,
  keepaspectratio]{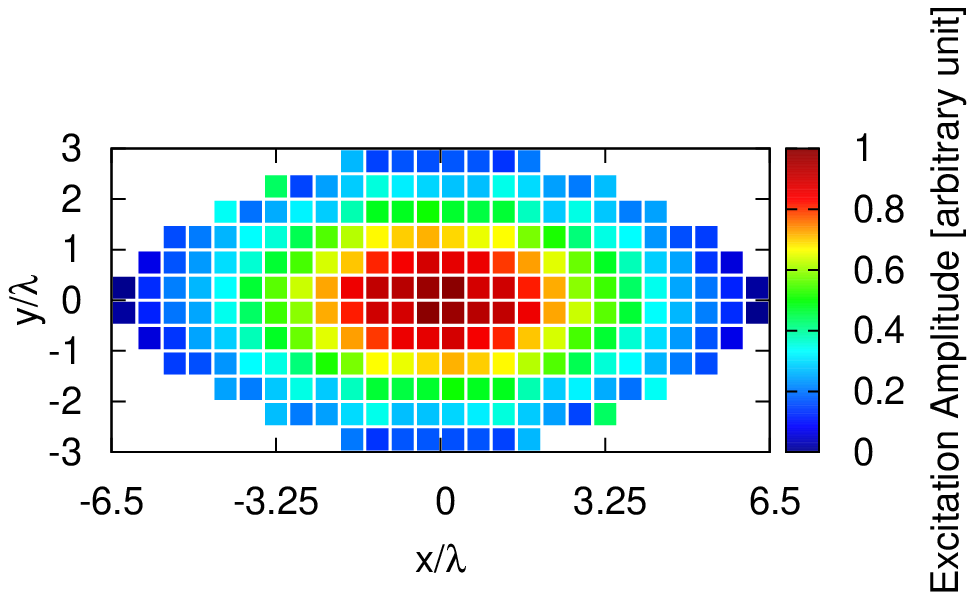}}\tabularnewline
\textcolor{black}{(}\textcolor{black}{\emph{a}}\textcolor{black}{)}\tabularnewline
\tabularnewline
\textcolor{black}{\includegraphics[%
  width=0.48\columnwidth,
  keepaspectratio]{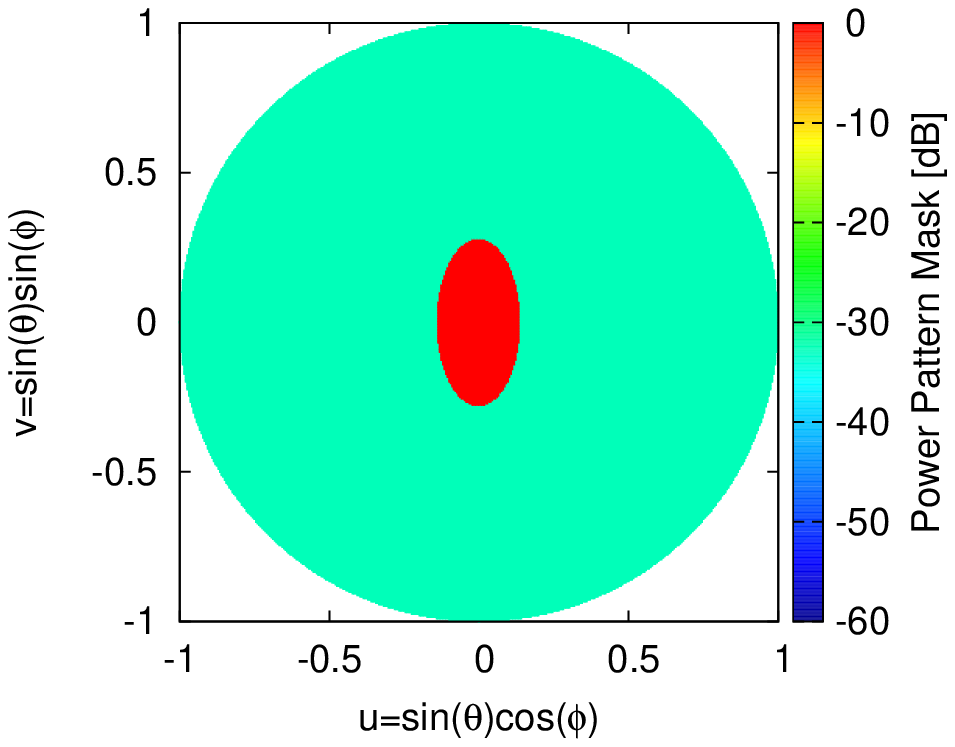}}\tabularnewline
\textcolor{black}{(}\textcolor{black}{\emph{b}}\textcolor{black}{)}\tabularnewline
\tabularnewline
\textcolor{black}{\includegraphics[%
  width=0.48\columnwidth,
  keepaspectratio]{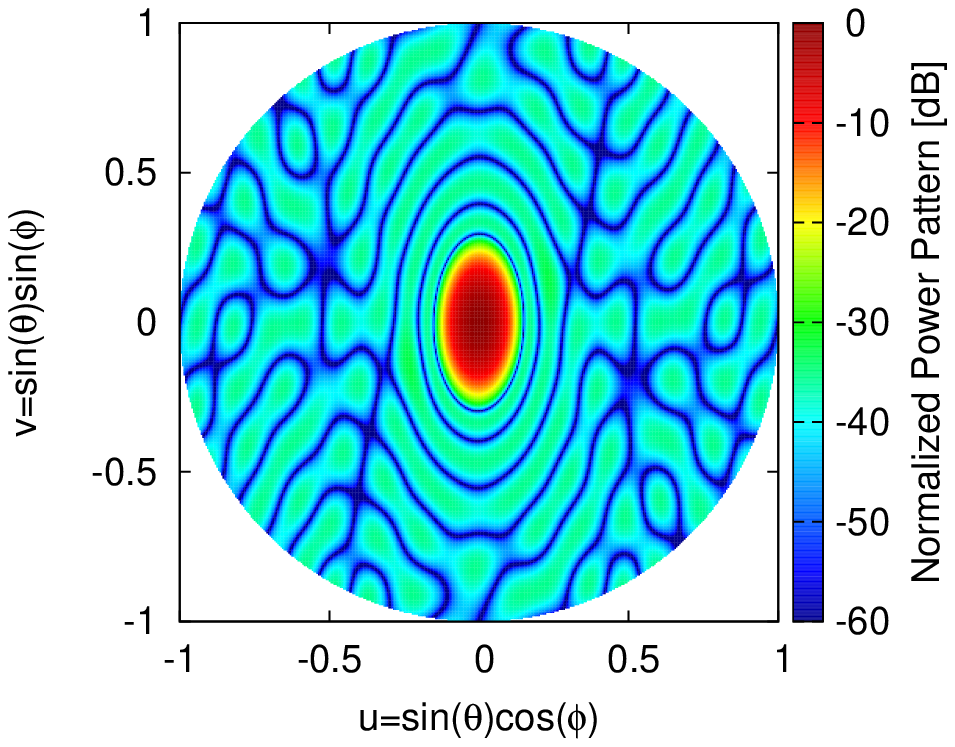}}\tabularnewline
\textcolor{black}{(}\textcolor{black}{\emph{c}}\textcolor{black}{)}\tabularnewline
\end{tabular}\end{center}

\begin{center}\textcolor{black}{~\vfill}\end{center}

\begin{center}\textbf{\textcolor{black}{Fig. 12 - P. Rocca et}} \textbf{\textcolor{black}{\emph{al.}}}\textbf{\textcolor{black}{,}}
\textbf{\textcolor{black}{\emph{{}``}}}\textcolor{black}{Pareto-Optimal
Domino-Tiling of ...''}\end{center}
\newpage

\begin{center}\textcolor{black}{~\vfill}\end{center}

\begin{center}\textcolor{black}{}\begin{tabular}{c}
\textcolor{black}{\includegraphics[%
  width=0.80\columnwidth,
  keepaspectratio]{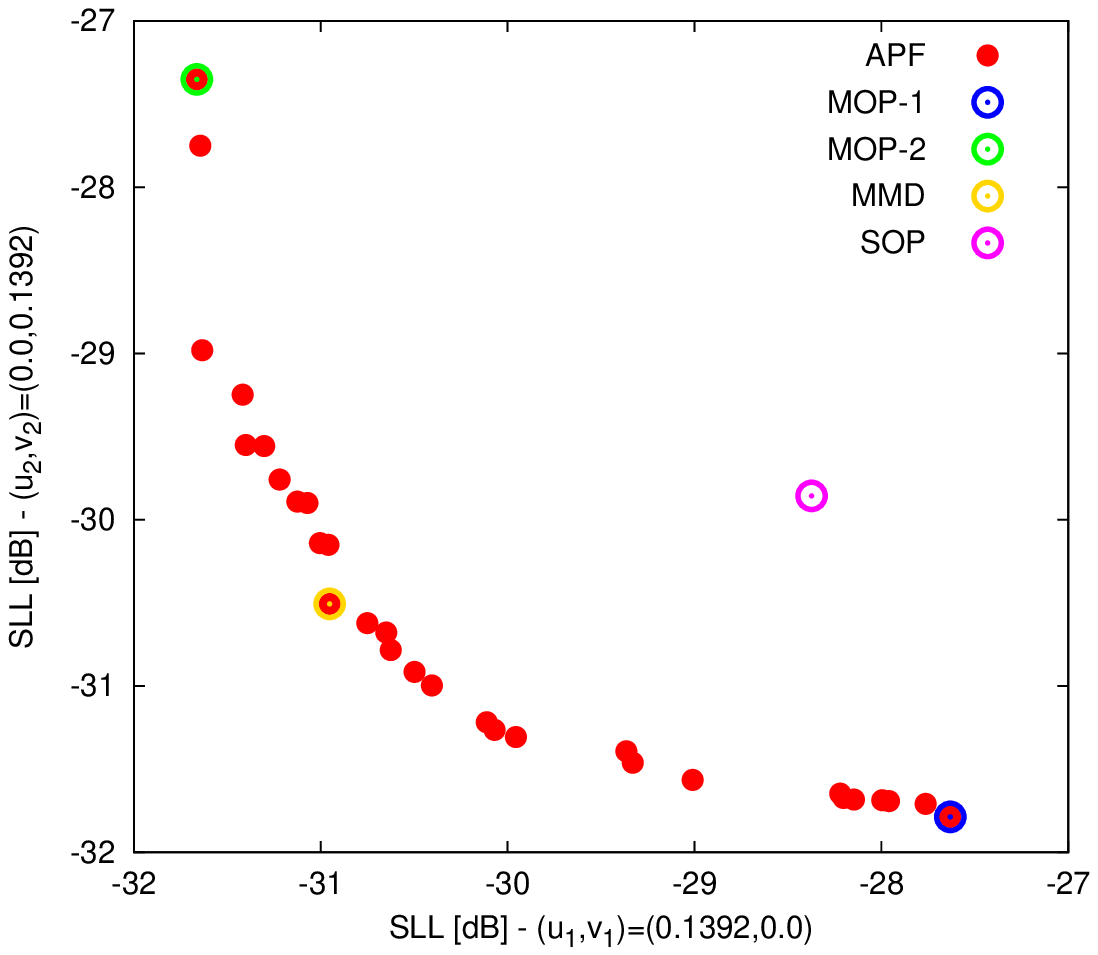}}\tabularnewline
\end{tabular}\end{center}

\begin{center}\textcolor{black}{~\vfill}\end{center}

\begin{center}\textbf{\textcolor{black}{Fig. 13 - P. Rocca et}} \textbf{\textcolor{black}{\emph{al.}}}\textbf{\textcolor{black}{,}}
\textbf{\textcolor{black}{\emph{{}``}}}\textcolor{black}{Pareto-Optimal
Domino-Tiling of ...''}\end{center}
\newpage

\begin{center}\textcolor{black}{~\vfill}\end{center}

\begin{center}\textcolor{black}{}\begin{tabular}{cc}
\multicolumn{2}{c}{\textcolor{black}{\includegraphics[%
  width=0.48\columnwidth,
  keepaspectratio]{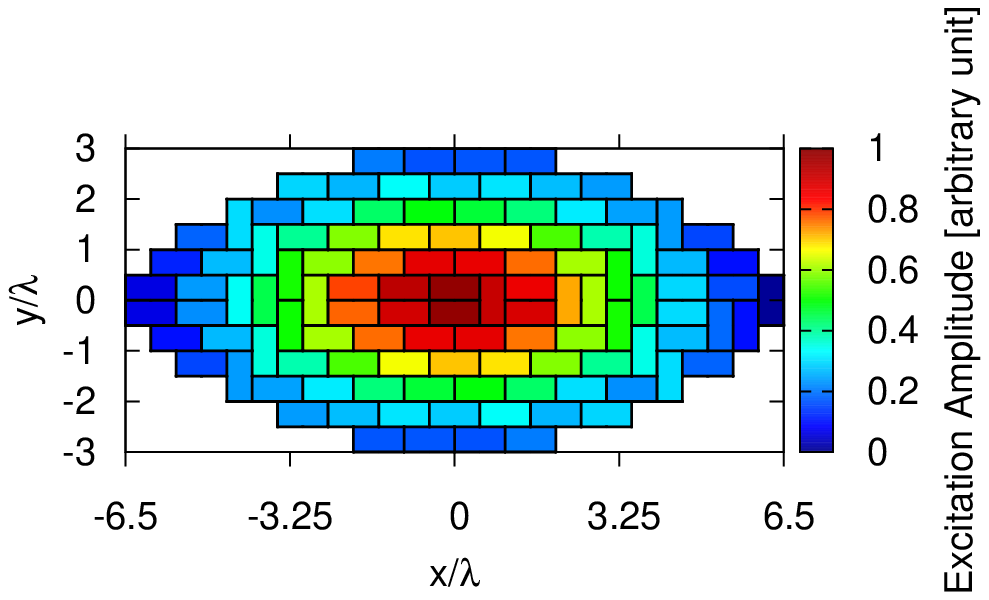}}}\tabularnewline
\multicolumn{2}{c}{\textcolor{black}{(}\textcolor{black}{\emph{a}}\textcolor{black}{)}}\tabularnewline
\multicolumn{2}{c}{}\tabularnewline
\textcolor{black}{\includegraphics[%
  width=0.48\columnwidth,
  keepaspectratio]{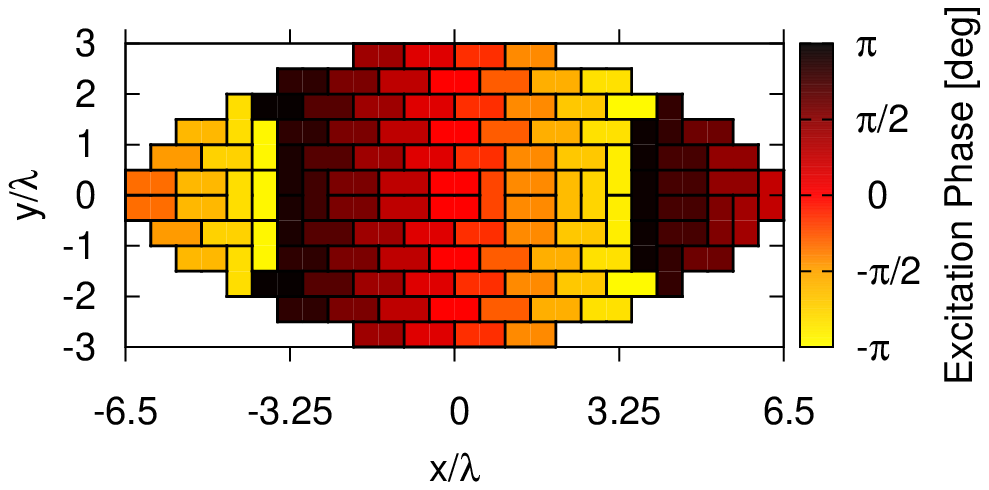}}&
\textcolor{black}{\includegraphics[%
  width=0.48\columnwidth,
  keepaspectratio]{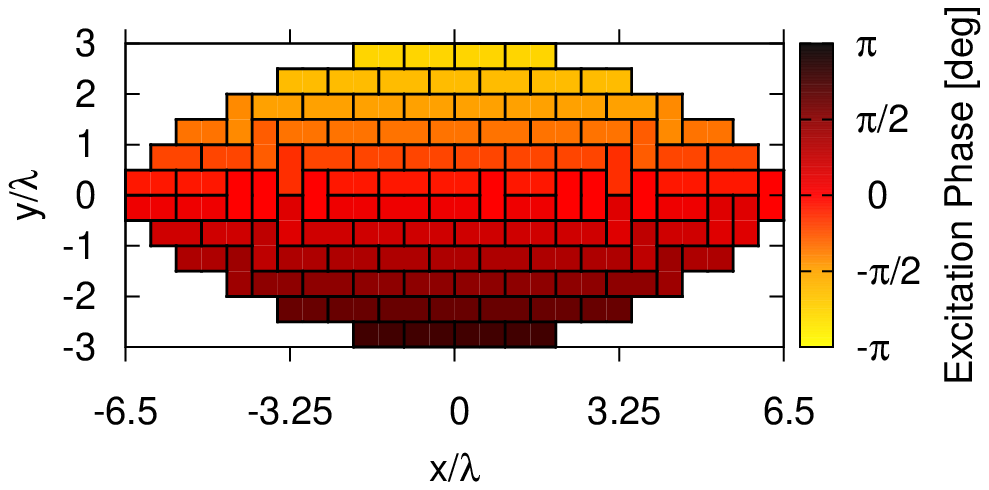}}\tabularnewline
\textcolor{black}{(}\textcolor{black}{\emph{b}}\textcolor{black}{)}&
\textcolor{black}{(}\textcolor{black}{\emph{c}}\textcolor{black}{)}\tabularnewline
&
\tabularnewline
\textcolor{black}{\includegraphics[%
  width=0.48\columnwidth,
  keepaspectratio]{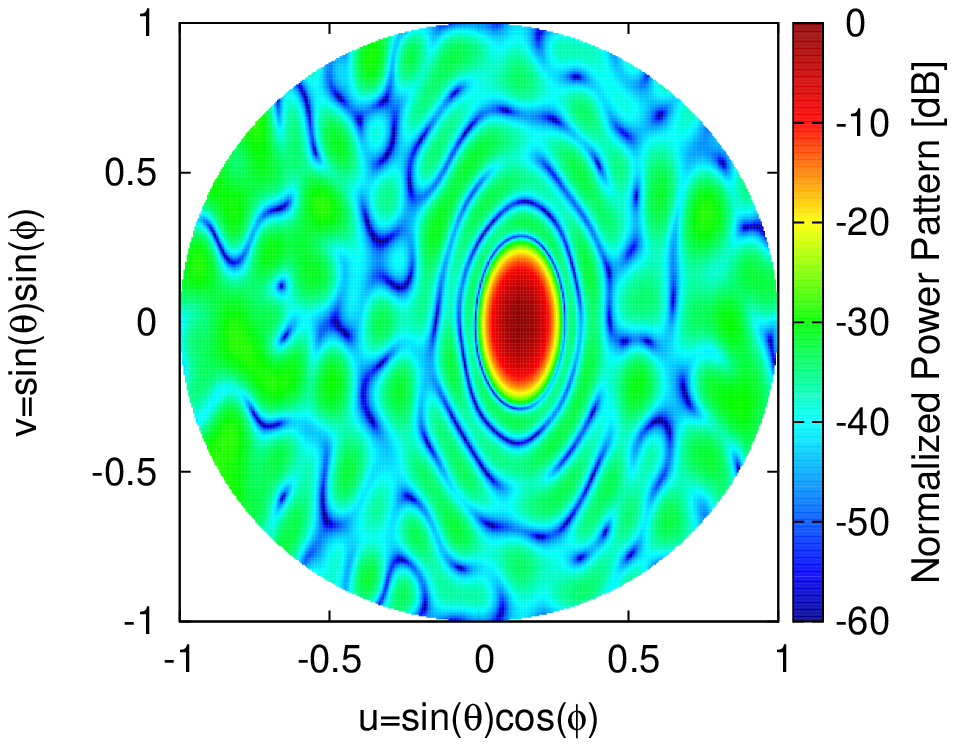}}&
\textcolor{black}{\includegraphics[%
  width=0.48\columnwidth,
  keepaspectratio]{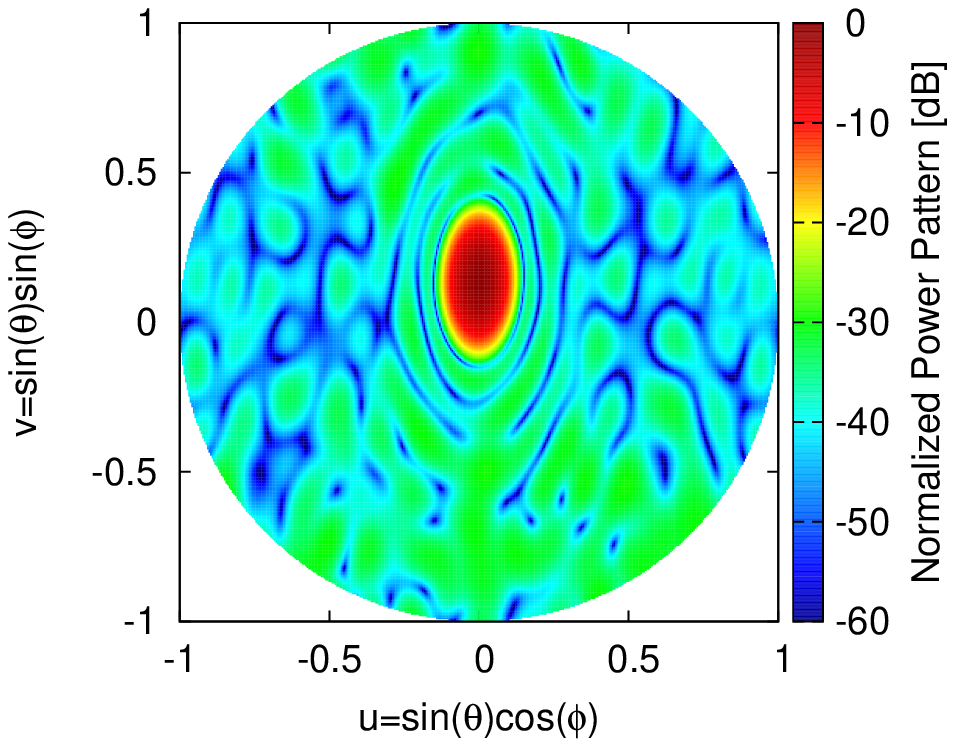}}\tabularnewline
\textcolor{black}{(}\textcolor{black}{\emph{d}}\textcolor{black}{)}&
\textcolor{black}{(}\textcolor{black}{\emph{e}}\textcolor{black}{)}\tabularnewline
\end{tabular}\end{center}

\begin{center}\textcolor{black}{~\vfill}\end{center}

\begin{center}\textbf{\textcolor{black}{Fig. 14 - P. Rocca et}} \textbf{\textcolor{black}{\emph{al.}}}\textbf{\textcolor{black}{,}}
\textbf{\textcolor{black}{\emph{{}``}}}\textcolor{black}{Pareto-Optimal
Domino-Tiling of ...''}\end{center}
\newpage

\begin{center}\textcolor{black}{~\vfill}\end{center}

\begin{center}\textcolor{black}{}\begin{tabular}{c}
\textcolor{black}{\includegraphics[%
  width=1.0\columnwidth,
  keepaspectratio]{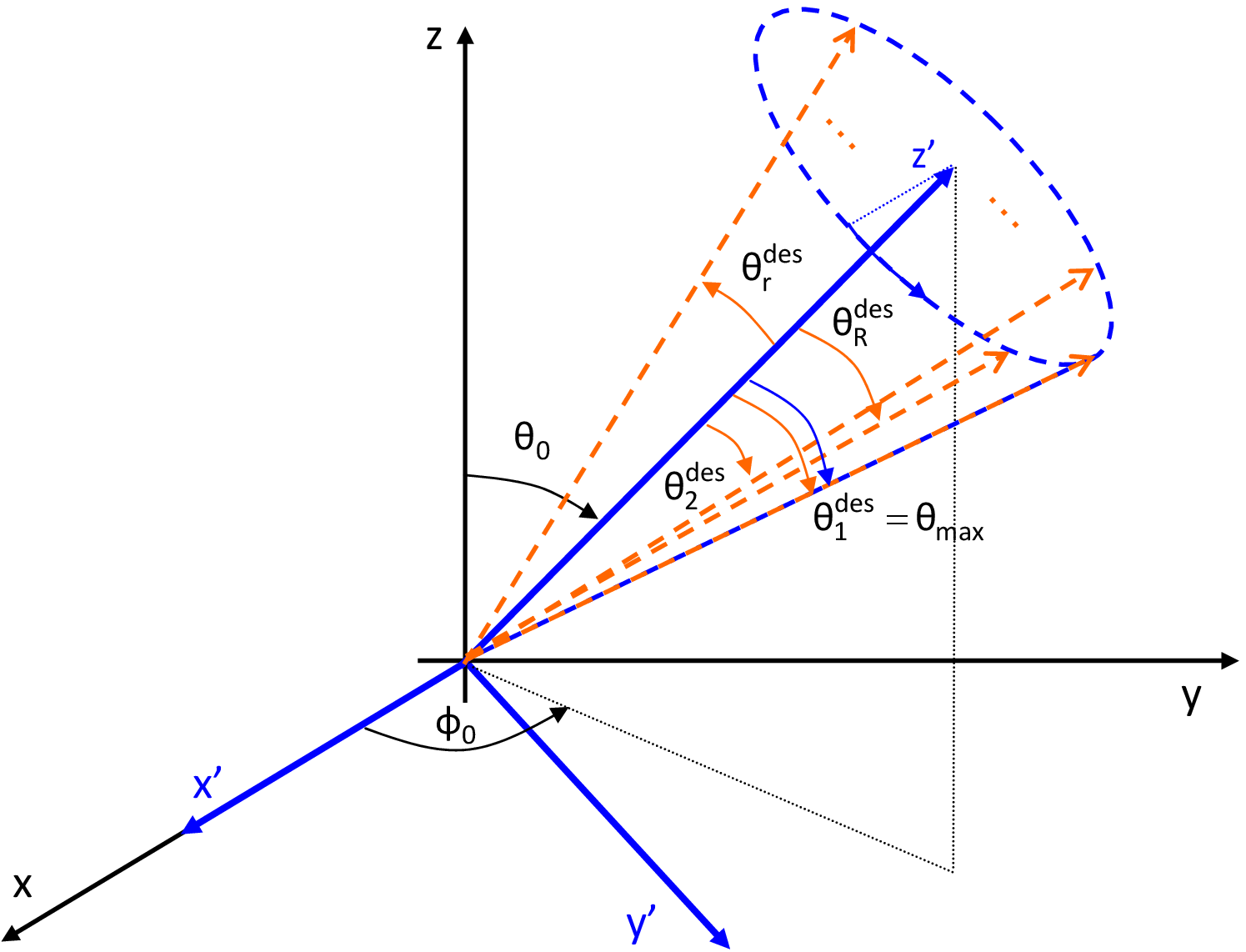}}\tabularnewline
\end{tabular}\end{center}

\begin{center}\textcolor{black}{~\vfill}\end{center}

\begin{center}\textbf{\textcolor{black}{Fig. 15 - N. Anselmi}} \textbf{\textcolor{black}{\emph{et
al.}}}\textbf{\textcolor{black}{,}} \textbf{\textcolor{black}{\emph{{}``}}}\textcolor{black}{Orthogonal
Polygon Array Design ...''}\end{center}
\newpage

\begin{center}\textcolor{black}{~\vfill}\end{center}

\begin{center}\textcolor{black}{}\begin{tabular}{c}
\textcolor{black}{\includegraphics[%
  width=0.80\columnwidth,
  keepaspectratio]{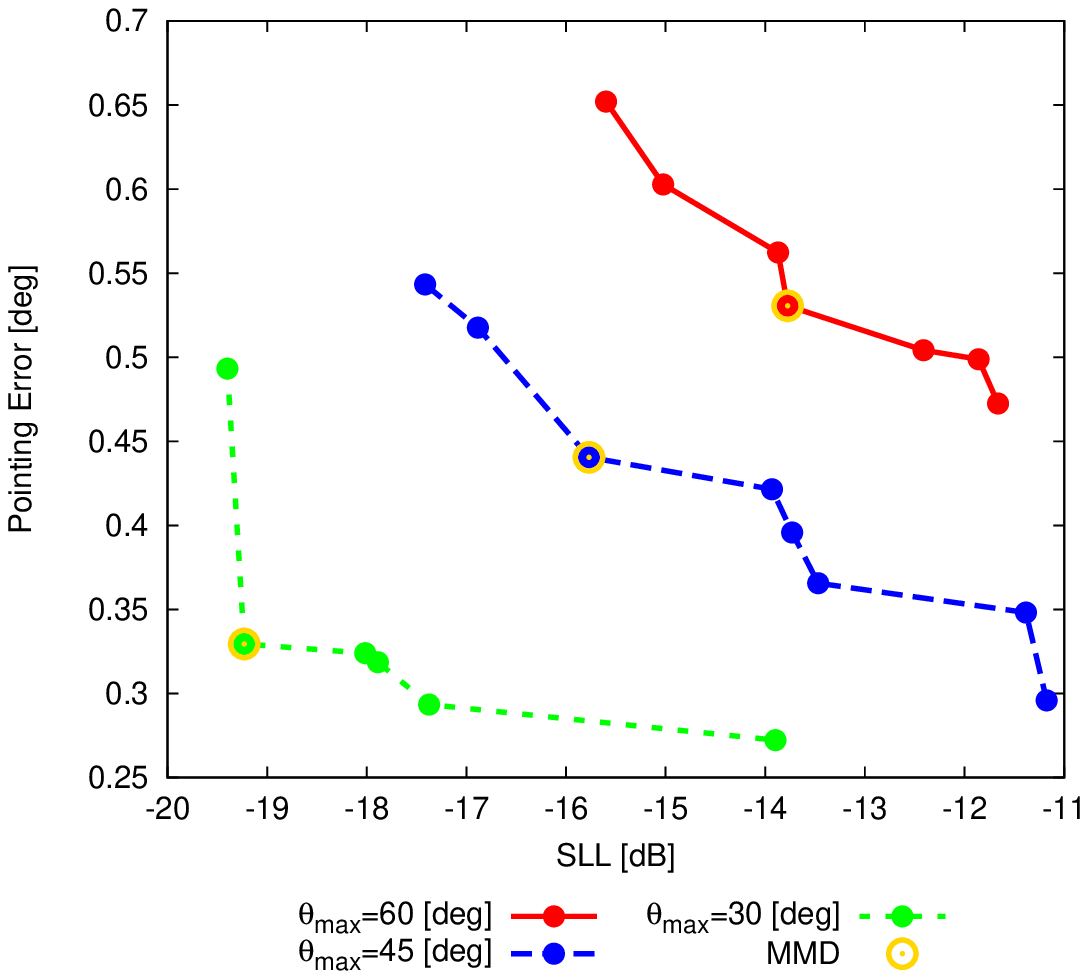}}\tabularnewline
\end{tabular}\end{center}

\begin{center}\textcolor{black}{~\vfill}\end{center}

\begin{center}\textbf{\textcolor{black}{Fig. 16 - P. Rocca et}} \textbf{\textcolor{black}{\emph{al.}}}\textbf{\textcolor{black}{,}}
\textbf{\textcolor{black}{\emph{{}``}}}\textcolor{black}{Pareto-Optimal
Domino-Tiling of ...''}\end{center}
\newpage

\begin{center}\textcolor{black}{}\begin{tabular}{c}
\textcolor{black}{\includegraphics[%
  width=0.50\columnwidth,
  keepaspectratio]{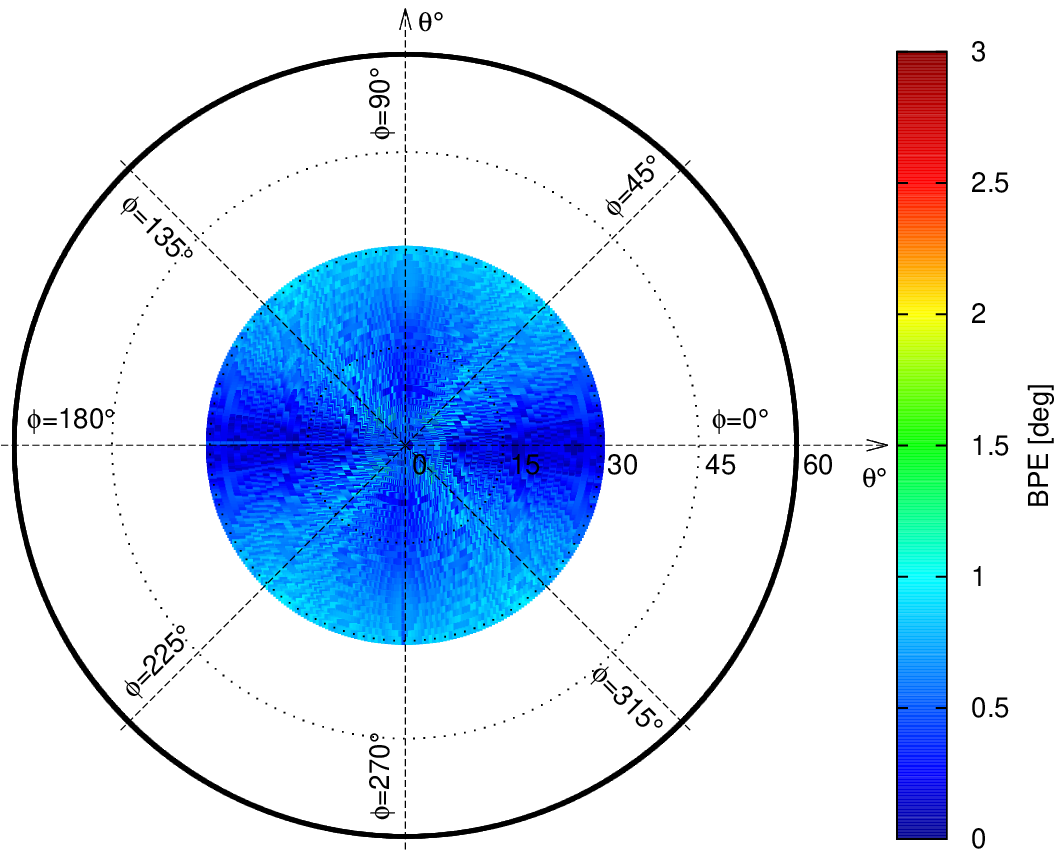}}\tabularnewline
(\emph{a})\tabularnewline
\textcolor{black}{\includegraphics[%
  width=0.50\columnwidth,
  keepaspectratio]{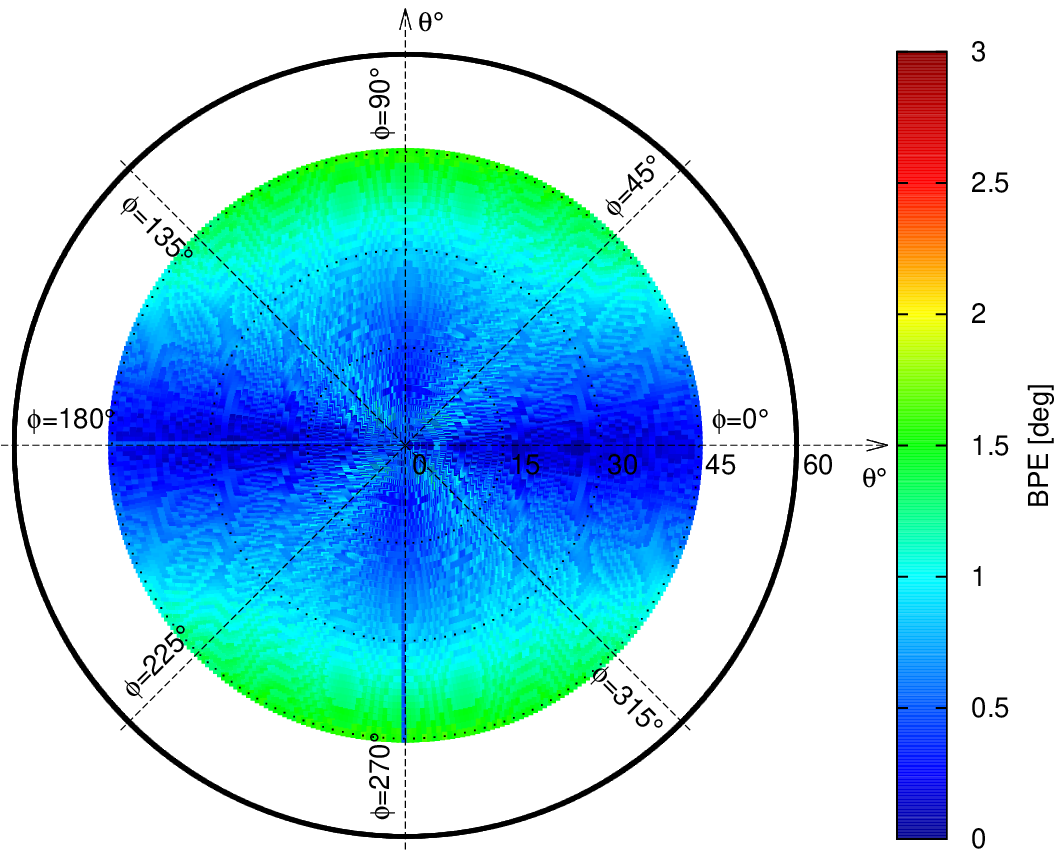}}\tabularnewline
(\emph{b})\tabularnewline
\textcolor{black}{\includegraphics[%
  width=0.50\columnwidth,
  keepaspectratio]{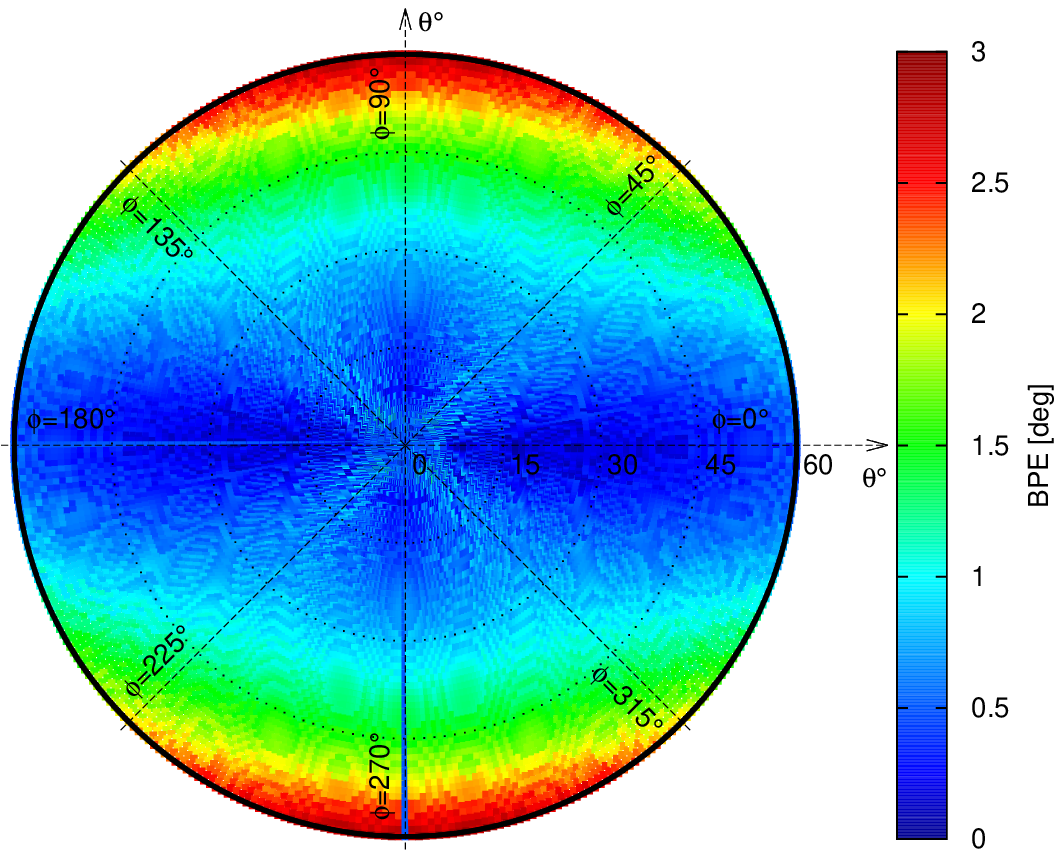}}\tabularnewline
(\emph{c})\tabularnewline
\end{tabular}\end{center}

\begin{center}\textcolor{black}{~\vfill}\end{center}

\begin{center}\textbf{\textcolor{black}{Fig. 18 - P. Rocca et}} \textbf{\textcolor{black}{\emph{al.}}}\textbf{\textcolor{black}{,}}
\textbf{\textcolor{black}{\emph{{}``}}}\textcolor{black}{Pareto-Optimal
Domino-Tiling of ...''}\end{center}
\newpage

\begin{center}\textcolor{black}{~\vfill}\end{center}

\begin{center}\textcolor{black}{}\begin{tabular}{cc}
\textcolor{black}{\includegraphics[%
  width=0.48\columnwidth,
  keepaspectratio]{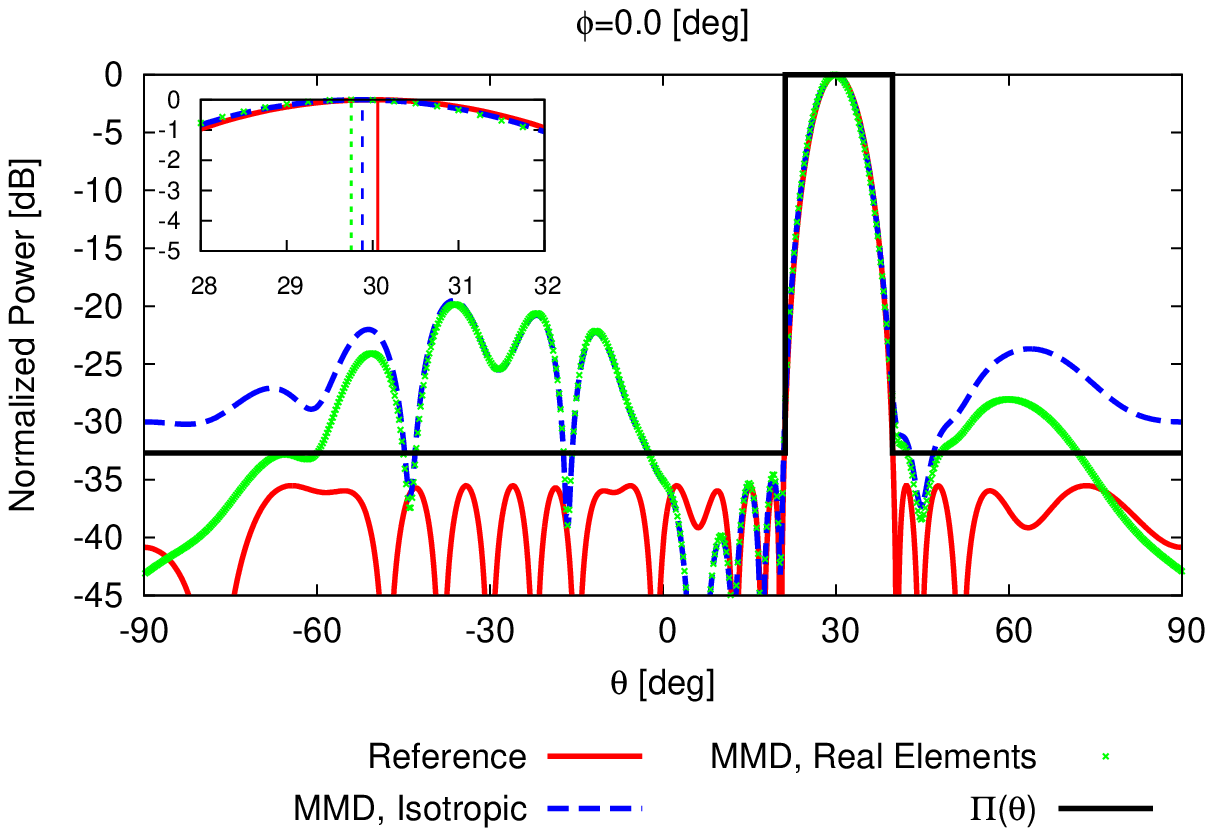}}&
\textcolor{black}{\includegraphics[%
  width=0.48\columnwidth,
  keepaspectratio]{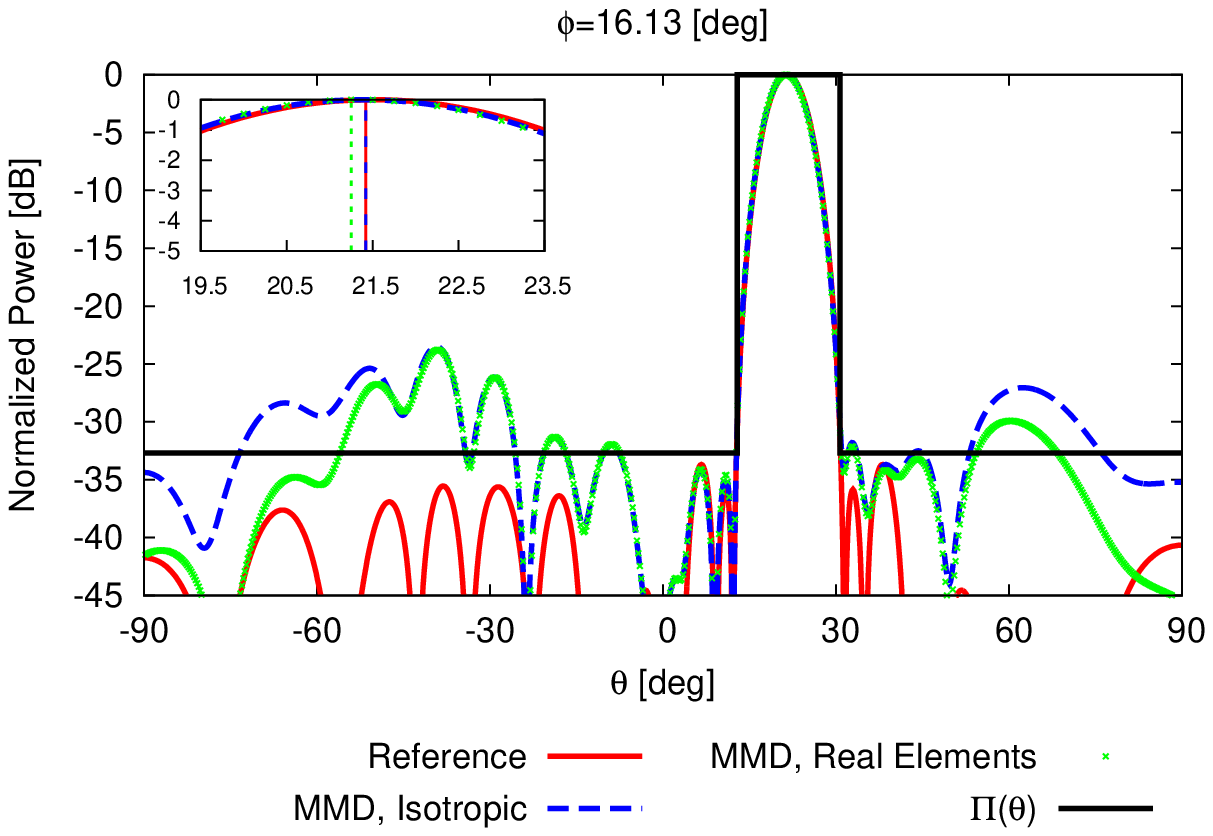}}\tabularnewline
\textcolor{black}{(}\textcolor{black}{\emph{a}}\textcolor{black}{)}&
\textcolor{black}{(}\textcolor{black}{\emph{b}}\textcolor{black}{)}\tabularnewline
&
\tabularnewline
\textcolor{black}{\includegraphics[%
  width=0.48\columnwidth,
  keepaspectratio]{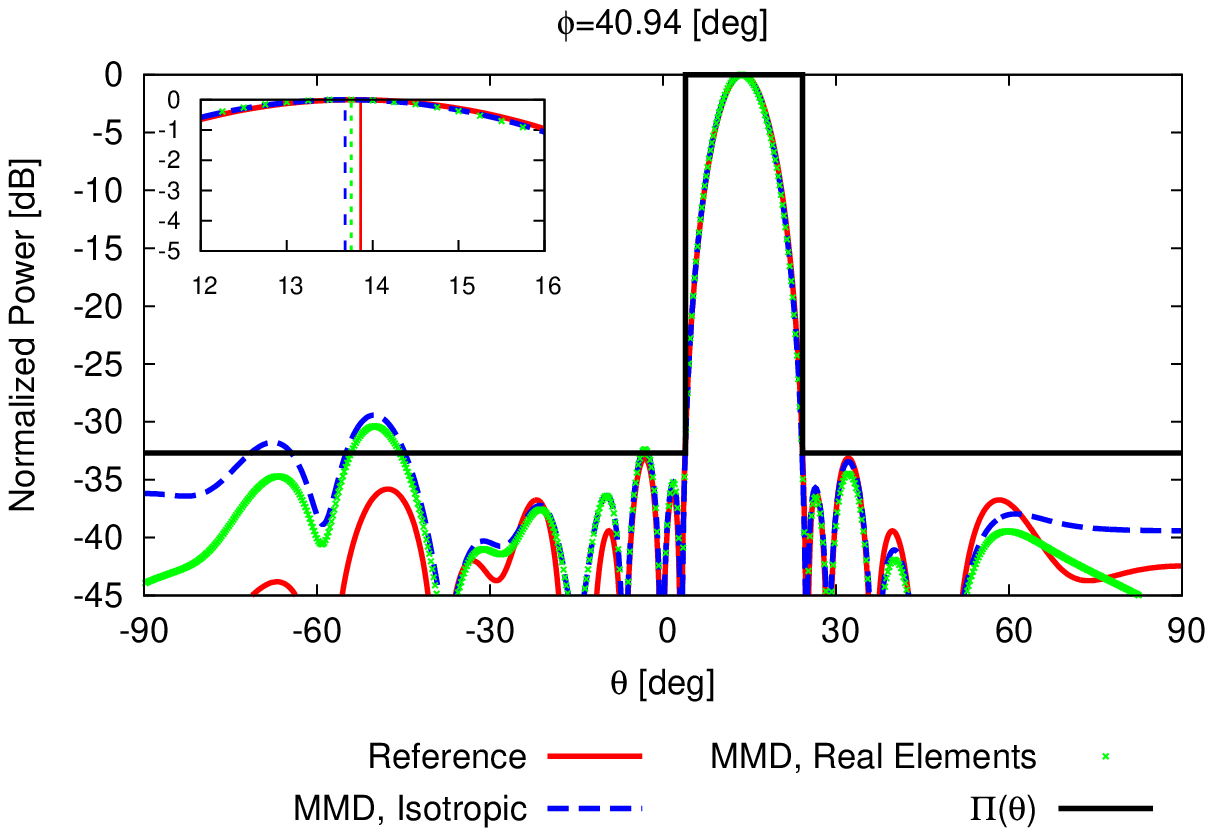}}&
\textcolor{black}{\includegraphics[%
  width=0.48\columnwidth,
  keepaspectratio]{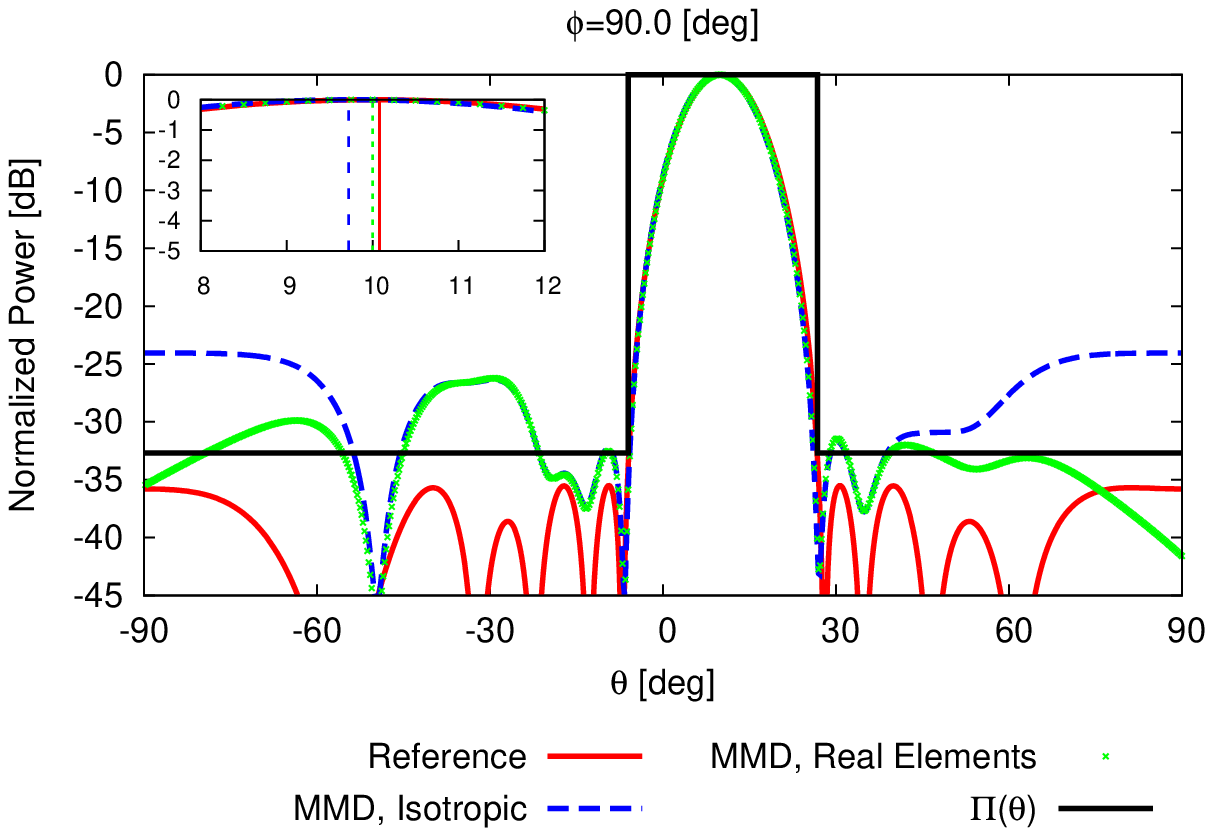}}\tabularnewline
\textcolor{black}{(}\textcolor{black}{\emph{c}}\textcolor{black}{)}&
\textcolor{black}{(}\textcolor{black}{\emph{d}}\textcolor{black}{)}\tabularnewline
\end{tabular}\end{center}

\begin{center}\textcolor{black}{~\vfill}\end{center}

\begin{center}\textbf{\textcolor{black}{Fig. 18 - P. Rocca et}} \textbf{\textcolor{black}{\emph{al.}}}\textbf{\textcolor{black}{,}}
\textbf{\textcolor{black}{\emph{{}``}}}\textcolor{black}{Pareto-Optimal
Domino-Tiling of ...''}\end{center}
\newpage

\begin{center}\textcolor{black}{~\vfill}\end{center}

\begin{center}\textcolor{black}{}\begin{tabular}{|c|c|c|c|c|}
\hline 
&
\textcolor{black}{$SLL$ }&
\textcolor{black}{$D$ }&
\textcolor{black}{$HPBW_{az}$ }&
\textcolor{black}{$HPBW_{el}$ }\tabularnewline
&
\textcolor{black}{{[}dB{]}}&
\textcolor{black}{{[}dBi{]}}&
\textcolor{black}{{[}deg{]} }&
\textcolor{black}{{[}deg{]}}\tabularnewline
\hline
\hline 
\textcolor{black}{\emph{Mask}}\textcolor{black}{~$\Pi$}&
\textcolor{black}{$-20.00$}&
\textcolor{black}{$-$}&
\textcolor{black}{$53.94$}&
\textcolor{black}{$34.92$}\tabularnewline
\hline
\textcolor{black}{\emph{Reference }}&
\textcolor{black}{$-20.00$}&
\textcolor{black}{$19.94$}&
\textcolor{black}{$21.12$}&
\textcolor{black}{$15.62$}\tabularnewline
\hline 
\textcolor{black}{\emph{MMD}}&
\textcolor{black}{$-19.62$}&
\textcolor{black}{$19.98$}&
\textcolor{black}{$21.19$}&
\textcolor{black}{$15.58$}\tabularnewline
\hline
\end{tabular}\end{center}

\begin{center}\textcolor{black}{~\vfill}\end{center}

\begin{center}\textbf{\textcolor{black}{Tab. I - P. Rocca et}} \textbf{\textcolor{black}{\emph{al.}}}\textbf{\textcolor{black}{,}}
\textbf{\textcolor{black}{\emph{{}``}}}\textcolor{black}{Pareto-Optimal
Domino-Tiling of ...''}\end{center}
\newpage

\begin{center}\textcolor{black}{~\vfill}\end{center}

\begin{center}\textcolor{black}{}\begin{tabular}{|c|c|c|c|c|}
\hline 
&
\textcolor{black}{$SLL$ }&
\textcolor{black}{$D$ }&
\textcolor{black}{$HPBW_{az}$ }&
\textcolor{black}{$HPBW_{el}$ }\tabularnewline
&
\textcolor{black}{{[}dB{]}}&
\textcolor{black}{{[}dBi{]}}&
\textcolor{black}{{[}deg{]} }&
\textcolor{black}{{[}deg{]}}\tabularnewline
\hline
\hline 
\textcolor{black}{\emph{Mask}}\textcolor{black}{~$\Pi$}&
\textcolor{black}{$-20.00$}&
\textcolor{black}{$-$}&
\textcolor{black}{$32.52$}&
\textcolor{black}{$32.53$}\tabularnewline
\hline
\hline 
&
\multicolumn{4}{c|}{\textcolor{black}{$\left(\theta_{1},\phi_{1}\right)=\left(0,0\right)$
{[}deg{]}}}\tabularnewline
\hline
\hline 
\textcolor{black}{\emph{Reference }}&
\textcolor{black}{$-20.60$}&
\textcolor{black}{$21.56$}&
\textcolor{black}{$14.92$}&
\textcolor{black}{$14.92$}\tabularnewline
\hline 
\textcolor{black}{\emph{MOP-1 }}&
\textcolor{black}{$-20.16$}&
\textcolor{black}{$21.57$}&
\textcolor{black}{$14.96$}&
\textcolor{black}{$14.96$}\tabularnewline
\hline 
\textcolor{black}{\emph{MOP-2}}&
\textcolor{black}{$-19.07$}&
\textcolor{black}{$21.73$}&
\textcolor{black}{$14.92$}&
\textcolor{black}{$15.09$}\tabularnewline
\hline 
\textcolor{black}{\emph{MMD}}&
\textcolor{black}{$-19.07$}&
\textcolor{black}{$21.73$}&
\textcolor{black}{$14.92$}&
\textcolor{black}{$15.09$}\tabularnewline
\hline
\hline 
&
\multicolumn{4}{c|}{\textcolor{black}{$\left(\theta_{2},\phi_{2}\right)=\left(30,0\right)$
{[}deg{]}}}\tabularnewline
\hline
\hline 
\textcolor{black}{\emph{Reference }}&
\textcolor{black}{$-20.60$}&
\textcolor{black}{$21.09$}&
\textcolor{black}{$17.31$}&
\textcolor{black}{$14.92$}\tabularnewline
\hline 
\textcolor{black}{\emph{MOP-1}}&
\textcolor{black}{$-10.65$}&
\textcolor{black}{$19.59$}&
\textcolor{black}{$16.54$}&
\textcolor{black}{$15.85$}\tabularnewline
\hline 
\textcolor{black}{\emph{MOP-2}}&
\textcolor{black}{$-19.07$}&
\textcolor{black}{$21.13$}&
\textcolor{black}{$17.31$}&
\textcolor{black}{$15.09$}\tabularnewline
\hline 
\textcolor{black}{\emph{MMD}}&
\textcolor{black}{$-19.07$}&
\textcolor{black}{$21.13$}&
\textcolor{black}{$17.31$}&
\textcolor{black}{$15.09$}\tabularnewline
\hline
\end{tabular}\end{center}

\begin{center}\textcolor{black}{~\vfill}\end{center}

\begin{center}\textbf{\textcolor{black}{Tab. II - P. Rocca et}} \textbf{\textcolor{black}{\emph{al.}}}\textbf{\textcolor{black}{,}}
\textbf{\textcolor{black}{\emph{{}``}}}\textcolor{black}{Pareto-Optimal
Domino-Tiling of ...''}\end{center}
\newpage

\begin{center}\textcolor{black}{~\vfill}\end{center}

\begin{center}\textcolor{black}{}\begin{tabular}{|c|c|c|c|c|}
\hline 
&
\textcolor{black}{$SLL$ }&
\textcolor{black}{$D$ }&
\textcolor{black}{$HPBW_{az}$ }&
\textcolor{black}{$HPBW_{el}$ }\tabularnewline
&
\textcolor{black}{{[}dB{]}}&
\textcolor{black}{{[}dBi{]} }&
\textcolor{black}{{[}deg{]} }&
\textcolor{black}{{[}deg{]} }\tabularnewline
\hline
\hline 
\textcolor{black}{\emph{Mask}}\textcolor{black}{~$\Pi$}&
\textcolor{black}{$-32.70$}&
\textcolor{black}{$-$}&
\textcolor{black}{$16.10$}&
\textcolor{black}{$24.24$}\tabularnewline
\hline
\hline 
&
\multicolumn{4}{c|}{\textcolor{black}{$\left(\theta_{1},\phi_{1}\right)=\left(8,0\right)$
{[}deg{]}}}\tabularnewline
\hline
\hline 
\textcolor{black}{$Reference$}&
\textcolor{black}{$-32.70$}&
\textcolor{black}{$27.15$}&
\textcolor{black}{$6.11$}&
\textcolor{black}{$12.03$}\tabularnewline
\hline 
\textcolor{black}{$SOP$}&
\textcolor{black}{$-28.37$}&
\textcolor{black}{$26.95$}&
\textcolor{black}{$6.15$}&
\textcolor{black}{$11.94$}\tabularnewline
\hline 
\textcolor{black}{$MOP-1$}&
\textcolor{black}{$-27.63$}&
\textcolor{black}{$27.05$}&
\textcolor{black}{$6.14$}&
\textcolor{black}{$12.03$}\tabularnewline
\hline 
\textcolor{black}{$MOP-2$}&
\textcolor{black}{$-31.66$}&
\textcolor{black}{$27.06$}&
\textcolor{black}{$6.14$}&
\textcolor{black}{$12.05$}\tabularnewline
\hline 
\textcolor{black}{$MMD$}&
\textcolor{black}{$-30.97$}&
\textcolor{black}{$27.03$}&
\textcolor{black}{$6.14$}&
\textcolor{black}{$12.04$}\tabularnewline
\hline
\hline 
&
\multicolumn{4}{c|}{\textcolor{black}{$\left(\theta_{2},\phi_{2}\right)=\left(8,90\right)$
{[}deg{]}}}\tabularnewline
\hline
\hline 
\textcolor{black}{$Reference$}&
\textcolor{black}{$-32.70$}&
\textcolor{black}{$27.11$}&
\textcolor{black}{$6.11$}&
\textcolor{black}{$12.15$}\tabularnewline
\hline 
\textcolor{black}{$SOP$}&
\textcolor{black}{$-29.86$}&
\textcolor{black}{$27.05$}&
\textcolor{black}{$6.10$}&
\textcolor{black}{$12.04$}\tabularnewline
\hline 
\textcolor{black}{$MOP-1$}&
\textcolor{black}{$-31.79$}&
\textcolor{black}{$27.11$}&
\textcolor{black}{$6.11$}&
\textcolor{black}{$12.13$}\tabularnewline
\hline 
\textcolor{black}{$MOP-2$}&
\textcolor{black}{$-27.35$}&
\textcolor{black}{$27.07$}&
\textcolor{black}{$6.10$}&
\textcolor{black}{$12.11$}\tabularnewline
\hline 
\textcolor{black}{$MMD$}&
\textcolor{black}{$-30.51$}&
\textcolor{black}{$27.10$}&
\textcolor{black}{$6.10$}&
\textcolor{black}{$12.13$}\tabularnewline
\hline
\end{tabular}\end{center}

\begin{center}\textcolor{black}{~\vfill}\end{center}

\begin{center}\textbf{\textcolor{black}{Tab. III - P. Rocca et}} \textbf{\textcolor{black}{\emph{al.}}}\textbf{\textcolor{black}{,}}
\textbf{\textcolor{black}{\emph{{}``}}}\textcolor{black}{Pareto-Optimal
Domino-Tiling of ...''}\end{center}
\newpage

\begin{center}\textcolor{black}{~\vfill}\end{center}

\begin{center}\textcolor{black}{}\begin{tabular}{|c|c|c|c|c|}
\hline 
\textcolor{black}{$r$}&
\multicolumn{3}{c|}{\textcolor{black}{$\theta_{r}$ {[}deg{]} }}&
\textcolor{black}{$\phi_{r}$ {[}deg{]}}\tabularnewline
\hline 
&
\textcolor{black}{($\theta_{max}=30$)}&
\textcolor{black}{($\theta_{max}=45$)}&
\textcolor{black}{($\theta_{max}=60$)}&
\textcolor{black}{-}\tabularnewline
\hline
\hline 
\textcolor{black}{$1$}&
\textcolor{black}{$30.00$}&
\textcolor{black}{$45.00$}&
\textcolor{black}{$60.00$}&
\textcolor{black}{$0.00$}\tabularnewline
\hline
\textcolor{black}{$2$}&
\textcolor{black}{$21.45$}&
\textcolor{black}{$32.18$}&
\textcolor{black}{$42.90$}&
\textcolor{black}{$16.13$}\tabularnewline
\hline
\textcolor{black}{$3$}&
\textcolor{black}{$13.82$}&
\textcolor{black}{$20.72$}&
\textcolor{black}{$27.63$}&
\textcolor{black}{$40.94$}\tabularnewline
\hline
\textcolor{black}{$4$}&
\textcolor{black}{$10.00$}&
\textcolor{black}{$15.00$}&
\textcolor{black}{$20.00$}&
\textcolor{black}{$90.00$}\tabularnewline
\hline
\textcolor{black}{$5$}&
\textcolor{black}{$13.82$}&
\textcolor{black}{$20.72$}&
\textcolor{black}{$27.63$}&
\textcolor{black}{$139.06$}\tabularnewline
\hline
\textcolor{black}{$6$}&
\textcolor{black}{$21.45$}&
\textcolor{black}{$32.18$}&
\textcolor{black}{$42.90$}&
\textcolor{black}{$163.87$}\tabularnewline
\hline
\textcolor{black}{$7$}&
\textcolor{black}{$30.00$}&
\textcolor{black}{$45.00$}&
\textcolor{black}{$60.00$}&
\textcolor{black}{$180.00$}\tabularnewline
\hline
\textcolor{black}{$8$}&
\textcolor{black}{$21.45$}&
\textcolor{black}{$32.18$}&
\textcolor{black}{$42.90$}&
\textcolor{black}{$196.13$}\tabularnewline
\hline
\textcolor{black}{$9$}&
\textcolor{black}{$13.82$}&
\textcolor{black}{$20.72$}&
\textcolor{black}{$27.63$}&
\textcolor{black}{$220.94$}\tabularnewline
\hline
\textcolor{black}{$10$}&
\textcolor{black}{$10.00$}&
\textcolor{black}{$15.00$}&
\textcolor{black}{$20.00$}&
\textcolor{black}{$270.00$}\tabularnewline
\hline
\textcolor{black}{$11$}&
\textcolor{black}{$13.82$}&
\textcolor{black}{$20.72$}&
\textcolor{black}{$27.63$}&
\textcolor{black}{$319.06$}\tabularnewline
\hline
\textcolor{black}{$12$}&
\textcolor{black}{$21.45$}&
\textcolor{black}{$32.18$}&
\textcolor{black}{$42.90$}&
\textcolor{black}{$343.87$}\tabularnewline
\hline
\end{tabular}\end{center}

\begin{center}\textcolor{black}{~\vfill}\end{center}

\begin{center}\textbf{\textcolor{black}{Tab. IV - P. Rocca et}} \textbf{\textcolor{black}{\emph{al.}}}\textbf{\textcolor{black}{,}}
\textbf{\textcolor{black}{\emph{{}``}}}\textcolor{black}{Pareto-Optimal
Domino-Tiling of ...''}\end{center}
\end{document}